\documentclass[authoryear,3p,11pt]{elsarticle}

\usepackage{amssymb}
\usepackage{rotating,longtable,boldline,amsmath}
\usepackage[header]{appendix}
\usepackage{mathrsfs}
\usepackage[normalem]{ulem}
\usepackage{color}

\newcommand \beq{\begin{equation}}
\newcommand \eeq{\end{equation}}
\newcommand \beqn{\begin{equation*}}
\newcommand \eeqn{\end{equation*}}

\newcommand{\id}{\mathrm{d}} %for integral d...
\renewcommand{\d}[2]{\frac{\id #1}{\id #2}} % for single derivatives
\newcommand{\dd}[2]{\frac{\id^2 #1}{\id #2^2}} % for double derivatives
\newcommand{\dt}[2]{\frac{\id^3 #1}{\id #2^3}} % for triple derivatives
\newcommand{\df}[2]{\frac{\id^4 #1}{\id #2^4}} % for fourth order derivatives

\newcommand{\pd}[2]{\frac{\partial #1}{\partial #2}} 
\newcommand{\pdd}[2]{\frac{\partial^2 #1}{\partial #2^2}} 
\newcommand{\pdt}[2]{\frac{\partial^3 #1}{\partial #2^3}} 
\newcommand{\pdf}[2]{\frac{\partial^4 #1}{\partial #2^4}} 
\journal{Journal of the Mechanics and Physics of Solids.}

\begin{document}

\begin{frontmatter}

%% Title, authors and addresses

%% use the tnoteref command within \title for footnotes;
%% use the tnotetext command for theassociated footnote;
%% use the fnref command within \author or \address for footnotes;
%% use the fntext command for theassociated footnote;
%% use the corref command within \author for corresponding author footnotes;
%% use the cortext command for theassociated footnote;
%% use the ead command for the email address,
%% and the form \ead[url] for the home page:
%% \title{Title\tnoteref{label1}}
%% \tnotetext[label1]{}
%% \author{Name\corref{cor1}\fnref{label2}}
%% \ead{email address}
%% \ead[url]{home page}
%% \fntext[label2]{}
%% \cortext[cor1]{}
%% \address{Address\fnref{label3}}
%% \fntext[label3]{}

\title{Dynamics of viscoelastic snap-through\footnote{\copyright ~2018. This manuscript version is made available under the CC-BY-NC-ND 4.0 license http://creativecommons.org/licenses/by-nc-nd/4.0/}}

%% use optional labels to link authors explicitly to addresses:
%% \author[label1,label2]{}
%% \address[label1]{}
%% \address[label2]{}

\author{Michael Gomez, Derek E. Moulton and Dominic Vella}

\address{Mathematical Institute, University of Oxford, Woodstock Rd, Oxford OX2 6GG, UK}

\begin{abstract}
%% Text of abstract
We study the dynamics of snap-through when viscoelastic effects are present. To gain analytical insight we analyse a modified form of the Mises truss, a single-degree-of-freedom structure, which features an `inverted' shape that snaps to a `natural' shape. Motivated by the anomalously slow snap-through shown by spherical elastic caps, we consider a thought experiment in which the truss is first indented to an inverted state and allowed to relax while a specified displacement is maintained; the constraint of an imposed displacement is then removed. Focussing on the dynamics for the limit in which the timescale of viscous relaxation is much larger than the characteristic elastic timescale, we show that two types of snap-through are possible: the truss either immediately snaps back over the elastic timescale or it displays `pseudo-bistability', in which it undergoes a slow creeping motion before rapidly accelerating. In particular, we demonstrate that accurately determining when pseudo-bistability occurs requires the consideration of inertial effects immediately after the indentation force is removed. Our analysis also explains many basic features of pseudo-bistability that have been observed previously in experiments and numerical simulations; for example, we show that pseudo-bistability occurs in a narrow parameter range at the bifurcation between bistability and monostability, so that the dynamics is naturally susceptible to critical slowing down. We then study an analogous thought experiment performed on a continuous arch, showing that the qualitative features of the snap-through dynamics are well captured by the truss model. In addition, we analyse experimental and numerical data of viscoelastic snap-through times reported previously in the literature. Combining these approaches suggests that our conclusions may also extend to more complex viscoelastic structures used in morphing applications. 
\end{abstract}

\begin{keyword}
%% keywords here, in the form: keyword \sep keyword
Snap-through \sep Buckling \sep Viscoelasticity \sep Bistability \sep Creep \sep Snap-through time
%% PACS codes here, in the form: \PACS code \sep code

%% MSC codes here, in the form: \MSC code \sep code
%% or \MSC[2008] code \sep code (2000 is the default)
\end{keyword}

\end{frontmatter}

%% \linenumbers

%% main text
\section{Introduction}
\label{sec:intro}

\subsection{Elastic snap-through}
Snap-through buckling is a striking instability in which an elastic object rapidly jumps from one state to another. Such instabilities are familiar from everyday life: umbrellas suddenly flip upwards on a windy day, while the leaves of the Venus flytrap store elastic energy slowly before abruptly snapping shut to catch prey unawares \citep{forterre2005}. Similarly, snap-through is harnessed to generate fast motions in technological applications ranging from fluidic actuators \citep{overvelde2015,gomez2017b,rothemund2018}, micro-scale switches \citep{krylov2008,ramachandran2016}, responsive surfaces \citep{holmes2007} and artificial heart valves \citep{goncalves2003}. In these applications, snap-through has proved to be particularly useful among other elastic instabilities, such as wrinkling and crumpling, due to its ability to convert energy stored slowly into fast motions in a highly reproducible way. 

Despite the ubiquity of snap-through in nature and engineering, its dynamics is not well understood, with classical work focussing on determining the onset of snap-through in simple elastic objects such as plates and shells \citep{bazant,patricio1998}. Because snap-through generically occurs when a system is initially in an equilibrium state that ceases to exist (a saddle-node/fold bifurcation), standard analytical techniques often cannot be used to study the dynamics. For example, it is generally not possible to perform a linear stability analysis to obtain an eigenvalue (natural frequency) that characterises the growth rate of the instability: beyond the fold point there ceases to be an equilibrium base state from which the system evolves. This is in contrast to the case when snap-through is caused by a bifurcation in which the equilibrium state becomes unstable without ceasing to exist  \citep{pandey2014,fargette2014}. The dynamics near a saddle-node bifurcation have been well studied in low dimensional systems, consisting of a few ordinary differential equations (ODEs), in various physical and biological settings \citep{strogatz1989,trickey1998,majumdar2013} including work on slow-fast systems --- see \cite{jones2012} and chapter $2$ of \cite{berglund2006} (and references therein). However, it is much more difficult to extend this to an elastic continuum described by partial differential equations (PDEs). For this reason, previous work has mainly relied on experiments and numerical simulations (e.g.~using commercially available finite element packages, or solutions of the governing PDEs using standard numerical methods) to quantitatively model the snap-through dynamics \citep{diaconu2009,santer2010,arrieta2011,brinkmeyer2012,loukaides2014}. Some progress has also been made using lumped mass-spring models \citep{carrella2008}, though there remains a general lack of analytical results in the literature, for example closed-form expressions for the time taken to snap-through in terms of the physical system parameters. Analytical insight would be of interest both from the perspective of fundamental science and also for applications of snap-through, as it provides a basis to control the dynamic response and guide more detailed simulations or experiments. 

\begin{figure}
\centering
\includegraphics[width=\textwidth]{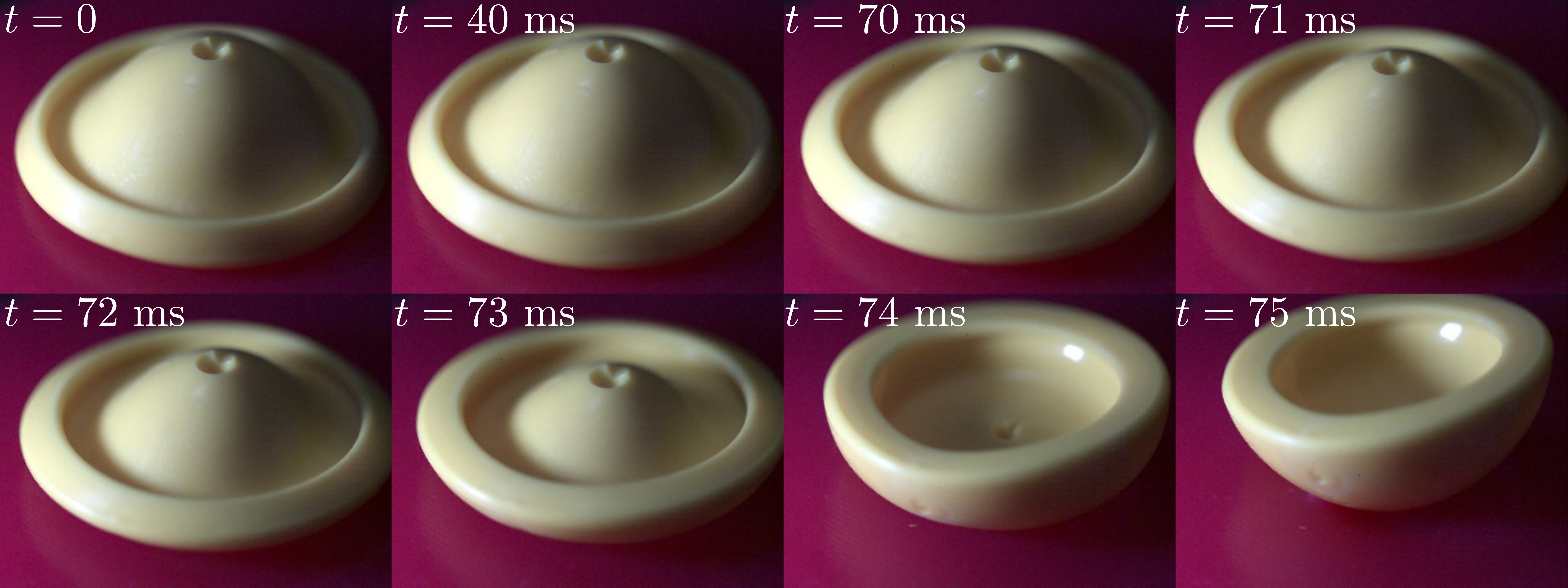} 
\caption{A jumping popper toy can be turned inside-out and released on a surface. It becomes unstable, and after a time delay ($\approx 70~\mathrm{ms}$ here) the popper rapidly snaps (in under $5~\mathrm{ms}$) back to its natural shape and leaps from the surface.}
\label{fig:poppersnaps}
\end{figure}

Moreover, some features of snap-through are not understood at a qualitative level, including delay phenomena: snap-through often occurs much more slowly than would be expected for an elastic instability. This slowness is illustrated by children's `jumping popper' toys, which resemble rubber spherical caps that can be turned inside-out. The inverted configuration remains stable while the cap is held at its edges, but leaving the popper on a surface causes it to snap back to its natural shape and leap upwards. As shown in figure \ref{fig:poppersnaps}, the snap back is not immediate: a time delay is observed during which the popper moves very slowly, apparently close to equilibrium, before rapidly accelerating. The delay can be several tens of seconds in duration --- much slower than the estimated elastic timescale, which is on the order of a millisecond \citep{gomezthesis}.

To explain such discrepancies between estimates of the speed of snap-through and that actually observed, it is commonly assumed that some dissipation mechanism must be present. For example in the Venus flytrap, the estimated elastic timescale is orders of magnitude faster than the observed snap-through time, and air damping is not enough to account for the discrepancy. In this case the proposed mechanism is poroelasticity \citep{forterre2005}: the snapping leaves are saturated with water and may dissipate energy via internal fluid flow. Similarly, morphing devices that demonstrate delayed snap-through are commonly composed of silicone-based  elastomers, which are known to exhibit viscoelastic behaviour \citep{brinkmeyer2012}.  It is also easily demonstrated that holding a popper toy for longer in its inverted state causes a slower snap-back --- an observation that is consistent with the importance of viscoelastic effects.

While attributing delayed snap-through to various dissipation mechanisms is natural, we recently demonstrated that anomalously slow dynamics are, in fact, possible in elastic systems with negligible dissipation \citep{gomez2017a}. In such scenarios, the time delay arises from the remnant or `ghost' of the snap-through bifurcation, reminiscent of the `critical slowing down' observed in other areas of physics such as phase transitions \citep{chaikin} and electrostatic `pull-in' instabilities \citep{gomez2018}: the saddle-node bifurcation continues to attract trajectories that are nearby in parameter space, producing a bottleneck whose duration increases without bound as the distance from the bifurcation decreases. Snap-through then appears to proceed much slower than the elastic timescale. In the process, we were able to propose new analytical formulae for the snap-through time as a function of the material parameters. Nevertheless, a key feature of this slowing down is that the system needs to be very close to the snap-through transition: the amount of delay that is experimentally attainable may in practice be small. Moreover, in viscoelastic systems, it is not clear what role viscoelastic effects play in obtaining anomalously slow snap-through dynamics, as opposed to the purely elastic slowing down. While we have previously considered the influence of external viscous damping (e.g.~due to air drag) \citep{gomezthesis}, viscoelastic behaviour is fundamentally different because it modifies the stability characteristics of structures. Here,  therefore, we seek to extend these studies to understand analytically how material viscosity affects the snap-through dynamics.

\subsection{Viscoelasticity and pseudo-bistability}
Unlike elastic solids, viscoelastic materials generally undergo stress relaxation when subject to a constant strain; this causes the effective stiffness of the structure to evolve in time. If a constant stress is imposed instead, the material may also exhibit a slow creeping motion \citep{howell}. \cite{santer2010} has demonstrated how these combined effects allow structures to exhibit `temporary bistability' or `pseudo-bistability' during snap-through. The idea of pseudo-bistability is that when a structure is held in a configuration that is near (but just beyond) a snap-through threshold, just as a popper toy may be held inside-out, the change in stiffness associated with stress relaxation may cause the structure to appear bistable (i.e.~an elastic structure with the same instantaneous stiffness would be bistable). When the structure is released, the stiffness recovers during a creeping motion, until eventually this bistability is lost and rapid snap-through occurs. Similar to the phenomenon of creep buckling \citep{hayman1978}, the total snap-through time is then governed by the viscous timescale of the material and can be very large. This phenomenon may be useful in morphing devices that are required to cycle continuously between two distinct states; for example, dimples proposed for aircraft wings that buckle in response to the air flow to reduce skin friction \citep{dearing2010,terwagne2014}, and ventricular assist devices which use snap-through of a spherical cap under a cyclic pneumatic load to pump blood \citep{goncalves2003}. In these applications, pseudo-bistability means that the actuation needed to move the structure between different states can be applied for a  shorter duration, which may lead to a significant reduction in the energy consumed \citep{santer2010}.

Using finite element simulations, \cite{santer2010} has demonstrated pseudo-bistability in a single-degree-of-freedom truss-like structure, as well as spherical caps similar to the jumping popper toy of figure \ref{fig:poppersnaps}. The phenomenon has been observed experimentally in spherical caps \citep{madhukar2014}  and truncated conical shells \citep{urbach2017}, and generically appears to occur only in a narrow parameter range near the transition to bistability, i.e.~the threshold at which snap-through no longer occurs. \cite{brinkmeyer2012} performed a systematic study of the snap-through dynamics of viscoelastic spherical caps, using a combination of finite element simulations and experiments. Continuing this work, \cite{brinkmeyer2013} studied the pseudo-bistable effect in viscoelastic arches. In these studies the phenomenon is found to have a number of common features, including (i) to obtain any time delay the structure needs to be held for a minimum period of time in an inverted state before release; and (ii) the resulting snap-through time depends sensitively on the parameters of the system and appears to diverge at the bistability transition. However, these basic features are not well understood quantitatively despite having important implications for applications of pseudo-bistability. The sensitivity of the snap-through time, for instance, means the system needs to be precisely tuned to obtain a desired response time. For this reason direct comparison between experiments and finite element simulations has revealed large quantitative errors \citep{brinkmeyer2012,brinkmeyer2013}. 

In addition, the numerical simulations referred to above are all based on two key assumptions regarding the viscoelastic response:  (i) the material behaves as though elastic with an effective stiffness that evolves in time, and (ii) the response during recovery is the reverse of the response during relaxation, i.e.~once the structure is released, the stiffness smoothly recovers to its initial, fully unrelaxed, value. These assumptions mean that modelling the dynamics is relatively simple compared to more general viscoelastic models, and the resulting equations are more easily implemented in commercially-available finite element packages. Furthermore, the different dynamical regimes can often be inferred by considering the elastic response in which the stiffness is fully unrelaxed and fully relaxed, as the instantaneous stiffness must be bounded between these two extremes \citep{santer2010}. However, the validity of these assumptions, and whether they can be justified from first principles, remains unclear.

An alternative approach is to start from the constitutive law of a viscoelastic solid, and derive the equations of motion that couple the stress to the deformation of the structure. While this approach is significantly more complicated, it eliminates the need to make any additional assumptions regarding the behaviour of the stiffness. This method has previously been used to obtain analytical expressions for the snap-through loads of simple viscoelastic structures \citep{nachbar1967}, as well as the conditions under which creep buckling occurs \citep{hayman1978}. More recently, \cite{urbach2018} developed a general theoretical framework for modelling viscoelastic snap-through based on a metric description of the constitutive equations. While this approach yields insight into the phenomenon of pseudo-bistability, the dynamics are modelled quasi-statically by neglecting the system inertia, so that it is unclear precisely when pseudo-bistable behaviour is obtained. Elsewhere, due to the inherent complexity of viscoelastic effects, it is unknown what role inertia plays in the dynamics and why the snap-through time appears to diverge near the snap-through transition. Are we simply observing another instance of critical slowing down, similar to the purely elastic dynamics studied by \cite{gomez2017a}? 

\subsection{Summary and structure of this paper}
In this paper, we aim to provide analytical understanding of the dynamics of viscoelastic snap-through, and in particular the features of pseudo-bistability. We consider a thought experiment in which a structure is indented to a specified displacement, and allowed to undergo stress relaxation before the indenter is abruptly removed. While we are motivated by continuous viscoelastic structures such as shells and arches, we first study a Mises truss for simplicity. This is a single-degree-of-freedom structure that exhibits bistability and snap-through, and enables us to make significant analytical progress. Focussing on the limit in which the timescale of viscous relaxation is much larger than the characteristic elastic timescale, we obtain three key results. (1) Inertial effects immediately after the indentation force is removed play an important role in determining when snap-through and pseudo-bistability occur. (2) While the intuitive picture of pseudo-bistability as being caused by a temporary change is stiffness is correct, the assumption of reversibility made in previous numerical studies (i.e.~that the stiffness smoothly reverses back to its fully unrelaxed value when the indenter is removed) leads to significantly different predictions of when snap-through occurs, compared to our first principles analysis. (3) Pseudo-bistability is a type of creeping motion governed by the viscous timescale, and so does not rely on critical slowing down to obtain slow dynamics, unlike  purely elastic snap-through \citep{gomez2017a}. Nevertheless, this creeping motion may be very slow indeed as the system may, in addition, be subject to critical slowing down in the pseudo-bistable regime. We then study a pre-buckled viscoelastic arch as an example of a more realistic structure that is used in morphing applications. Using direct numerical solutions, we show that the predictions of the truss model are qualitatively accurate for the arch system. This suggests that the analytical insight gained from the truss model may apply more broadly to the complex viscoelastic structures used in applications of snap-through.

The remainder of this paper is organised as follows. We begin in \S\ref{CH5sec:formulation} by deriving the equations governing the motion of the Mises truss. We then discuss the equilibrium states and the stress relaxation during indentation. In \S\ref{CH5sec:dynamicslargeDe}, we analyse the snap-through dynamics when the indenter is released, focussing on the limit in which the timescale of viscous relaxation is much longer than the characteristic elastic timescale. Using direct numerical solutions, we identify the different dynamical regimes and explain these asymptotically using the method of multiple scales. In \S\ref{sec:arch}, we perform simulations of a viscoelastic arch, showing that its snap-through behaviour is well captured by the truss model. In \S\ref{sec:datacompare}, we compare our predictions to experimental and numerical data of pseudo-bistable snap-through times reported in the literature. Finally, in \S\ref{CH5sec:conclusions}, we summarise our findings and conclude.

\section{A simple model system: Mises truss}
\label{CH5sec:formulation}
\begin{figure}
\centering
\includegraphics[width=\textwidth]{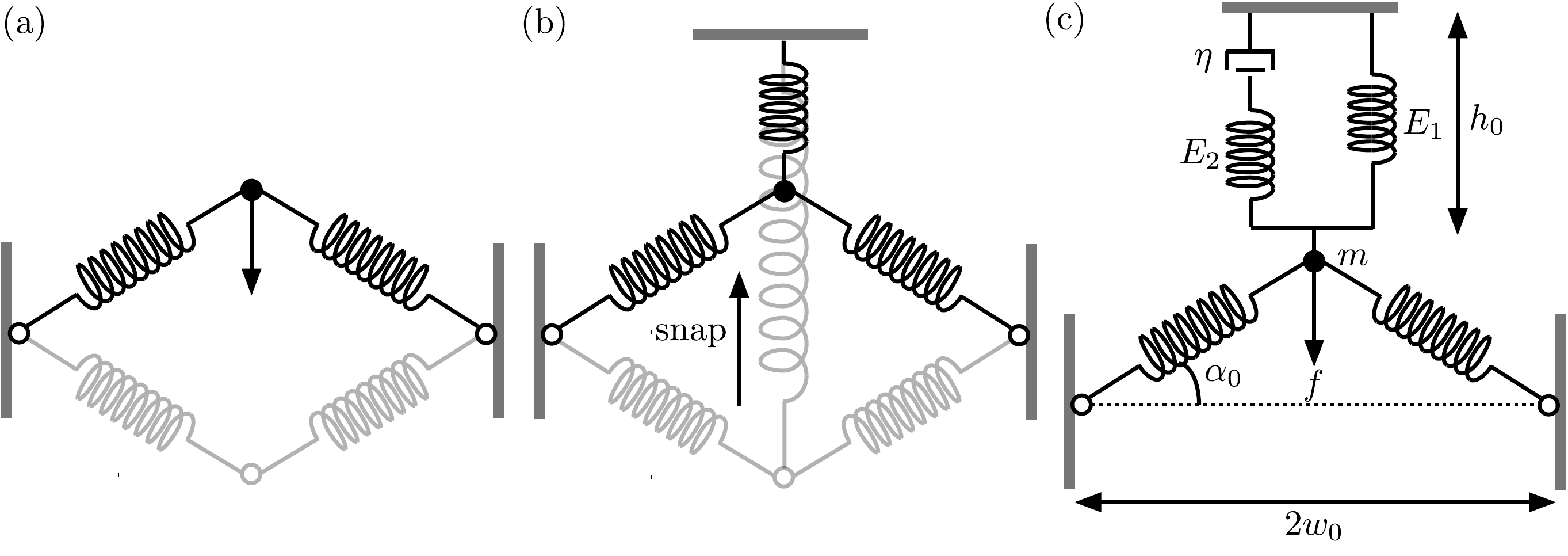} 
\caption{(a) The simplest form of the Mises truss, which features bistable `natural' (highlighted) and `inverted' (lightly shaded) equilibrium states. (b) This bistability is lost when an additional, linearly elastic, spring of sufficient stiffness is attached vertically to the point mass. (c) Replacing the vertical spring by a viscoelastic element, modelled as a standard linear solid (SLS), maintains the bistable--monostable behaviour.}
\label{CH5fig:trusssetup}
\end{figure}

As a first step towards understanding the dynamics of viscoelastic snap-through, we follow \cite{brinkmeyer2013} and consider a Mises truss (also referred to as a `von Mises truss'). In its simplest (elastic) form, this features two central springs, assumed to be linearly elastic, that are pin-jointed at their ends and inclined at a non-zero angle to the horizontal in their natural state. To give the system inertia, we place a point mass where the springs meet, and restrict the mass to move only in the vertical direction; see figure \ref{CH5fig:trusssetup}a.
%\noteMG{I think it's better to keep the truss first: the SLS element helps to develop intuition for stress relaxation, and the force-displacement curve helps to understand  pseudo-bistability and why we expect different regimes, etc. We also show that the nai\"{i}ve prediction of pseudo-bistability is invalid due to inertia, and it is not possible to obtain an equivalent analytical prediction for the arch. The arch simulations also have more parameters and technical details that get in the way, and the SLS element is more abstract.}

The truss in its current form is bistable: as well as the undeformed or `natural' state, the truss may be in equilibrium in a reflected state, where the length of each spring is unchanged from its rest length. However, by connecting an additional spring of sufficient stiffness to the point mass (figure \ref{CH5fig:trusssetup}b) \citep{krylov2008,panovko}, the inverted state ceases to be a stable equilibrium: in an experiment in which the truss is held fixed in an inverted position using an indenter, the truss will immediately snap back to its natural state when the indenter is released, independently of how long it is held. This snap-through is reminiscent of a spherical cap; in fact, we may consider the truss as a lumped model for a generic, continuous elastic structure that features an `inverted'  state that snaps back to a `natural' state. The central springs represent the membrane (stretching) stiffness, since these springs can be viewed as corresponding to the midsurface of the structure. The vertical spring models the bending stiffness \citep{krylov2008}: this spring penalises rotating the truss about its pin-jointed ends, mimicking bending the structure about its edges as it is loaded to an inverted position.

We now suppose that the vertical spring is viscoelastic. To choose an appropriate viscoelastic model, we note that a typical snap-through experiment includes both displacement-control and force-control: during indentation we impose a given displacement, but releasing the structure corresponds to imposing zero indentation force. It is therefore insufficient to describe the viscoelastic response using a Kelvin-Voigt or Maxwell model, since these fail to accurately capture both stress relaxation (under displacement-control) and creep (under force-control) behaviour. Instead, we shall use the constitutive law of a standard linear solid (SLS), which is the simplest model that describes both of these effects \citep{lakes}.  Physically, the SLS model is equivalent to placing a linear spring in parallel with a Maxwell element that features a second spring and a dashpot in series; see figure \ref{CH5fig:trusssetup}c. While we could also incorporate viscoelasticity of the central springs in our formulation, this would introduce additional viscous timescales and hence make it much more difficult to make analytical progress; we will show that our simpler model successfully captures snap-through and pseudo-bistability without additional complexity.

When the truss is indented and held for a specified time, stress relaxation causes the effective stiffness of the SLS element to decrease, so that the behaviour upon release is no longer obvious: the truss may immediately snap back, or it may initially creep in an inverted state for a period of time. In particular, these regimes cannot be inferred by only considering the equilibrium states of the system. We note that \cite{nachbar1967} have analysed a similar truss using a Kelvin-Voigt model, and determined the onset of snap-through to an inverted state when a constant indentation force is suddenly applied. Here, we are interested in the dynamics of the snap-back when the indentation force is removed.
 
\subsection{Governing equations} 
\label{sec:governingeqns}
As shown in figure \ref{CH5fig:trusssetup}c, in the natural state the central springs are assumed to be inclined at an angle $\alpha_0 > 0$ to the horizontal, and the springs are at their rest length; the distance between the pin joints at each base is $2 w_0$. We assume that the central springs are linearly elastic with constant stiffness $k$. The rest length of the vertical SLS element is $h_0$, the dashpot has viscosity $\eta$, and the upper springs have modulus $E_1$ (for the spring in parallel with the dashpot) and $E_2$ (for the spring in series). 

Let $x$ be the downward displacement of the point mass $m$ from the natural state. To obtain an equation of motion for $x$, we calculate the various forces exerted on the point mass. We write $\alpha$ and  $\Delta l$ for the corresponding inclination angle and change in length of the central springs, respectively. For simplicity, we assume that the truss remains shallow in shape, i.e.~$\alpha_0 \ll 1$ and $|\alpha| \ll 1$. Neglecting terms of $O(\alpha_0^3,\alpha^3)$, simple geometry gives that \citep{gomezthesis}
\beqn
\Delta l \approx \frac{x}{2 w_0}\left(x-2\alpha_0 w_0\right), \quad \alpha \approx \frac{1}{w_0}(\alpha_0 w_0 - x).
\eeqn
Because the central springs are linearly elastic, the vertical component of the total force exerted on the point mass by the central springs (directed downwards) is approximately $2 k\alpha \Delta l$. 

%To obtain an equation of motion for $x$, we calculate the various forces exerted on the point mass. Writing $\Delta l$ for the change in length of the central springs, simple geometry gives that
%\begin{eqnarray*}
%x & = & w_0\left(\tan\alpha_0-\tan\alpha\right), \\
%\Delta l & = & w_0\left(\frac{1}{\cos\alpha}-\frac{1}{\cos\alpha_0}\right). 
%\end{eqnarray*}
%For simplicity, we assume that the truss remains shallow in shape, i.e.~$\alpha_0 \ll 1$ and $|\alpha| \ll 1$. Neglecting terms of $O(\alpha_0^3,\alpha^3)$, we then have 
%\begin{eqnarray*}
%x & = & w_0\left(\alpha_0-\alpha\right), \\
%\Delta l & = & \frac{w_0}{2}\left(\alpha^2-\alpha_0^2\right). 
%\end{eqnarray*}
%The first equation shows that, for small angles, the vertical displacement $x$ is proportional to the change in the inclination angle $\alpha$. Hence, the vertical SLS element is analogous to a viscoelastic torsional spring at the base of each central spring, as was considered by \cite{brinkmeyer2013}. We also note that it is necessary to retain the quadratic nonlinearity in the second equation above; this provides the geometric nonlinearity necessary to obtain bistability and snap-through behaviour. Combining these relations, we obtain
%\beqn
%\Delta l = \frac{x}{2 w_0}\left(x-2\alpha_0 w_0\right).
%\eeqn
%\beqn
%2 k \Delta l \sin\alpha \approx \frac{k x}{w_0^2}\left(x-2\alpha_0w_0\right)\left(\alpha_0 w_0-x\right).
%\eeqn

The displacement $x$ also leads to a strain in the upper SLS element of size $e = x/h_0$. The  corresponding stress $\sigma$ satisfies the constitutive law of a standard linear solid \citep{lakes} (with $t$ denoting time)
\beq
\frac{(E_1+E_2)}{E_2} \d{e}{t} + \frac{E_1}{\eta}e = \frac{1}{E_2}\d{\sigma}{t} + \frac{1}{\eta}\sigma. \label{CH5eqn:SLSdim}
\eeq
Note that the limiting case of a Maxwell material is recovered by setting $E_1 = 0$, while the constitutive law for a Kelvin-Voigt material is recovered in the limit $E_2 \to \infty$. The stress $\sigma$ in the SLS element leads to a vertical force $-A \sigma$ exerted on the point mass (directed downwards), where $A$ is the cross-sectional area of each element. 

Combining the above, and also accounting for a downwards indentation force $f$, conservation of momentum gives
\beq
m\dd{x}{t} = \frac{k x}{w_0^2}\left(x-2\alpha_0w_0\right)\left(\alpha_0 w_0-x\right)-A\sigma + f.
 \label{CH5eqn:forcebalancedim}
\eeq
Together with appropriate initial conditions specified below, the coupled ODEs \eqref{CH5eqn:SLSdim}--\eqref{CH5eqn:forcebalancedim} (together with the relation $e = x/h_0$) provide a closed system to determine the trajectory $x(t)$ and stress $\sigma(t)$. 

\subsection{Non-dimensionalisation}
\label{sec:trussnondim}
To make the problem dimensionless, it is natural to scale the displacement with the initial height of the truss in the small-angle approximation, i.e.~$x \sim \alpha_0 w_0$. We scale time with the characteristic timescale of stress relaxation, $t \sim \eta/E_2$, obtained by balancing the final two terms in \eqref{CH5eqn:SLSdim}. Balancing the remaining terms in equations \eqref{CH5eqn:SLSdim}--\eqref{CH5eqn:forcebalancedim}, we introduce the dimensionless variables
\beqn
x = \alpha_0 w_0 X, \quad t = \frac{\eta}{E_2}T, \quad \sigma = \frac{E_1\alpha_0 w_0}{h_0}\Sigma, \quad f = k\alpha_0^3 w_0 F.
\eeqn
Here we have chosen the stress scale so that the constitutive equation for an elastic solid is simply $\Sigma = X$ in dimensionless variables. Inserting these scalings into \eqref{CH5eqn:SLSdim}, and eliminating the strain for the dimensionless displacement $X$, we obtain
\beq
\frac{1}{1-\beta}\d{X}{T} + X = \d{\Sigma}{T}+\Sigma, \label{CH5eqn:SLS}
\eeq
where we define
\beq
\beta = \frac{E_2}{E_1+E_2}. \label{CH5eqn:defnbeta}
\eeq
The parameter $\beta \in (0,1)$ plays a key role in the stability of viscoelastic structures \citep{urbach2018}, as it measures the degree of stress relaxation that occurs in response to a step increase in strain; we shall discuss this further below when considering the behaviour of the truss when the indenter is applied. The parameter $(1-\beta)$ may also be interpreted as the ratio of the timescale of stress relaxation ($t \sim \eta/E_2$) to the characteristic timescale over which creep occurs ($t \sim \eta[E_1+E_2]/[E_1 E_2]$), obtained by balancing the first two terms in equation \eqref{CH5eqn:SLSdim}. The related parameter $E_2/E_1 = \beta/(1-\beta)$ is also referred to as the relaxation strength. The value of $\beta$ is governed by the physical mechanisms causing viscoelastic behaviour, such as molecular processes (e.g.~molecular rearrangement in polymers) or the effects of coupled field variables (e.g.~fluid flow in poroelastic materials); for a detailed discussion see \cite{lakes}. Here we assume that $\beta$ is a known material constant, which can be measured experimentally using relaxation tests \citep{urbach2017}. We also note that in this non-dimensionalisation, the limit $\beta \to 1$ corresponds to a Maxwell material, since this is equivalent to setting $E_1 = 0$ in equation \eqref{CH5eqn:SLSdim}. This limiting case more closely resembles fluid-like behaviour in which the material has no preferred natural state and simply relaxes to the current configuration \citep{urbach2018}. The opposite limit $\beta \to 0$ corresponds to a purely elastic material, in which stress relaxation does not occur and the solution of \eqref{CH5eqn:SLS} is simply $\Sigma = X$ for all times. Note that the Kelvin-Voigt model (i.e.~sending $E_2 \to \infty$ in  \eqref{CH5eqn:SLSdim}) cannot be obtained in this non-dimensionalisation, since we have scaled time by the relaxation timescale $\eta/E_2$. (This limiting case can only be obtained by first rescaling time by the creep timescale before sending $\beta \to 1$.)

In terms of dimensionless variables, the momentum equation \eqref{CH5eqn:forcebalancedim} can be written as
\beq
\mathrm{De}^{-2}\dd{X}{T}  = X(X-2)(1-X)-\lambda \Sigma + F,  \label{CH5eqn:forcebalance}
\eeq
where we have introduced the dimensionless parameters
\beqn
\mathrm{De} = \alpha_0 \frac{\eta/E_2}{\sqrt{m/k}}, \quad \lambda = \frac{A E_1}{k h_0 \alpha_0^2}.
\eeqn
Here, the Deborah number $\mathrm{De}$ measures the ratio of the timescale of stress relaxation to the characteristic timescale of the experiment  \citep{howell}, which here is the timescale of elastic oscillations ($\sim \alpha_0^{-1}\sqrt{m/k}$). (Hence, in this non-dimensionalisation, the viscous timescale is $T = O(1)$ while the elastic timescale is $T = O(\mathrm{De}^{-1})$.) We may interpret $\lambda$ as the relative stiffness of the upper SLS element compared to the central springs. The cubic term on the right-hand side of \eqref{CH5eqn:forcebalance} represents the dimensionless force due to the central springs. As expected, this vanishes in the undeformed state $X = 0$, the reflected state $X = 2$ (when the central springs are also at their natural length), and the intermediate displacement $X  = 1$ when the springs are aligned horizontally --- in this state they are compressed but do not contribute any vertical force. 

\subsection{Steady solutions}
\label{sec:trusssteady}
\begin{figure}
\centering
\includegraphics[width=\textwidth]{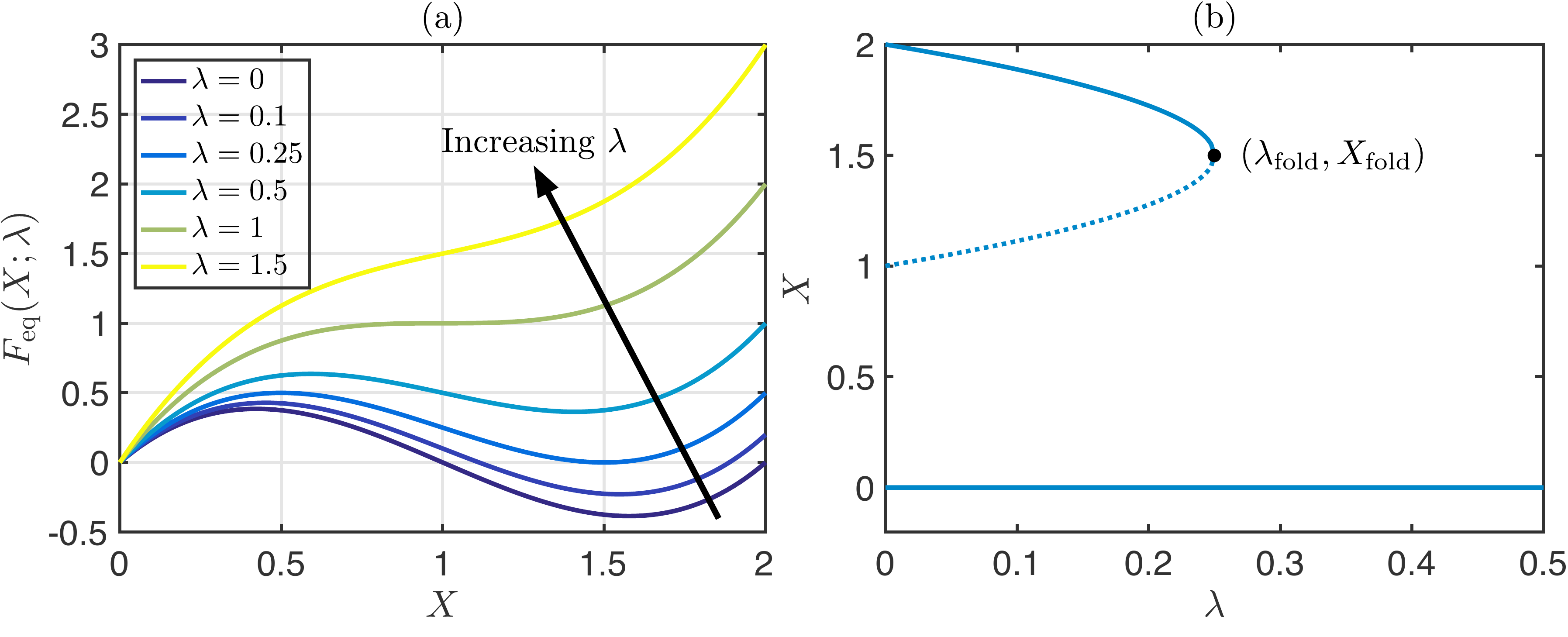} 
\caption{(a) The force-displacement curve for a truss in equilibrium: plotting the indentation force required to impose a steady displacement $X$ (coloured curves; see legend). At zero force, the truss is bistable for $\lambda < \lambda_{\mathrm{fold}} = 1/4$ and monostable for $\lambda > \lambda_{\mathrm{fold}}$. (b) Response diagram for the steady roots of $F_{\mathrm{eq}}(X;\lambda) = 0$ as $\lambda$ varies. At the critical value  $\lambda = \lambda_{\mathrm{fold}} = 1/4$, the stable non-zero root (upper solid curve) meets an unstable root (dotted curve) and disappears at a saddle-node bifurcation.}
\label{CH5fig:steadyresponse}
\end{figure}

When the system is in equilibrium with elastic constitutive law $\Sigma = X$, the momentum equation \eqref{CH5eqn:forcebalance} implies that the indentation force $F$ must balance the total force exerted by the central springs and the SLS element, which we label $F_{\mathrm{eq}}$. In particular, the force associated with a steady displacement $X$ is
\beqn
F_{\mathrm{eq}}(X;\lambda) \equiv -X(X-2)(1-X)+\lambda X.
\eeqn
When the indentation force is removed, any equilibria must satisfy $F_{\mathrm{eq}} = 0$, which has roots
\beqn
X = 0,\quad X = \frac{3\pm\sqrt{1-4\lambda}}{2}.
\eeqn
For $\lambda < 1/4$, there are two real non-zero solutions, which coincide and disappear at a saddle-node bifurcation when $\lambda = \lambda_{\mathrm{fold}} = 1/4$; the corresponding displacement at this point is $X = X_{\mathrm{fold}} = 3/2$. For $\lambda > \lambda_{\mathrm{fold}}$, the only real solution is the undeformed state, $X = 0$. This behaviour is apparent in figure \ref{CH5fig:steadyresponse}a, which plots the force-displacement curve for different values of $\lambda$; we see that increasing $\lambda$ (corresponding to a stiffer SLS element) acts to rotate the curve anti-clockwise about the origin, until eventually the turning point of the cubic lies above the line $F_{\mathrm{eq}} = 0$. The corresponding behaviour of the roots to $F_{\mathrm{eq}} = 0$ is shown in figure \ref{CH5fig:steadyresponse}b. It can be shown that the roots in which $F_{\mathrm{eq}}'(X;\lambda) > 0$ where $' = \id/\id X$ (solid branches on  figure \ref{CH5fig:steadyresponse}b) are linearly stable, while the root in which $F_{\mathrm{eq}}'(X;\lambda) < 0$ (dotted branch) is linearly unstable \citep{panovko}.

\subsection{Indentation response}
\label{CH5sec:indentation}
In a snap-through thought experiment, we imagine indenting the truss to an inverted state by imposing the constant displacement $X = X_{\mathrm{ind}} \geq 1$, for a time interval $-T_{\mathrm{ind}} < T < 0$ of duration $T_{\mathrm{ind}} > 0$ (for later convenience, we define $T = 0$ to be the time at which the indenter is released). To avoid introducing additional timescales into the problem, we suppose that the indentation is suddenly applied at $T = -T_{\mathrm{ind}}$, i.e.~over a timescale much faster than the viscous timescale $\eta/E_2$. We can then approximate the behaviour for $T < 0$ as
\beqn
X = X_{\mathrm{ind}}H(T+T_{\mathrm{ind}}),
\eeqn
where $H(\cdot)$ is the Heaviside step function. Substituting this into the constitutive equation  \eqref{CH5eqn:SLS}, and solving for the stress in the upper SLS element, we obtain
%\beqn
%X_{\mathrm{ind}}\left[\frac{1}{1-\beta}\delta(T+T_{\mathrm{ind}})+H(T+T_{\mathrm{ind}})\right] = \d{\Sigma}{T}+\Sigma, \qquad T < 0,
%\eeqn
%where $\delta(\cdot)$ is the Dirac delta-function. The solution is
\beqn
\Sigma = X_{\mathrm{ind}}\left[1+\frac{\beta}{1-\beta}e^{-(T+T_{\mathrm{ind}})}\right], \qquad -T_{\mathrm{ind}} < T < 0.
\eeqn
This solution is classical in the literature and represents the stress relaxation of a standard linear solid under a step increase in strain \citep{lakes,santer2010}: the stress initially (i.e.~at $T = -T_{\mathrm{ind}}$) jumps instantaneously to a fully unrelaxed value $\Sigma = X_{\mathrm{ind}}/(1-\beta)$ when the indentation is applied, and then decays exponentially to the fully relaxed value $\Sigma =  X_{\mathrm{ind}}$ associated with an elastic material. 

Inspecting the momentum equation \eqref{CH5eqn:forcebalance}, we see that the effect of this relaxation is to give an effective value of $\lambda$ that changes in time. The corresponding indentation force can be written as 
\beqn
F = F_{\mathrm{eq}}\left(X_{\mathrm{ind}};\lambda_{\mathrm{eff}}(T)\right), \qquad -T_{\mathrm{ind}} < T < 0,
\eeqn
where the effective value of $\lambda$ is 
\beqn
\lambda_{\mathrm{eff}}= \lambda\left[1+\frac{\beta}{1-\beta}e^{-(T+T_{\mathrm{ind}})}\right].
\eeqn
Note that $\lambda_{\mathrm{eff}}$ decreases from $\lambda_{\mathrm{eff}}(-T_{\mathrm{ind}}) = \lambda/(1-\beta)$ to $\lambda_{\mathrm{eff}}(\infty) = \lambda$ during relaxation. In terms of the force-displacement curve in figure \ref{CH5fig:steadyresponse}a, this corresponds to rotating the curve \emph{clockwise} as stress relaxation occurs, so that the indentation force decreases in time. From this picture, we anticipate that there are different dynamical regimes when the indenter is released, depending on the values of $\lambda$ and $T_{\mathrm{ind}}$. For $\lambda < (1-\beta)\lambda_{\mathrm{fold}}$, the turning point on the cubic lies below the line $F_{\mathrm{eq}} = 0$ in the fully unrelaxed state (since $\lambda_{\mathrm{eff}}(-T_{\mathrm{ind}}) < \lambda_{\mathrm{fold}}$), and moves further below this line as relaxation occurs. Hence the truss is bistable at the moment when the indenter is released, and we do not expect snap-through to occur if the indentation displacement is sufficiently close to the stable non-zero root of  $F_{\mathrm{eq}} = 0$. Similarly, for $\lambda = \lambda_{\mathrm{eff}}(\infty)  > \lambda_{\mathrm{fold}}$, the turning point lies above the line $F_{\mathrm{eq}} = 0$ when the structure is fully relaxed, and so the truss is always monostable; we expect snap-through to occur for any value of $X_{\mathrm{ind}}$ and $T_{\mathrm{ind}}$. For $(1-\beta)\lambda_{\mathrm{fold}} < \lambda < \lambda_{\mathrm{fold}}$, the turning point lies above the line $F_{\mathrm{eq}} = 0$ when the structure is fully unrelaxed (since $\lambda_{\mathrm{eff}}(-T_{\mathrm{ind}}) > \lambda_{\mathrm{fold}}$), but eventually decreases below this line as stress relaxation occurs. In particular, the truss is effectively bistable when the indenter is released (i.e.~$T = 0$) provided that  $\lambda_{\mathrm{eff}}(0) < \lambda_{\mathrm{fold}}$, which can be re-arranged to give
\beq
T_{\mathrm{ind}} > T_{\mathrm{ind}}^{\mathrm{crit}} =  \log\left[\frac{\beta \lambda}{(1-\beta)(\lambda_{\mathrm{fold}}-\lambda)}\right]. \label{CH5eqn:naivebdry}
\eeq
We then expect snap-through to generally not occur if the inequality \eqref{CH5eqn:naivebdry} is satisfied, and to occur otherwise. We will show that while this na\"{i}ve argument correctly accounts for different dynamical regimes, it fails to quantitatively predict when snap-through occurs because of the effects of inertia.

For later reference, we shall write $F_{\mathrm{ind}}$ for the value of the indentation force $F$ just before the indenter is released, i.e.~at $T = 0^-$. From above, this is given by
\beq
F_{\mathrm{ind}}  = F_{\mathrm{eq}}\left(X_{\mathrm{ind}};\lambda_{\mathrm{eff}}(0)\right). \label{CH5eqn:indentationstress,force}
\eeq
% \Sigma_{\mathrm{ind}} = X_{\mathrm{ind}}\left(1+e^{-T_{\mathrm{ind}}}\right),

\subsection{Dynamics of release}
At $T = 0$, the indenter is suddenly released so that the indentation force
\beqn
F = F_{\mathrm{ind}}\left[1-H(T)\right].
\eeqn
We solve the momentum equation \eqref{CH5eqn:forcebalance} for the corresponding stress $\Sigma$ and substitute this into the constitutive equation \eqref{CH5eqn:SLS}. After re-arranging, we obtain
\beq
\mathrm{De}^{-2}\left(\dt{X}{T}+\dd{X}{T}\right) + F_{\mathrm{eq}}'\left(X;\frac{\lambda}{1-\beta}\right)\d{X}{T} +F_{\mathrm{eq}}(X;\lambda) = 0, \quad T > 0. \label{CH5eqn:releaseode}
\eeq
Due to the presence of inertia, $X$ and $\dot{X}$ must be continuous across $T = 0$ (writing $\dot{} = \id/\id T$), giving the initial conditions
\beq
X(0+) = X_{\mathrm{ind}}, \quad \dot{X}(0+) = 0, \quad \mathrm{De}^{-2}\ddot{X}(0+) = -F_{\mathrm{ind}}.  \label{CH5eqn:releaseics}
\eeq
The jump in acceleration here is necessary to balance the discontinuity in the applied indentation force. 

%After re-arranging we obtain
%\beqn
%\mathrm{De}^{-2}\left(\dt{X}{T}+\dd{X}{T}\right) + F_{\mathrm{eq}}'\left(X;\frac{\lambda}{1-\beta}\right)\d{X}{T} +F_{\mathrm{eq}}(X;\lambda) = -F_{\mathrm{ind}}\left[\delta(T)+H(T)-1\right],
%\eeqn
%where $' = \id/\id X$. 

Currently, we have five dimensionless parameters in the problem: the Deborah number $\mathrm{De}$, relative stiffness $\lambda$, relaxation parameter $\beta$, indentation depth $X_{\mathrm{ind}}$, and indentation time $T_{\mathrm{ind}}$. Throughout this paper, we restrict our attention to indentation depths $1 \leq X_{\mathrm{ind}} \leq 2$. As baseline values we use  $\beta = 1/2$ (i.e.~both springs in the SLS element have equal modulus, $E_1 = E_2$) and $X_{\mathrm{ind}} = X_{\mathrm{fold}} = 3/2$, i.e.~the displacement at the saddle-node bifurcation; in this case the initial conditions  \eqref{CH5eqn:releaseics} are analogous to those studied by \cite{gomez2017a} for purely elastic snap-through. We expect to recover similar behaviour here in the elastic limit $\beta \to 0$, i.e.~we expect the dynamics are governed by the elastic timescale and only slow down considerably near the saddle-node bifurcation at $\lambda = \lambda_{\mathrm{fold}}$. However, for values $\beta > 0$, it is not clear when the dynamics are instead governed by viscous relaxation. To gain insight, we focus on the limit $\mathrm{De} \gg 1$, which corresponds to a relaxation timescale that is much slower than the elastic timescale. This is the relevant regime for many structures composed of rubbery polymers, such as silicone-based elastomers typically used in morphing devices \citep{brinkmeyer2012,brinkmeyer2013,urbach2017}, where molecular rearrangement underlying viscoelastic behaviour occurs over slow timescales \citep{lakes}.

\section{Snap-through dynamics: $\mathrm{De} \gg 1$}
\label{CH5sec:dynamicslargeDe}

\subsection{Numerical solution}
\begin{figure}
\centering
\includegraphics[width=0.7\textwidth]{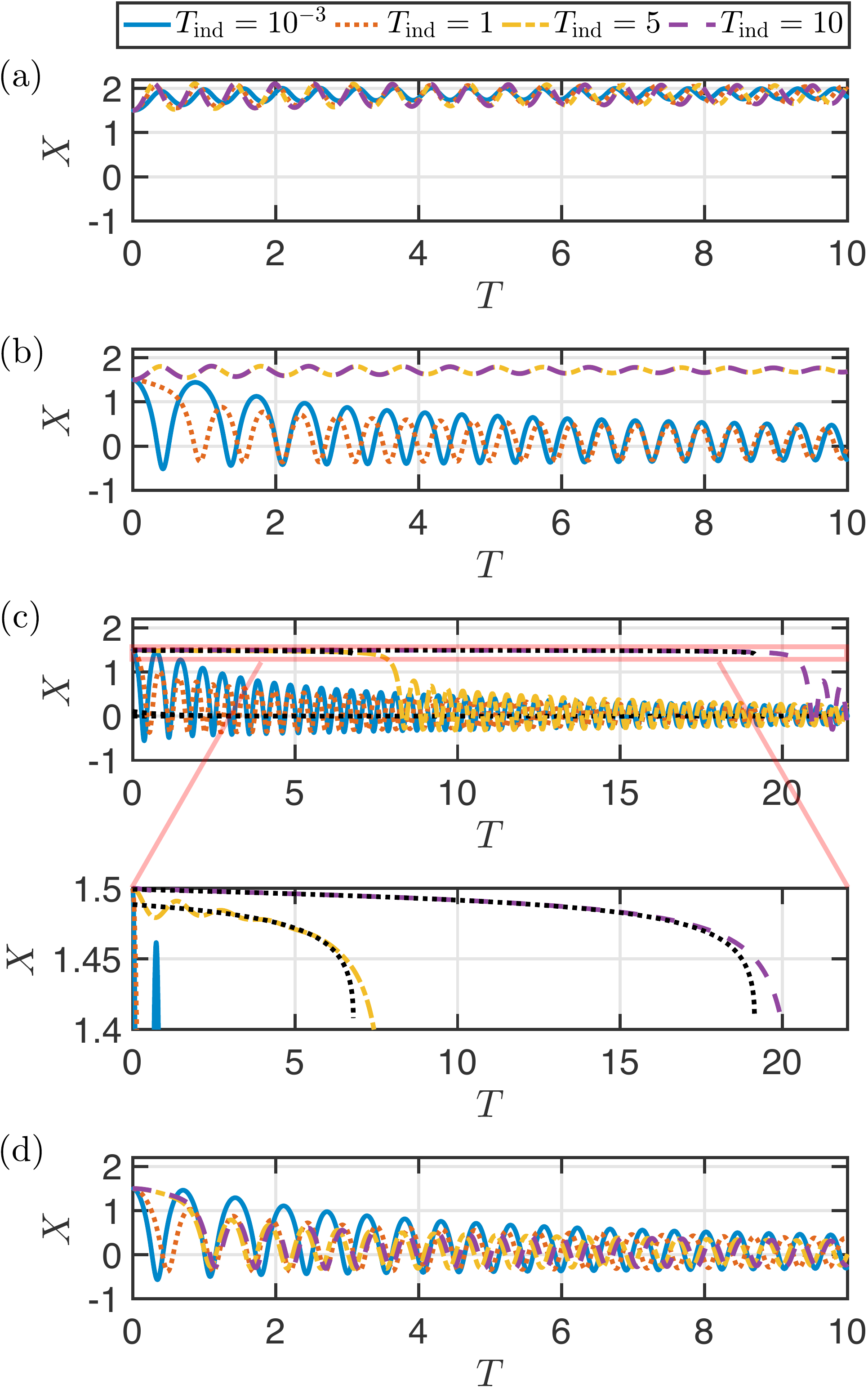} 
\caption{Dimensionless trajectories obtained by numerical integration of \eqref{CH5eqn:releaseode} with initial conditions  \eqref{CH5eqn:releaseics} (coloured curves). Here $X_{\mathrm{ind}} = X_{\mathrm{fold}}$, $\mathrm{De} = 10$, $\beta = 1/2$ and data is shown for (a) $\lambda = 0.1<(1-\beta)\lambda_{\mathrm{fold}}$, (b) $\lambda = 0.2\in\left((1-\beta)\lambda_{\mathrm{fold}},\lambda_{\mathrm{fold}}\right)$, (c) $\lambda = 0.2501=\lambda_{\mathrm{fold}}+10^{-4}$ (both panels; the lower panel displays a zoom), and (d) $\lambda = 0.26=\lambda_{\mathrm{fold}}+10^{-2}$. In each panel, trajectories associated with four different values of the indentation time $T_{\mathrm{ind}}$ (given in the upper legend) are shown. Note that in panels (c) the range of times plotted is larger and, for later reference, the predictions from the multiple-scale analysis are shown as black dotted curves.}
\label{CH5fig:largeDetrajectories}
\end{figure}

Typical dimensionless trajectories $X(T)$ in the limit $\mathrm{De} \gg 1$ are shown in figures \ref{CH5fig:largeDetrajectories}a--d. These are obtained by integrating the ODE \eqref{CH5eqn:releaseode} with initial conditions  \eqref{CH5eqn:releaseics} numerically in \textsc{matlab} (routine \texttt{ode45}, error tolerances $10^{-10}$ here and throughout). Figure \ref{CH5fig:largeDetrajectories} shows that the initial jump in acceleration causes oscillations to occur on the fast elastic timescale $T = O(\mathrm{De}^{-1})$; these oscillations persist due to the absence of external damping in our model. As anticipated from the discussion in \S\ref{CH5sec:indentation},  there are different regimes depending on the size of $\lambda$. For $\lambda \lesssim (1-\beta)\lambda_{\mathrm{fold}}$ ($= 1/8$ with $\beta = 1/2$), the truss appears never to snap and instead approaches the stable non-zero root of $F_{\mathrm{eq}} = 0$ (figure \ref{CH5fig:largeDetrajectories}a). For $(1-\beta)\lambda_{\mathrm{fold}} \lesssim \lambda \lesssim \lambda_{\mathrm{fold}}$, the truss snaps back to the natural state for small enough values of $T_{\mathrm{ind}}$, but remains in an inverted state indefinitely for larger $T_{\mathrm{ind}}$ (figure \ref{CH5fig:largeDetrajectories}b). For $\lambda \gtrsim \lambda_{\mathrm{fold}}$, the truss appears to snap for any value of $T_{\mathrm{ind}}$. However, the dynamics slow down considerably when $0 < \lambda-\lambda_{\mathrm{fold}} \ll 1$ and $T_{\mathrm{ind}}$ is sufficiently large; see figure \ref{CH5fig:largeDetrajectories}c. In this regime the oscillations are rapidly damped out, and the trajectory features an initial plateau before abruptly accelerating towards the natural configuration (highlighted in the lower panel of figure \ref{CH5fig:largeDetrajectories}c), reminiscent of the dynamical bottleneck caused by a saddle-node ghost \citep{gomez2017a}. For larger values of $\lambda$, the dependence on $T_{\mathrm{ind}}$ decreases and this initial bottleneck phase is not observed (figure \ref{CH5fig:largeDetrajectories}d).

\begin{figure}
\centering
\includegraphics[width=\textwidth]{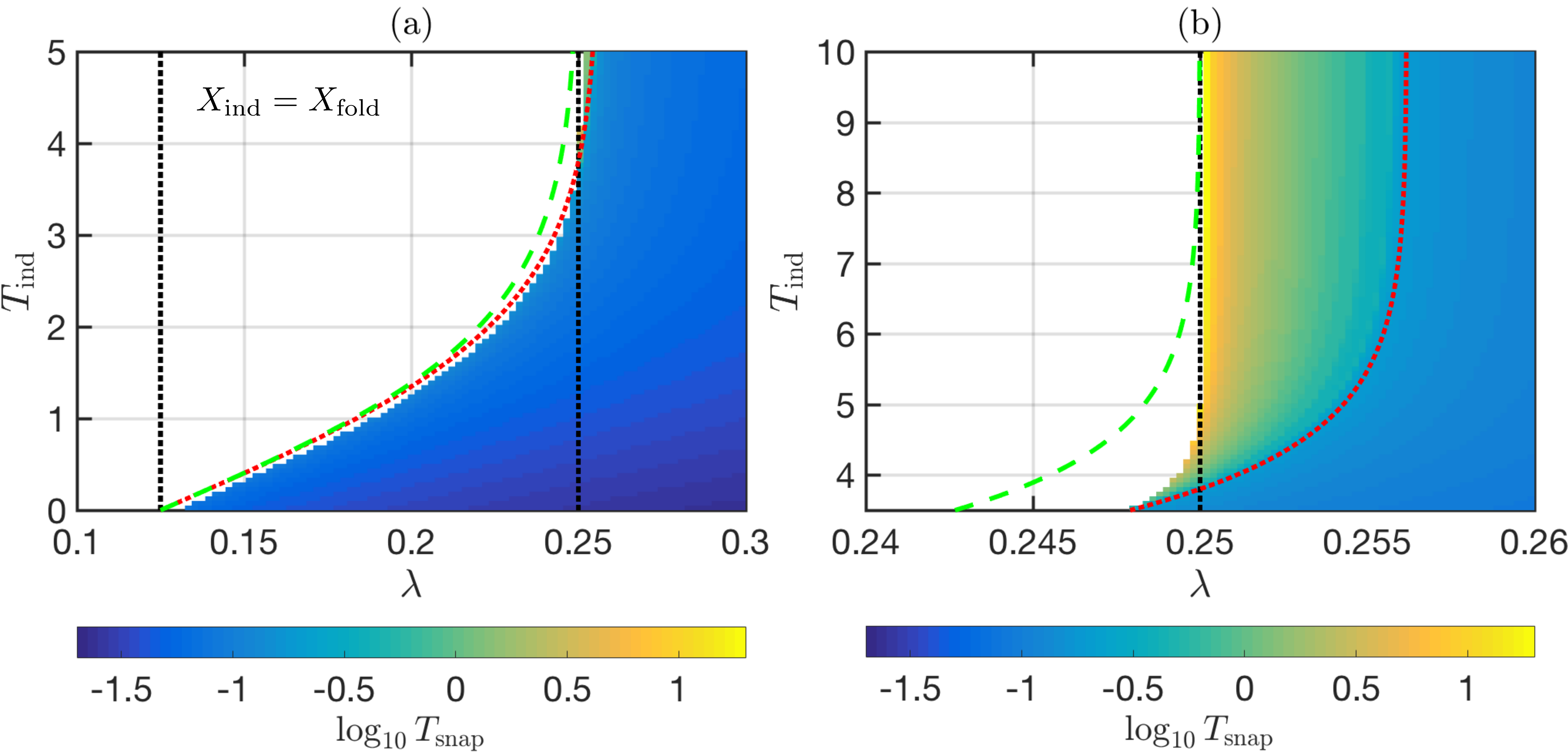} 
\caption{Snap-through times in the limit $\mathrm{De} \gg 1$ ($X_{\mathrm{ind}} = X_{\mathrm{fold}}$, $\mathrm{De} = 100$, $\beta = 1/2$). (a) Numerical results obtained by integrating  \eqref{CH5eqn:releaseode}--\eqref{CH5eqn:releaseics} until the point where $X = 0$ (see colourbar). The critical values $\lambda = (1-\beta)\lambda_{\mathrm{fold}}$ and $\lambda = \lambda_{\mathrm{fold}}$ are plotted as vertical black dotted lines. Also shown is the boundary separating snap-through/no snap-through predicted by \eqref{CH5eqn:naivebdry} (green dashed curve), and, for later reference, the boundary predicted by equation \eqref{CH5eqn:largeXindboundary} in \S\ref{CH5sec:multiplescale} (red dotted curve). (b) A close-up of the region where the dynamics slow down significantly. In each panel, the snap-through times have been computed on a $100\times 100$ grid of equally spaced values in the region displayed.}
\label{CH5fig:largeDesnapsurface}
\end{figure}

%(While this definition allows us to measure the dynamics in a consistent way, we note that the value is somewhat influenced by variations in the oscillations that occur on the $O(\mathrm{De}^{-1})$ timescale.)
These regimes are confirmed when we analyse the snap-through time, $T_{\mathrm{snap}}$ (defined as the time at which the displacement first crosses the natural displacement, $X = 0$); the computed times are shown on the  $(\lambda,T_{\mathrm{ind}})$-plane for the baseline parameter values in figure \ref{CH5fig:largeDesnapsurface}a. The blank regions on the figure correspond to regions where snap-through does not occur (after integrating the equations up to $T = 50$, which was found to be sufficient due to the limited amount of slowing down in figure \ref{CH5fig:largeDesnapsurface}a). This shows that the critical value of  $T_{\mathrm{ind}}$ at which snap-through no longer occurs with $(1-\beta)\lambda_{\mathrm{fold}} \lesssim \lambda \lesssim \lambda_{\mathrm{fold}}$ increases nonlinearly as $\lambda$ increases, and appears to approach a finite value $T_{\mathrm{ind}} \approx 4$ as $\lambda \to \lambda_{\mathrm{fold}}$. For comparison, we have also plotted the na\"{i}ve prediction \eqref{CH5eqn:naivebdry} based on whether the truss is effectively bistable at the moment when the indenter is released (green dashed curve). This provides a good approximation for smaller values of $\lambda$, but increasingly over-predicts the critical value of $T_{\mathrm{ind}}$ as $\lambda$ increases, with the predicted value diverging as $\lambda \to \lambda_{\mathrm{fold}}$. (For later comparison, the boundary predicted by the multiple-scale analysis in \S\ref{CH5sec:multiplescale} is shown as a red dotted curve).

Another key feature of figure \ref{CH5fig:largeDesnapsurface}a is that the snap-through time is very small throughout most of the parameter space. In fact, we will show that here the elastic oscillations cause the truss to immediately cross $X = 0$, so that $T_{\mathrm{snap}} = O(\mathrm{De}^{-1}) \ll 1$. Figure \ref{CH5fig:largeDesnapsurface}a also confirms that the snap-through time only becomes $O(1)$ or larger  in a very narrow region of the parameter space, where $0 < \lambda - \lambda_{\mathrm{fold}} \ll 1$ and $T_{\mathrm{ind}} \gtrsim 4$.  A zoom of this region is provided in figure \ref{CH5fig:largeDesnapsurface}b, which shows that considerable slowing down can occur. In fact, the snap-through time appears to increase without bound as we take $\lambda \searrow \lambda_{\mathrm{fold}}$ in this region. We will show that this is precisely the pseudo-bistable regime: here the displacement initially oscillates around an inverted state and does not immediately cross $X = 0$. As with the trajectories in figure \ref{CH5fig:largeDetrajectories}c (lower panel), this inverted state also undergoes a slow creeping motion until the truss rapidly accelerates towards the natural state, so that $T_{\mathrm{snap}} \gtrsim O(1)$.  This difference in timescales (i.e.~a slow creep followed by a rapid snap-through event) is considered to be a distinguishing feature of pseudo-bistable behaviour \citep{brinkmeyer2012,brinkmeyer2013}.

\begin{figure}
\centering
\includegraphics[width=\textwidth]{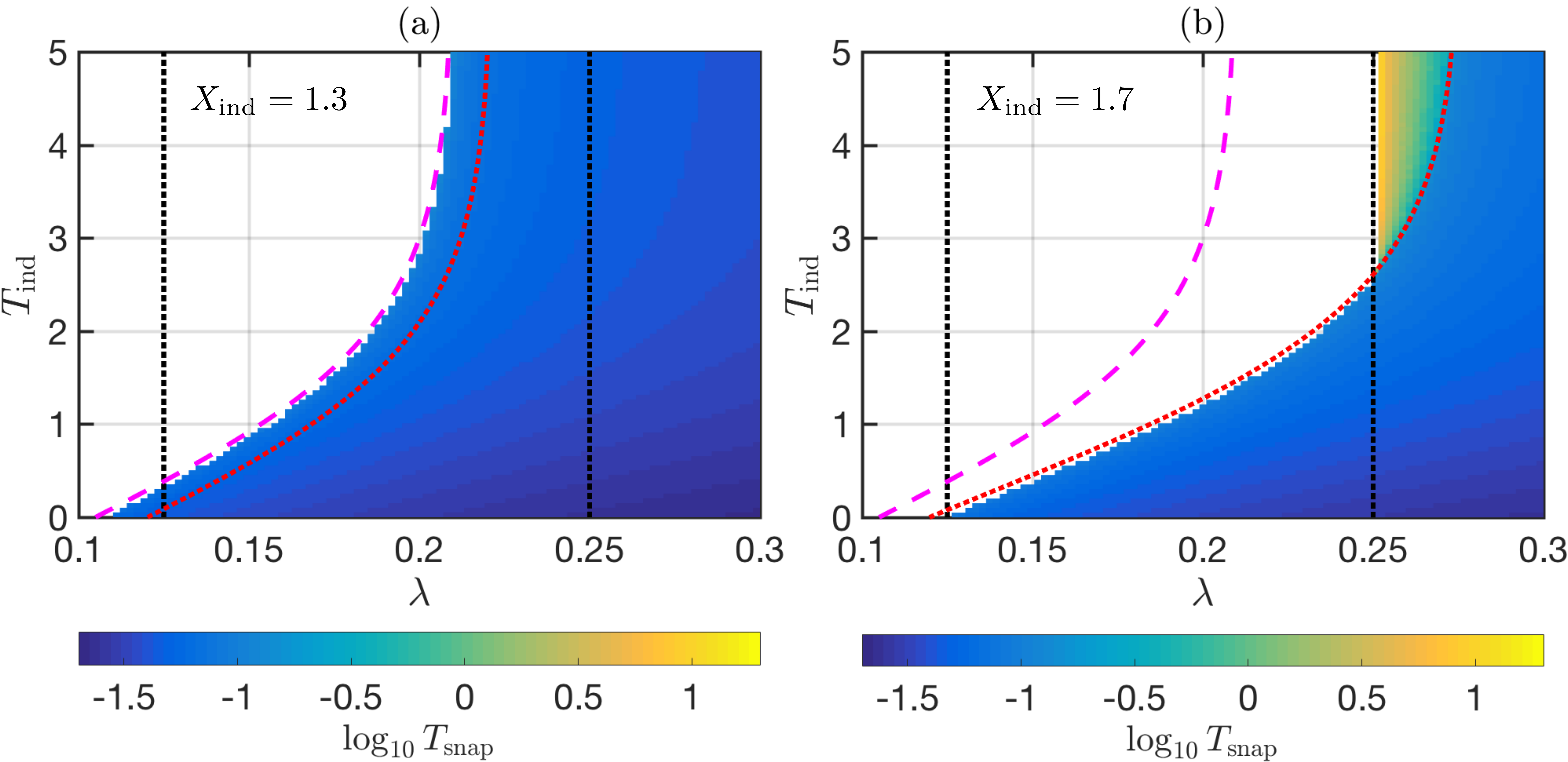} 
\caption{Snap-through times when $\mathrm{De} \gg 1$ for different indentation depths ($\mathrm{De} = 100$, $\beta = 1/2$). Numerical results are shown for (a)  $X_{\mathrm{ind}}=1.3<X_{\mathrm{fold}}$ and (b) $X_{\mathrm{ind}}=1.7>X_{\mathrm{fold}}$. The critical values $\lambda = (1-\beta)\lambda_{\mathrm{fold}}$ and $\lambda = \lambda_{\mathrm{fold}}$ are plotted as vertical black dotted lines.  For later reference, also shown is the boundary predicted by equation \eqref{CH5eqn:smallXindboundary} relevant for $X_{\mathrm{ind}} < X_{\mathrm{fold}}$ (purple dashed curve), and the boundary predicted by equation \eqref{CH5eqn:largeXindboundary} relevant for $X_{\mathrm{ind}} \geq X_{\mathrm{fold}}$ (red dotted curve). In each panel, the snap-through times have been computed on a $100\times 100$ grid of equally spaced values. For ease of comparison the range of the colourbar is the same in both panels here and in figures \ref{CH5fig:largeDesnapsurface}a--b.}
\label{CH5fig:largeDesnapsotherXind}
\end{figure}

Computed snap-through times for different values of $X_{\mathrm{ind}} $ are shown in figures \ref{CH5fig:largeDesnapsotherXind}a--b. These show that the boundary at which snap-through no longer occurs is qualitatively different depending on whether  $X_{\mathrm{ind}} < X_{\mathrm{fold}}$ or  $X_{\mathrm{ind}} \geq X_{\mathrm{fold}}$. For a shallower indentation $X_{\mathrm{ind}} < X_{\mathrm{fold}}$, the boundary appears to be shifted entirely to the left of the line $\lambda = \lambda_{\mathrm{fold}}$, and there is no longer a region where the dynamics slow down considerably (figure \ref{CH5fig:largeDesnapsotherXind}a). The truss also snaps at values $\lambda < (1-\beta)\lambda_{\mathrm{fold}}$ when $T_{\mathrm{ind}}$ is sufficiently small. In contrast, for deeper indentations $X_{\mathrm{ind}} \geq X_{\mathrm{fold}}$, the boundary intercepts the line $\lambda = \lambda_{\mathrm{fold}}$ and the size of the pseudo-bistable region may be significantly larger compared to the case $X_{\mathrm{ind}} = X_{\mathrm{fold}}$ (figure \ref{CH5fig:largeDesnapsotherXind}b). For different values of the relaxation parameter $\beta$, we observe a qualitatively similar picture provided $\beta \lesssim 1/2$; see Appendix A. 

\subsection{Multiple-scale analysis}
\label{CH5sec:multiplescale}
To understand the above observations, we now perform a detailed analysis of the dynamics in the limit $\mathrm{De} \gg 1$. The trajectories in figures \ref{CH5fig:largeDetrajectories}a--d indicate that the displacement undergoes fast oscillations (on an $O(\mathrm{De}^{-1})$ timescale) around a value that varies on an $O(1)$ timescale. This suggests that the dynamics can be understood asymptotically using the method of multiple scales \citep{hinch}. We introduce the fast timescale $\mathcal{T}$ defined by $T = \mathrm{De}^{-1}\mathcal{T}$. 
%\beqn
%T = \mathrm{De}^{-1}\mathcal{T},
%\eeqn
%On this timescale, equation \eqref{CH5eqn:releaseode} becomes
%\beq
%\dt{X}{\mathcal{T}}+\mathrm{De}^{-1}\dd{X}{\mathcal{T}} + F_{\mathrm{eq}}'\left(X;\frac{\lambda}{1-\beta}\right)\d{X}{\mathcal{T}} + \mathrm{De}^{-1}F_{\mathrm{eq}}(X;\lambda) = 0, \quad \mathcal{T} > 0,  \label{CH5eqn:fastode}
%\eeq
%while the initial conditions \eqref{CH5eqn:releaseics} modify to (subscripts denoting differentiation)
%\beq
%X(0+) = X_{\mathrm{ind}}, \quad X_{\mathcal{T}}(0+) = 0, \quad X_{\mathcal{TT}}(0+) = -F_{\mathrm{ind}}.  \label{CH5eqn:fastics}
%\eeq
Treating $T$ and $\mathcal{T}$ as independent, the chain rule implies that
\beq
\d{}{\mathcal{T}} = \pd{}{\mathcal{T}} + \mathrm{De}^{-1}\pd{}{T}. \label{CH5eqn:chainrule}
\eeq
We seek an asymptotic expansion of the solution in the form
\beq
X \sim X_0(\mathcal{T},T)+\mathrm{De}^{-1}X_1(\mathcal{T},T)+\ldots. \label{CH5eqn:multscaleexpansion}
\eeq

\subsubsection{Leading order problem}
We insert the expansion \eqref{CH5eqn:multscaleexpansion} into the ODE \eqref{CH5eqn:releaseode}, re-scaling time in terms of $\mathcal{T}$. After Taylor expanding the $F_{\mathrm{eq}}'$ and $F_{\mathrm{eq}}$ force terms about $X_0$, and expanding the derivatives using \eqref{CH5eqn:chainrule}, we obtain to leading order in $\mathrm{De}^{-1}$ the homogeneous problem
\beqn
\pdt{X_0}{\mathcal{T}}+F_{\mathrm{eq}}'\left(X_0;\frac{\lambda}{1-\beta}\right)\pd{X_0}{\mathcal{T}} = 0.
\eeqn
The initial conditions \eqref{CH5eqn:releaseics} become
\beq
X_0(0+) = X_{\mathrm{ind}}, \quad \pd{X_0}{\mathcal{T}}(0+) = 0, \quad \pdd{X_0}{\mathcal{T}}(0+) = -F_{\mathrm{ind}}.  \label{CH5eqn:fastics}
\eeq
Integrating the above equation for $X_0$ with these conditions, and simplifying using the expression \eqref{CH5eqn:indentationstress,force} for $F_{\mathrm{ind}}$, we obtain
\beq
\pdd{X_0}{\mathcal{T}}+F_{\mathrm{eq}}\left(X_0;\frac{\lambda}{1-\beta}\right) = \frac{\beta \lambda X_{\mathrm{ind}}}{1-\beta}\left(1-e^{-T_{\mathrm{ind}}}\right)+ A(T), \label{CH5eqn:multiscaleleading}
\eeq
where $A(T)$ is unknown and satisfies $A(0+) = 0$.
%We have used the identity
%\beqn
%-F_{\mathrm{ind}}+F_{\mathrm{eq}}\left(X_{\mathrm{ind}};\frac{\lambda}{1-\beta}\right) =  \frac{\beta\lambda X_{\mathrm{ind}}}{1-\beta}\left(1-e^{-T_{\mathrm{ind}}}\right),
%\eeqn
%which follows from the expression \eqref{CH5eqn:indentationstress,force} for $F_{\mathrm{ind}}$. Currently, the function $A(T)$ is unknown. As with other problems in elasticity, such as Euler buckling of a straight beam \citep{howell}, we expect to determine $A(T)$ as a result of a solvability condition on a higher-order problem. 

To reveal the role that $A(T)$ plays in the dynamics, we decompose the leading order solution into a ``slow part'' and a ``fast part'' respectively:
\beq
X_0(\mathcal{T},T) = \mathcal{X}(T) + \mathscr{X}(\mathcal{T},T). \label{CH5eqn:multiscaledecompose}
\eeq
We specify that $\mathcal{X}$ satisfies the ``slow part'' of the leading order equation \eqref{CH5eqn:multiscaleleading}, i.e.~
\beq
F_{\mathrm{eq}}\left(\mathcal{X};\frac{\lambda}{1-\beta}\right) = \frac{\beta\lambda X_{\mathrm{ind}}}{1-\beta}\left(1-e^{-T_{\mathrm{ind}}}\right)+ A(T). \label{CH5eqn:multscaleleadingslowpart}
\eeq
Under the assumption that the ``fast part'' $\vert\mathscr{X}\vert \ll 1$ (see Appendix B), we Taylor expand the $F_{\mathrm{eq}}$ force term in \eqref{CH5eqn:multiscaleleading} about $\mathcal{X}$ and retain only linear terms in $\mathscr{X}$ to obtain
\beq
\pdd{\mathscr{X}}{\mathcal{T}}+F_{\mathrm{eq}}'\left(\mathcal{X};\frac{\lambda}{1-\beta}\right)\mathscr{X} = 0.
\label{CH5eqn:multscaleleadingfastpartsimplified}
\eeq
Provided that $\omega^2 = F_{\mathrm{eq}}'(\mathcal{X};\frac{\lambda}{1-\beta}) > 0$ (as justified later in Appendix C), the solutions are periodic; we denote the period by $L = 2\pi/\omega$, which will vary on the slow timescale as $\mathcal{X}$ varies. Integrating the  equation from $\mathcal{T} = 0$ to $\mathcal{T} = L$ then shows that for each $T$
\beq
\int_0^L \mathscr{X}\id\mathcal{T} = 0. \label{CH5eqn:zeromean}
\eeq
Returning to the way we decomposed the solution in \eqref{CH5eqn:multiscaledecompose}, we see that $\mathcal{X}$ corresponds to the mean value of $X_0$ and varies on the slow timescale $T$. This evolution is captured by the variable $A(T)$. The variable $\mathscr{X}$ describes the oscillations around this mean displacement that occur on the fast timescale $\mathcal{T}$; the property \eqref{CH5eqn:zeromean} guarantees that these oscillations do not influence the mean value if their amplitude is small. We now show that it is possible to obtain an evolution equation for $A(T)$ without requiring detailed knowledge of $\mathscr{X}$, using only the zero-mean property \eqref{CH5eqn:zeromean}. (While it is possible to obtain an analytical expression for $\mathscr{X}$ using the simplified equation \eqref{CH5eqn:multscaleleadingfastpartsimplified}, we do not pursue this here, as this introduces a further unknown function of the slow timescale $T$ --- knowledge of $\mathcal{X}$ will be sufficient to determine when snap-through occurs and the associated snap-through time.)

\subsubsection{First order problem}
At $O(\mathrm{De}^{-1})$, the ODE \eqref{CH5eqn:releaseode} in terms of $\mathcal{T}$ can be written
%\begin{eqnarray*}
%\pdt{X_1}{\mathcal{T}}+F_{\mathrm{eq}}'\left(X_0;\frac{\lambda}{1-\beta}\right)\pd{X_1}{\mathcal{T}} +F_{\mathrm{eq}}''\left(X_0;\frac{\lambda}{1-\beta}\right)\pd{X_0}{\mathcal{T}}X_1 &=& -\pdd{X_0}{\mathcal{T}}-3\frac{\partial^3 X_0}{\partial\mathcal{T}^2\partial T} -F_{\mathrm{eq}}(X_0;\lambda) \\
%&& \: -F_{\mathrm{eq}}'\left(X_0;\frac{\lambda}{1-\beta}\right)\pd{X_0}{T}.
%\end{eqnarray*}
%Re-writing the final two terms on the right-hand side using  the leading order equation \eqref{CH5eqn:multiscaleleading}, this can be written as 
\beqn
\pdt{X_1}{\mathcal{T}}+\pd{}{\mathcal{T}}\left[F_{\mathrm{eq}}'\left(X_0;\frac{\lambda}{1-\beta}\right)X_1\right] = -2\frac{\partial^3 X_0}{\partial\mathcal{T}^2\partial T}+\frac{\beta\lambda}{1-\beta} X_0 -\frac{\beta\lambda X_{\mathrm{ind}}}{1-\beta}\left(1-e^{-T_{\mathrm{ind}}}\right) -\d{A}{T}-A.
\eeqn
This represents a linear, inhomogeneous problem for $X_1$. Setting the right-hand side to zero, we see that the homogeneous problem can be solved approximately whenever $|\mathscr{X}|\ll 1$ by taking $X_1 = X_1(T)$, as $X_0 \approx \mathcal{X}(T)$ in this case and so all $\mathcal{T}$ derivatives vanish. 
%This homogeneous solution is associated with the freedom we had to incorporate the function $A(T)$ in the leading order problem; recall that $A(T)$ entered when we integrated the problem once with respect to $\mathcal{T}$. 
The Fredholm Alternative Theorem then implies that we determine $A(T)$ from the solvability condition associated with the approximate homogeneous solution $X_1 = X_1(T)$ \citep{keener}. To formulate this condition, we simply integrate the first order problem from $\mathcal{T} = 0$ to  $\mathcal{T} = L$. We assume that for each fixed $T$, the solution $X_1$ is also a periodic function with period $L$; this is reasonable, since $X_1$ is forced by the $X_0$ terms that have period $L$. It follows that all $\partial/\partial\mathcal{T}$ terms vanish in the integration and we are left with
\beqn
\frac{\beta\lambda}{1-\beta}\frac{1}{L}\int_0^L X_0\:\id\mathcal{T} - \frac{\beta\lambda X_{\mathrm{ind}}}{1-\beta}\left(1-e^{-T_{\mathrm{ind}}}\right) -\d{A}{T}-A = 0.
\eeqn
Using the zero-mean property \eqref{CH5eqn:zeromean}, the first term can be evaluated as $\beta\lambda\mathcal{X}/(1-\beta)$. Eliminating $A(T)$ for $\mathcal{X}$ using the relation \eqref{CH5eqn:multscaleleadingslowpart}, we arrive at
\beq
F_{\mathrm{eq}}'\left(\mathcal{X};\frac{\lambda}{1-\beta}\right)\d{\mathcal{X}}{T} +F_{\mathrm{eq}}(\mathcal{X};\lambda) = 0.
\label{CH5eqn:odeslowX}
\eeq
This equation is exactly the original ODE \eqref{CH5eqn:releaseode} in the limit $\mathrm{De} \to \infty$, i.e.~neglecting the terms associated with inertia. This is perhaps not surprising: when the zero-mean property \eqref{CH5eqn:zeromean} holds, the fast elastic oscillations `cancel out' on the slow viscous timescale $T$ and so do not affect the leading order dynamics. However, the above analysis does show that the correct initial condition is \emph{not} the indentation displacement $\mathcal{X}(0+) = X_{\mathrm{ind}}$, as might be expected. Instead, from \eqref{CH5eqn:multscaleleadingslowpart}, $\mathcal{X}(0+)$ satisfies
\beq
F_{\mathrm{eq}}\left(\mathcal{X}(0+);\frac{\lambda}{1-\beta}\right) = \frac{\beta\lambda X_{\mathrm{ind}}}{1-\beta}\left(1-e^{-T_{\mathrm{ind}}}\right) = -F_{\mathrm{ind}}+F_{\mathrm{eq}}\left(X_{\mathrm{ind}};\frac{\lambda}{1-\beta}\right),
\label{CH5eqn:slowXic}
\eeq
%\beqn
%F_{\mathrm{eq}}\left(\mathcal{X}(0+);\frac{\lambda}{1-\beta}\right) - F_{\mathrm{eq}}\left(X_{\mathrm{ind}};\frac{\lambda}{1-\beta}\right) =  -F_{\mathrm{ind}}. 
%\eeqn
where the second equality follows from \eqref{CH5eqn:indentationstress,force}. This correction arises from the initial transient around $T = 0$ in which inertia is always important; physically, the above equation states that the change in spring force in moving from $X_{\mathrm{ind}}$ to $\mathcal{X}(0+)$ (when the SLS element is fully unrelaxed with effective stiffness $\lambda/(1-\beta)$) must balance the discontinuity in the indentation force. When viewed on the slow timescale, the mean value $\mathcal{X}$ then appears to change discontinuously from the indentation displacement $ X_{\mathrm{ind}}$. 

To check our multiple-scale analysis, we integrate the simplified ODE \eqref{CH5eqn:odeslowX} numerically subject to the initial condition \eqref{CH5eqn:slowXic}. In figure \ref{CH5fig:largeDetrajectories}c solutions are superimposed (as black dotted curves) onto the trajectories obtained by integrating the full ODE \eqref{CH5eqn:releaseode} for the baseline parameter values. We see that the agreement is excellent, with the multiple-scale solution indeed capturing the average behaviour of the displacement during snap-through (see lower panel in figure \ref{CH5fig:largeDetrajectories}c). (The slight disagreement when the mean value changes rapidly on a timescale comparable to $\mathcal{T}$ is because the multiple-scale analysis is no longer applicable.) Figure \ref{CH5fig:largeDetrajectories}c also shows that the initial value $\mathcal{X}(0+)$ may be much smaller than $X_{\mathrm{ind}}$ depending on the indentation time $T_{\mathrm{ind}}$. We postpone a detailed analysis of $\mathcal{X}(0+)$ to section \S\ref{sec:dynamicsdiscuss} below and Appendix C.

\subsubsection{Snap-through dynamics}
\label{sec:dynamicsdiscuss}
We have shown that while the amplitude of the oscillations is small compared to the mean displacement, the leading order behaviour is given by 
\beq
X \sim \mathcal{X}(T) \quad \mathrm{where} \quad \d{\mathcal{X}}{T} = -\frac{F_{\mathrm{eq}}(\mathcal{X};\lambda)}{F_{\mathrm{eq}}'(\mathcal{X};\frac{\lambda}{1-\beta})},
\label{CH5eqn:multiplescalesummary}
\eeq
subject to the initial condition \eqref{CH5eqn:slowXic}. Since \eqref{CH5eqn:multiplescalesummary} represents a first-order autonomous ODE for $\mathcal{X}$, the dynamics can be understood by considering the $(\mathcal{X},\id\mathcal{X}/\id T)$ phase plane, for which the qualitative features are determined by the roots of $F_{\mathrm{eq}}(\mathcal{X};\lambda) = 0$ and $F_{\mathrm{eq}}'(\mathcal{X};\frac{\lambda}{1-\beta}) = 0$. For $\lambda < \lambda_{\mathrm{fold}}$, there are two distinct non-zero stationary points, which correspond to the stable and unstable roots of $F_{\mathrm{eq}}(\mathcal{X};\lambda) = 0$; for $\lambda > \lambda_{\mathrm{fold}}$, the only stationary point is $\mathcal{X} = 0$. The roots of $F_{\mathrm{eq}}'(\mathcal{X};\frac{\lambda}{1-\beta}) = 0$ correspond to vertical asymptotes where $|\id\mathcal{X}/\id T| = \infty$. Denoting these roots by $\mathcal{X}_{\pm}$, we find that 
\beqn 
\mathcal{X}_{\pm} = 1\pm\frac{\sqrt{3}}{3}\left[1-\frac{\lambda}{1-\beta}\right]^{1/2}, \quad F_{\mathrm{eq}}\left(\mathcal{X}_{\pm};\frac{\lambda}{1-\beta}\right) = \frac{\lambda}{1-\beta}\mp \frac{2\sqrt{3}}{9}\left[1-\frac{\lambda}{1-\beta}\right]^{3/2}.
\eeqn
These roots are real and distinct if and only if $\lambda < 1-\beta$.

\begin{figure}
\centering
\includegraphics[width=0.95\textwidth]{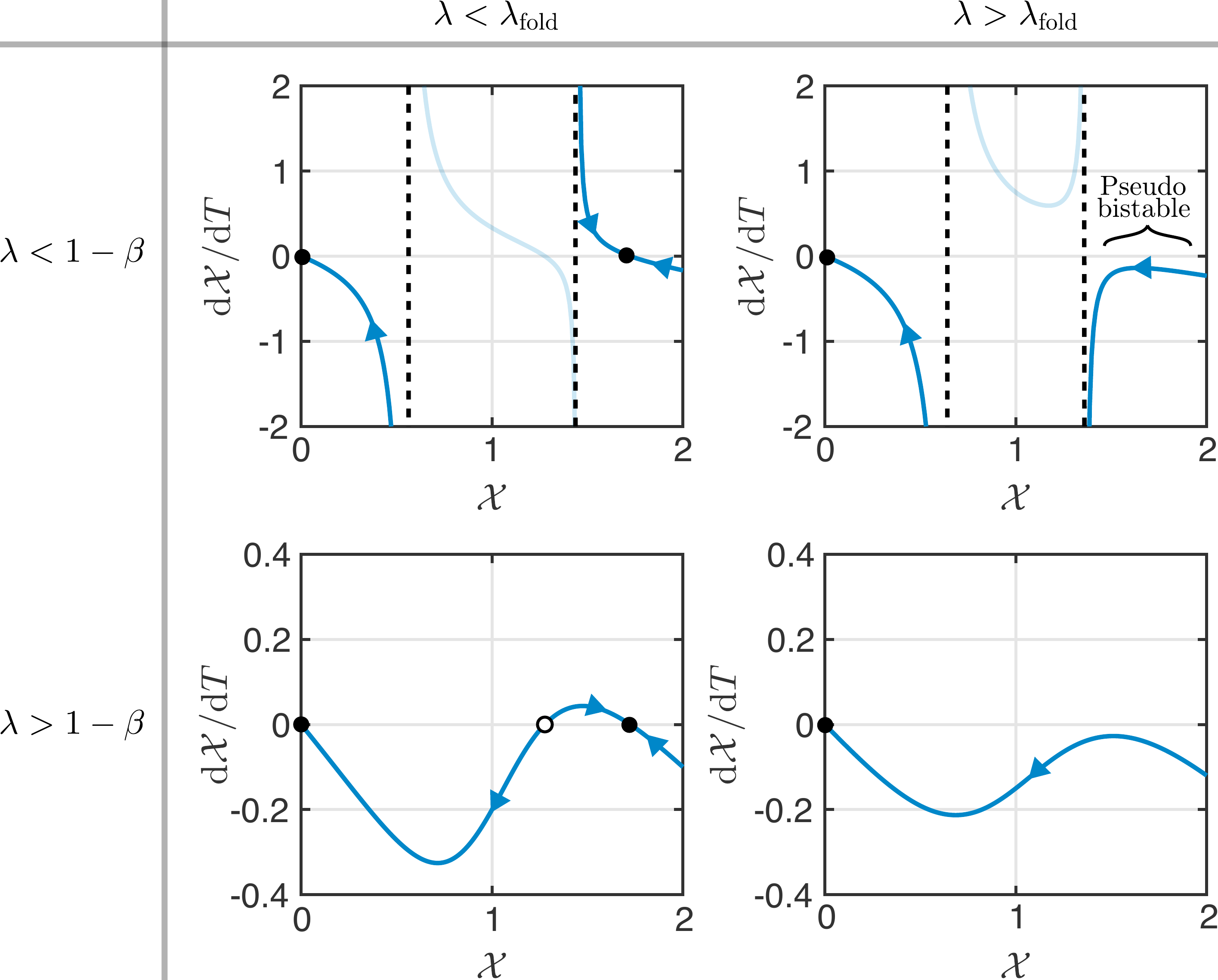} 
\caption{The phase plane of the simplified ODE \eqref{CH5eqn:multiplescalesummary}. Results are shown for $\lambda = 0.2$, $ \beta = 0.5$ (top left),  $\lambda = 0.3$, $ \beta = 0.5$ (top right),  $\lambda = 0.2$, $ \beta = 0.9$ (bottom left) and  $\lambda = 0.3$, $ \beta = 0.9$ (bottom right). In each panel, arrows indicate the direction of motion. The stable/unstable roots of $F_{\mathrm{eq}}(\mathcal{X};\lambda) = 0$ are shown as filled/unfilled circles respectively. The vertical asymptotes at $\mathcal{X}_{\pm}$ are plotted as black dotted lines. The initial value $\mathcal{X}(0+)$ is determined by the transient around $T = 0$ in which inertia is important --- this is the solution of the cubic \eqref{CH5eqn:slowXic}. Note that pseudo-bistability occurs for $\lambda>\lambda_{\mathrm{fold}}$, $\lambda<1-\beta$ (top right).}
\label{CH5fig:multiplescales}
\end{figure}

Different cases for the phase plane are illustrated in figure \ref{CH5fig:multiplescales}. In Appendix C, we show that when $\lambda < 1-\beta$ (so that $\id\mathcal{X}/\id T$ may diverge), the initial value $\mathcal{X}(0+)$ must satisfy either $\mathcal{X}(0+) <  \mathcal{X}_{-}$ or $\mathcal{X}(0+) > \mathcal{X}_{+}$, i.e.~$\mathcal{X}$ always starts outside the interval between the two vertical asymptotes. We have therefore lightly shaded this region in figure \ref{CH5fig:multiplescales}. Figure \ref{CH5fig:multiplescales} shows how when $\lambda > 1-\beta$, there are no vertical asymptotes and the trajectories smoothly approach the natural state; this explains the observation that pseudo-bistable behaviour does not occur for larger values of $\beta$, even though the snap-through time $T_{\mathrm{snap}} \gtrsim O(1)$ (discussed more in Appendix A). We see that pseudo-bistability is only possible if $\lambda < 1-\beta$: the asymptotes on the phase plane correspond to rapid snap-through events, which occur after a slow creeping motion provided $\mathcal{X}(0+)$ is sufficiently large (top right panel in figure \ref{CH5fig:multiplescales}). Since we are primarily interested in this pseudo-bistable regime, we restrict our attention to the case  $\lambda < 1-\beta$ in the following analysis; in fact, we will restrict to $\beta \leq 1/2$ so that we are safely in this regime in considering values of $\lambda$ up to and slightly beyond $ \lambda_{\mathrm{fold}}$\footnote{As indicated by figure \ref{CH5fig:largeDesnapsotherbeta}b in Appendix A, the truss may show different behaviour when $\lambda$ is slightly below the critical value $1-\beta$. Further analysis shows that as the left asymptote $\mathcal{X}_{-}$ on the $(\mathcal{X},\id\mathcal{X}/\id T)$ phase plane approaches $\mathcal{X}_{-} \nearrow 1$ (i.e.~as $\lambda \nearrow 1-\beta$), the precise definition of $T_{\mathrm{snap}}$ becomes important; for example, the mean value $\mathcal{X}(0+)$ may start to the left of both asymptotes so that the truss immediately snaps to near the natural configuration, but if $\mathcal{X}(0+)$ is close to $1$ the truss does not cross $X = 0$ (our definition of $T_{\mathrm{snap}}$) on the elastic timescale. By restricting to $\beta \leq 1/2$, we are able to bypass these technicalities for values of $\lambda$ in the interval of interest for pseudo-bistable behaviour.}.

Focussing on $\lambda < 1-\beta$, we deduce from figure \ref{CH5fig:multiplescales} that there are three possibilities depending on where the solution starts in the phase plane: 
\begin{itemize}
\item{If $\mathcal{X}(0+) > \mathcal{X}_{+}$ and $\lambda < \lambda_{\mathrm{fold}}$, then the mean value starts to the right of both vertical asymptotes and approaches the stable non-zero root of $F_{\mathrm{eq}}(\mathcal{X};\lambda) = 0$. Because $\mathcal{X}_{+} > 1$, the truss remains in an inverted position and there is no snap-through\footnote{An additional case is possible in which the unstable root of $F_{\mathrm{eq}}(\mathcal{X};\lambda) = 0$ is larger than the vertical asymptote at $\mathcal{X}_+$, so that two stationary points lie to the right of both asymptotes on the phase plane (unlike the top left panel of figure \ref{CH5fig:multiplescales}, in which the unstable root lies in the lightly shaded region between the asymptotes). However, in focussing on values $\beta \leq 1/2$, this regime is found to occur only for values of $\lambda < \lambda_{\mathrm{fold}}$ extremely close to $\lambda_{\mathrm{fold}}$ and so can generally be ignored.}.}
\item{If $\mathcal{X}(0+) > \mathcal{X}_{+}$ and $\lambda > \lambda_{\mathrm{fold}}$, then the mean value starts to the right of both vertical asymptotes. It then decreases to the vertical asymptote at $\mathcal{X}_{+}$ where rapid snap-through occurs and inertial effects again become significant --- because this approach to the asymptote occurs on the slow timescale $T$, the total time taken to snap-through is at least $O(1)$. This regime corresponds to pseudo-bistable behaviour, in which the snap-through time is governed by the timescale of viscous relaxation.}
\item{If $\mathcal{X}(0+) < \mathcal{X}_{-}$, then the mean value starts to the left of the vertical asymptotes and smoothly decays to zero. Because $ \mathcal{X}_{-} < 1$, the truss immediately snaps back to near its natural configuration during the initial transient in which inertia is important. The amplitude of the elastic oscillations will therefore be large compared to the mean value in this case, and our assumption $|\mathscr{X}|\ll 1$ is no longer valid; nevertheless, we expect the truss to pass $X = 0$ on an $O(\mathrm{De}^{-1})$ timescale so that $T_{\mathrm{snap}} = O(\mathrm{De}^{-1})$.}
\end{itemize} 

The final task is to determine when the initial value satisfies $\mathcal{X}(0+) > \mathcal{X}_{+}$. This is not obvious because \eqref{CH5eqn:slowXic} implies that $\mathcal{X}(0+)$ is the root of a cubic polynomial, for which there may be multiple real solutions. The relevant solution can be found by analysing the phase portrait of equation \eqref{CH5eqn:multiscaleleading}: setting $A(T) = 0$, this equation governs the elastic behaviour of the truss at very early times, and hence determines which root of  \eqref{CH5eqn:slowXic} the solution approaches. When viewed on the slow timescale, this root corresponds to the relevant value of $\mathcal{X}(0+)$. The full analysis is provided in Appendix C. The key result is that if $1 \leq X_{\mathrm{ind}} < X_{\mathrm{fold}}$, then $\mathcal{X}(0+) > \mathcal{X}_{+}$ if and only if $F_{\mathrm{ind}} < 0$; otherwise we have $\mathcal{X}(0+) < \mathcal{X}_{-}$. Physically, this states that the indentation force needs to be adhesive for the truss to remain in an inverted state. This is intuitive: if the truss has to be `pulled' upwards to the imposed indentation depth, it should move further downwards (increasing $X$) when the indenter is released. Using the expression \eqref{CH5eqn:indentationstress,force} for $F_{\mathrm{ind}}$, the condition  $F_{\mathrm{ind}} < 0$ can be expressed as
\beq
\mathcal{X}(0+) > \mathcal{X}_{+} \iff T_{\mathrm{ind}} > \log\left[\frac{-\beta\lambda X_{\mathrm{ind}}}{(1-\beta)F_{\mathrm{eq}}(X_{\mathrm{ind}};\lambda)}\right]. \label{CH5eqn:smallXindboundary}
\eeq
In the alternative case $X_{\mathrm{fold}} \leq X_{\mathrm{ind}} \leq 2$, the condition $F_{\mathrm{ind}} < 0$ turns out to no longer be relevant. Instead, in Appendix C we show that
\beq
\mathcal{X}(0+) > \mathcal{X}_{+} \iff T_{\mathrm{ind}} > \log\left[\frac{\beta\lambda X_{\mathrm{ind}}}{\beta\lambda X_{\mathrm{ind}} - (1-\beta)F_{\mathrm{eq}}(X^{*};\frac{\lambda}{1-\beta})}\right],
 \label{CH5eqn:largeXindboundary}
\eeq
where
\beqn
X^{*} = -\frac{1}{3}\left(X_{\mathrm{ind}}-4\right)+\frac{\sqrt{6}}{3}\left[1-\frac{\lambda}{1-\beta}-\frac{\left(X_{\mathrm{ind}}-1\right)^2}{3}\right]^{1/2}.
\eeqn
%\beq
%G(\lambda,X_{\mathrm{ind}}) = \frac{1}{3}\lambda \left(X_{\mathrm{ind}}-4\right) + \frac{\sqrt{2}}{27}\left[3-6\lambda-\left(X_{\mathrm{ind}}-2\right)^2\right]^{3/2}-\frac{1}{27}\left(X_{\mathrm{ind}}-1\right)\left(5 X_{\mathrm{ind}}^2-10X_{\mathrm{ind}}-4\right).
%\eeq
Physically, this condition arises from bounding the amplitude of the elastic oscillations at very early times, so that these do not `push' the truss sufficiently far from the inverted state and cause an immediate snap-back.

\begin{figure}
\centering
\includegraphics[width=\textwidth]{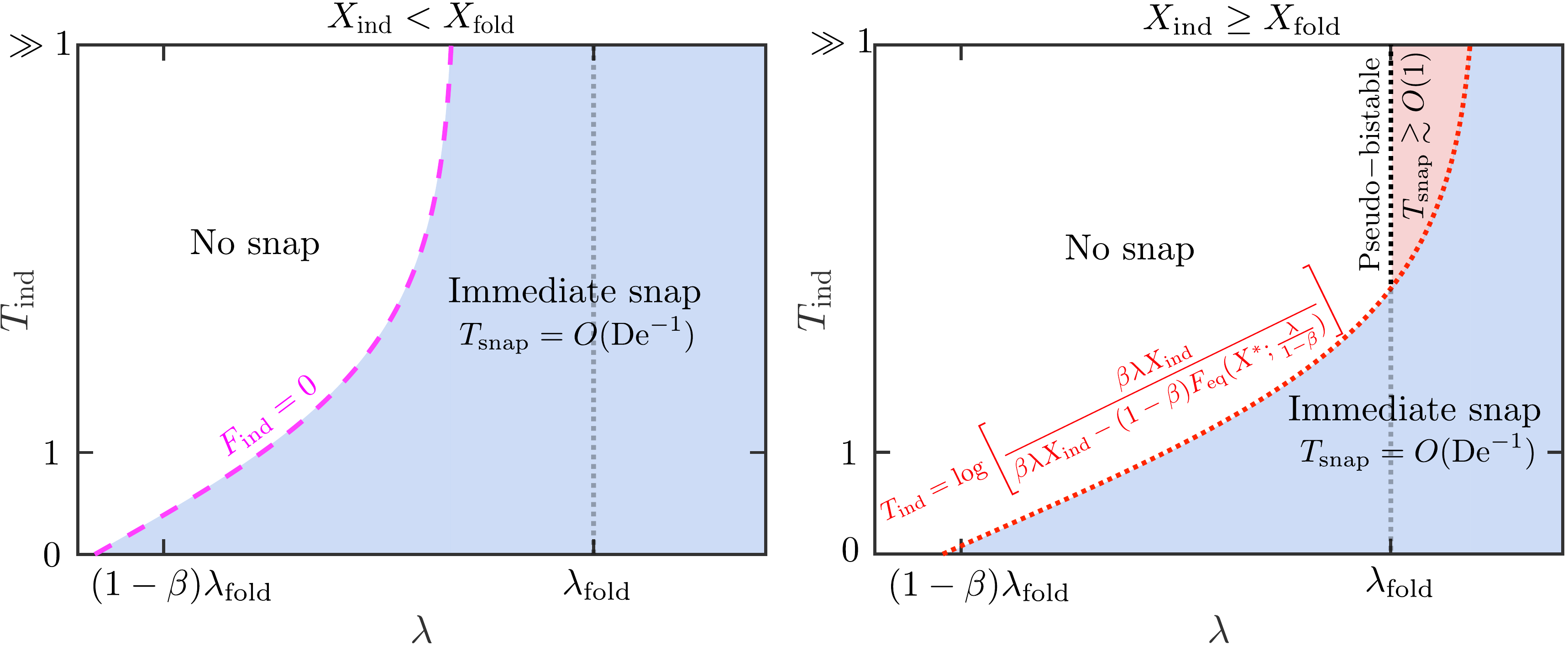} 
\caption{The different dynamical regimes in the limit $\mathrm{De} \gg 1$ predicted by our multiple-scale analysis: we combine the phase plane of figure \ref{CH5fig:multiplescales} with the analysis of $\mathcal{X}(0+)$ in Appendix C (restricting to $\beta \leq 1/2$ and $\lambda < 1-\beta$).}
\label{CH5fig:largeDeschematic}
\end{figure}

Combining this with the phase-plane discussion above, the predicted dynamical regimes are shown schematically in figure \ref{CH5fig:largeDeschematic}. This explains how the qualitative features of the dynamics are very different in the two cases $X_{\mathrm{ind}} < X_{\mathrm{fold}}$ and  $X_{\mathrm{ind}} \geq X_{\mathrm{fold}}$. When $X_{\mathrm{ind}} < X_{\mathrm{fold}}$, it may be shown that the boundary predicted by \eqref{CH5eqn:smallXindboundary} reaches a vertical asymptote on the $(\lambda,T_{\mathrm{ind}})$-plane when $\lambda < \lambda_{\mathrm{fold}}$. For values $T_{\mathrm{ind}}$ below the boundary we have $\mathcal{X}(0+) < \mathcal{X}_{-}$, and the above discussion implies that the truss immediately snaps with $T_{\mathrm{snap}}= O(\mathrm{De}^{-1})$ (shaded blue in figure \ref{CH5fig:largeDeschematic}). Above the boundary, $\mathcal{X}(0+) > \mathcal{X}_{+}$ and, because $\lambda < \lambda_{\mathrm{fold}}$ here, the truss does not snap-through. Pseudo-bistable behaviour therefore cannot be obtained when $X_{\mathrm{ind}} < X_{\mathrm{fold}}$. Conversely, when  $X_{\mathrm{fold}} \leq X_{\mathrm{ind}} \leq 2$, the boundary predicted by \eqref{CH5eqn:largeXindboundary} reaches a vertical asymptote when $\lambda > \lambda_{\mathrm{fold}}$.  Hence, there is a region where  $\mathcal{X}(0+) > \mathcal{X}_{+}$ and $\lambda > \lambda_{\mathrm{fold}}$, in which pseudo-bistable behaviour occurs (shaded red in figure \ref{CH5fig:largeDeschematic}). We deduce that $T_{\mathrm{snap}} \gtrsim O(1)$ precisely when
\beqn
X_{\mathrm{fold}} \leq X_{\mathrm{ind}} \leq 2, \quad  \lambda >  \lambda_{\mathrm{fold}}, \quad T_{\mathrm{ind}} > \log\left[\frac{\beta\lambda X_{\mathrm{ind}}}{\beta\lambda X_{\mathrm{ind}} - (1-\beta)F_{\mathrm{eq}}(X^{*};\frac{\lambda}{1-\beta})}\right].
\eeqn
%As $\lambda \to 1/4$, the critical value of $T_{\mathrm{ind}}$ approaches the $O(1)$ value
%\beqn
%T_{\mathrm{ind}} = \log\left[\frac{-27 X_{\mathrm{ind}}}{20 X_{\mathrm{ind}}^3 - 60X_{\mathrm{ind}}^2+15X_{\mathrm{ind}}+52-2\left[3-2(X_{\mathrm{ind}}-1)^2\right]^{3/2}}\right].
%\eeqn

To check the validity of the picture presented in figure \ref{CH5fig:largeDeschematic}, we have superimposed the boundaries predicted by \eqref{CH5eqn:smallXindboundary}--\eqref{CH5eqn:largeXindboundary} (purple dashed curves, red dotted curves respectively) onto the numerical snap-through times in figures \ref{CH5fig:largeDesnapsurface}--\ref{CH5fig:largeDesnapsotherXind} (and figure \ref{CH5fig:largeDesnapsotherbeta} of Appendix A). We observe that the agreement with the numerics is excellent when $\lambda < 1-\beta$, despite the fact that the assumption $| \mathscr{X}| \ll 1$ made in the multiple-scale analysis is not formally valid throughout the range of values shown. 

Figure \ref{CH5fig:largeDeschematic} explains many basic features of pseudo-bistability that have been observed previously in experiments and numerical simulations \citep{santer2010,brinkmeyer2012,brinkmeyer2013,madhukar2014,urbach2017,urbach2018}. We see that pseudo-bistability occurs only in a narrow parameter range, near the threshold at which snap-through no longer occurs (i.e.~$\lambda = \lambda_{\mathrm{fold}} = 1/4$ here) and the width of the pseudo-bistable region grows as the amount of stress relaxation increases (increasing $T_{\mathrm{ind}}$). Pseudo-bistable behaviour is not obtained if  $T_{\mathrm{ind}}$ is too small, nor if the indentation depth $X_{\mathrm{ind}}$ is below a critical value. In addition, the phase plane in figure \ref{CH5fig:multiplescales} explains how the truss initially creeps in an inverted state before abruptly accelerating, leading to the difference in timescales that is characteristic of pseudo-bistable behaviour \citep{brinkmeyer2012,brinkmeyer2013}. We emphasise that the analytical understanding of these features presented here is, to the best of our knowledge, new.

The importance of inertial effects immediately after the indenter is released has not been appreciated previously. Because this causes the displacement of the truss to change rapidly from the indentation displacement, the effective stiffness will also change rapidly from its value just before the indenter is released. This is in direct contrast to the viscoelastic models used by \cite{santer2010} and \cite{brinkmeyer2012,brinkmeyer2013}, which assume that (i) the stiffness reverses back to its fully unrelaxed value when the indenter is released, and (ii) there is no rapid change in the stiffness caused by the discontinuity in the applied indentation force. In Appendix D we show that when we make assumptions (i)--(ii) in the framework of the truss model, we obtain radically different predictions of when snap-through and pseudo-bistability occur. In the  type of snap-through experiment considered here (a structure is allowed to relax in a specified displacement before being abruptly released), assumptions (i)--(ii) cannot therefore be derived from first principles starting from the constitutive law for a standard linear solid. Instead, we believe it is necessary to couple the stress  within the structure to its displacement and account for inertial effects when the indenter is removed.

%However,  When $\beta = 0.8$, observe disagreement slightly before $\lambda$ reaches $1-\beta = 0.2$ (similar with other values $\beta > 1/2$). We attribute this to asymptotes on phase plane moving closer together as $\lambda$ gets near to $1-\beta$. In particular, it becomes possible for solution to start to the left of the asymptotes but to still not cross $X = 0$ immediately - need some viscous relaxation first. However, it still immediately crosses $X = 1$. 

\subsubsection{Snap-through time in the pseudo-bistable regime}
\label{CH5sec:largeDebottleneck}
Another key feature of the dynamics is that the snap-through time increases considerably as $\lambda \searrow \lambda_{\mathrm{fold}}$ in the pseudo-bistable regime, becoming much larger than $O(1)$ (figure \ref{CH5fig:largeDesnapsurface}b). This slowing down does not require $X_{\mathrm{ind}} \approx X_{\mathrm{fold}}$ (since it can be observed when $X_{\mathrm{ind}} = 1.7$, figure \ref{CH5fig:largeDesnapsotherXind}b). The phase plane in figure \ref{CH5fig:multiplescales} (top right panel) suggests that this slowing down is due to a saddle-node ghost: when $\lambda > \lambda_{\mathrm{fold}}$ the non-zero stationary point no longer exists, but as $\lambda \searrow \lambda_{\mathrm{fold}}$ the trajectory passes increasingly close to the line $\id\mathcal{X}/\id T = 0$ at $\mathcal{X} \approx X_{\mathrm{fold}}$. Because the velocity becomes very small but non-zero, this will lead to a slow passage through a bottleneck.

To analyse this slowing down in detail, we set
\beq
\lambda = \lambda_{\mathrm{fold}}+\epsilon, \quad \mathcal{X} = X_{\mathrm{fold}}-\epsilon^{1/2}\chi, \label{CH5eqn:pseudobistableexpansions}
\eeq
where $0 < \epsilon \ll 1$ and $|\chi | \ll \epsilon^{-1/2}$; here we anticipate an $\epsilon^{1/2}$ scaling for the change in displacement during the bottleneck phase, which is the generic scaling for overdamped dynamics near a saddle-node bifurcation \citep{strogatz}. (We have also introduced a minus sign since we expect the displacement to decrease during snap-through.) 
%We expand the force terms appearing in the simplified ODE \eqref{CH5eqn:multiplescalesummary} as
%\begin{eqnarray}
%F_{\mathrm{eq}}(\mathcal{X};\lambda) & = & \frac{3}{2}\epsilon\left(1+\chi^2\right)+O(\epsilon^{3/2}\chi,\epsilon^{3/2}\chi^3), \nonumber \\
%F_{\mathrm{eq}}'\left(\mathcal{X};\frac{\lambda}{1-\beta}\right) & = &\frac{\beta}{4(1-\beta)}+O(\epsilon,\epsilon^{1/2}\chi).
%\label{CH5eqn:bottleneckexpandforceterms}
%\end{eqnarray}
Expanding the force terms appearing in the simplified ODE \eqref{CH5eqn:multiplescalesummary}, we obtain
\beq
\d{\chi}{T} \sim \frac{6\epsilon^{1/2}(1-\beta)}{\beta}\left(1+\chi^2\right), \label{CH5eqn:largeDebottleneckeqn}
\eeq
which (up to numerical pre-factors) is the normal form for overdamped dynamics near a saddle-node bifurcation \citep{strogatz}; here the neglected terms are small compared to at least one retained term provided $|\chi | \ll \epsilon^{-1/2}$. The solution is
\beq
\chi \sim \tan\left[ \frac{6\epsilon^{1/2}(1-\beta)}{\beta}T +\arctan\chi(0+)\right],
\label{CH5eqn:bottlenecksoln}
\eeq
where
\beqn
\chi(0+) =\epsilon^{-1/2}\left[ X_{\mathrm{fold}}-\mathcal{X}(0+)\right].
\eeqn

The snap-through time is dominated by the time spent passing through the bottleneck, $T_b$, which can be determined by finding the time at which $\chi$ first reaches $O(\epsilon^{-1/2})$; after this point, we no longer have $\mathcal{X} \approx X_{\mathrm{fold}}$ and so the truss is moving rapidly. Using the expansion $\tan x \sim (\pi/2-x)^{-1}$ as $x \to \pi/2$, we have that $\chi = O(\epsilon^{-1/2})$ when
\beq
T_b = \frac{\pi \beta}{12(1-\beta)}\epsilon^{-1/2}-\frac{\beta}{6(1-\beta)}\epsilon^{-1/2}\arctan\chi(0+) + O(1). \label{CH5eqn:bottleneckduration}
\eeq
It follows that there are different distinguished limits, depending on the size and sign of $\chi(0+)$; these are discussed in Appendix E. For example, for the baseline case $X_{\mathrm{ind}}= X_{\mathrm{fold}}$, we insert the expansions \eqref{CH5eqn:pseudobistableexpansions} into the initial condition \eqref{CH5eqn:slowXic}. Upon neglecting terms quadratic in $\epsilon$ and $e^{-T_{\mathrm{ind}}}$ (e.g.~with $\beta = 1/2$, figure \ref{CH5fig:largeDesnapsurface}b implies we have $T_{\mathrm{ind}} \gtrsim 4$ in the pseudo-bistable regime, so that $e^{-T_{\mathrm{ind}}}\lesssim 0.02 \ll 1$), we obtain 
\beqn
\chi(0+) \sim \frac{6(1-\beta)}{\beta}\epsilon^{1/2} +\frac{3}{2}\epsilon^{-1/2}e^{-T_{\mathrm{ind}}}. 
\eeqn
The above expression for the bottleneck duration then gives the leading order estimate
\beq
T_{\mathrm{snap}} = \frac{\pi \beta}{12(1-\beta)}\epsilon^{-1/2}-\frac{\beta}{6(1-\beta)}\epsilon^{-1/2}\arctan\left[\frac{6(1-\beta)}{\beta}\epsilon^{1/2} +\frac{3}{2}\epsilon^{-1/2}e^{-T_{\mathrm{ind}}}\right]+O(1).
\label{CH5eqn:pseudobistablesnaptime}
\eeq
The distinguished limits correspond to $\epsilon^{-1/2}e^{-T_{\mathrm{ind}}} \ll 1$ and $\epsilon^{-1/2}e^{-T_{\mathrm{ind}}} \gg 1$, and we obtain
\beqn
 T_{\mathrm{snap}} = \begin{cases} \frac{\pi\beta}{12(1-\beta)}\epsilon^{-1/2} + O(\epsilon^{-1}e^{-T_{\mathrm{ind}}},1) \quad \mathrm{if} \quad T_{\mathrm{ind}} \gg \log\left(\epsilon^{-1/2}\right), \\ \frac{\beta}{9(1-\beta)}e^{T_{\mathrm{ind}}} + O(1) \quad \mathrm{if} \quad T_{\mathrm{ind}} \ll \log\left(\epsilon^{-1/2}\right). \end{cases} 
\eeqn
 
 \begin{figure}
\centering
\includegraphics[width= \textwidth]{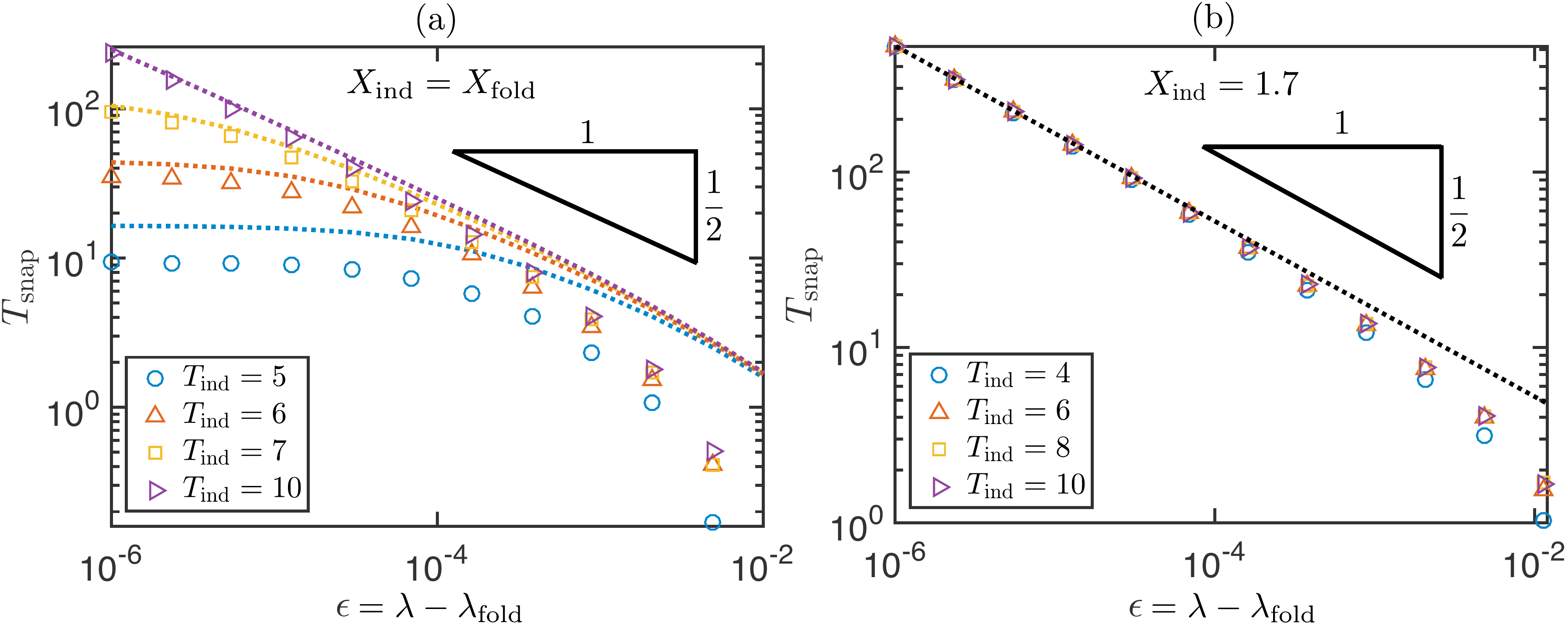} 
\caption{Snap-through times in the pseudo-bistable regime ($\mathrm{De} = 100$, $\beta = 1/2$). (a) Numerical results for (a) $X_{\mathrm{ind}} = X_{\mathrm{fold}} = 3/2$ and (b) $X_{\mathrm{ind}} = 1.7$ (symbols; see legends). Also shown for comparison in (a) are the predictions \eqref{CH5eqn:pseudobistablesnaptime} (coloured dotted curves) valid for $X_{\mathrm{ind}} = X_{\mathrm{fold}}$ and in (b) the asymptotic prediction \eqref{CH5eqn:largeXindpseudobistablesnaptime} computed in Appendix E (black dotted line), valid when $X_{\mathrm{ind}} - X_{\mathrm{fold}} \gg \epsilon^{1/2}$.}
\label{CH5fig:pseudobistabletimes}
\end{figure}

Figure \ref{CH5fig:pseudobistabletimes}a shows that the prediction \eqref{CH5eqn:pseudobistablesnaptime} approximates the numerically-computed snap-through times reasonably well, with the data indeed obeying an inverse square-root scaling law associated with an overdamped saddle-node ghost ($T_{\mathrm{snap}} \propto \epsilon^{-1/2}$) only when $T_{\mathrm{ind}}$ is sufficiently large compared to $\log\left(\epsilon^{-1/2}\right)$. However, the $O(1)$ error in \eqref{CH5eqn:pseudobistablesnaptime} becomes significant if $T_{\mathrm{ind}} \lesssim 5$ or $\epsilon \gtrsim 10^{-4}$ is only moderately small.  While it is possible to obtain the $O(1)$ correction analytically by integrating the ODE \eqref{CH5eqn:multiplescalesummary} directly, we do not compute this here. For deeper indentations $X_{\mathrm{ind}} > X_{\mathrm{fold}}$, we observe a similar picture, though the snap-through times are generally larger so that the $O(1)$ correction is less significant; see figure \ref{CH5fig:pseudobistabletimes}b.

A factor of $\beta/(1-\beta) = E_2/E_1$ (corresponding to the relaxation strength of the material) consistently appears in the expression \eqref{CH5eqn:pseudobistablesnaptime} for the snap-through time. This means that the dimensional snapping time scales as $\eta/E_1$ rather than the timescale of stress relaxation, $\eta/E_2$. To understand the origin of this timescale, we note that $\eta/E_1$  corresponds to the creep timescale for a Kelvin-Voigt material, i.e.~if we neglect the $E_2$ spring in the SLS element. Physically, this arises because the solution is close to the equilibrium at $X = X_{\mathrm{fold}}$ as it passes through the bottleneck.  In equilibrium, the $E_2$ spring (which is in series with the dashpot) will be relaxed at its natural length. Hence, sufficiently close to these solutions we do not expect this spring to play an important role compared to the $E_1$ spring, so the SLS element acts analogously to a Kelvin-Voigt element. Alternatively, we note that close to equilibrium the stress $\Sigma$ is nearly equal to its elastic value, $\Sigma = X$. If we write $\Sigma = X + \xi$ for some small perturbation $\xi$ and substitute this into equation \eqref{CH5eqn:SLS}, at leading order we obtain $[\beta/(1-\beta)]\dot{X} \sim \xi$ (assuming $\dot{\xi} \ll \xi$ due to the slow bottleneck timescale), which gives the same factor of $\beta/(1-\beta)$; this shows that the timescale $\eta/E_1$ is the natural timescale over which the system undergoes bottleneck behaviour near an equilibrium\footnote{Because pseudo-bistability only occurs when $\beta$ is not close to zero or one (Appendix A), which requires that both the $\eta/E_1$ and $\eta/E_2$ timescales be of the same order,  the order-of-magnitude estimate of the snapping time in the pseudo-bistable regime (figure 8) remains valid.}.

\section{A continuous model system: viscoelastic arch}
\label{sec:arch}
Our analysis of the Mises truss indicates that in the limit of large Deborah number, only two types of snap-through are possible: the system either immediately snaps on the elastic timescale, or it  is pseudo-bistable and first undergoes a slow creeping motion governed by the viscous timescale. This creeping motion may be very slow indeed, as we found that the snap-through time is subject to critical slowing down in this regime. 
%The snap-through time then inherits the inverse square-root scaling law for overdamped dynamics near a saddle-node bifurcation, i.e.~
%\beqn
%T_{\mathrm{snap}} = \frac{t_{\mathrm{snap}}}{\eta/ E_2} \propto \epsilon^{-1/2} \qquad \mathrm{where} \qquad \epsilon = \lambda - \lambda_{\mathrm{fold}},
%\eeqn
%valid when $0 < \epsilon \ll 1$ and $T_{\mathrm{ind}}$ is sufficiently large. 
As discussed in \S\ref{CH5sec:formulation}, we may also consider the truss as a lumped model for a continuous viscoelastic structure (such as those used in morphing applications), when we identify $\lambda$ with the analogous parameter that determines whether the continuous structure is bistable or monostable; for example the  F\"{o}ppl-von-K\'{a}rm\'{a}n number for complete spherical shells \citep{knoche2014b} and the analogous parameter for spherical caps \citep{taffetani2018}. Hence, we expect that our conclusions for the truss model may hold more generically and explain the features of pseudo-bistability observed in more complex systems. 

\begin{figure}
\centering
\includegraphics[width=\textwidth]{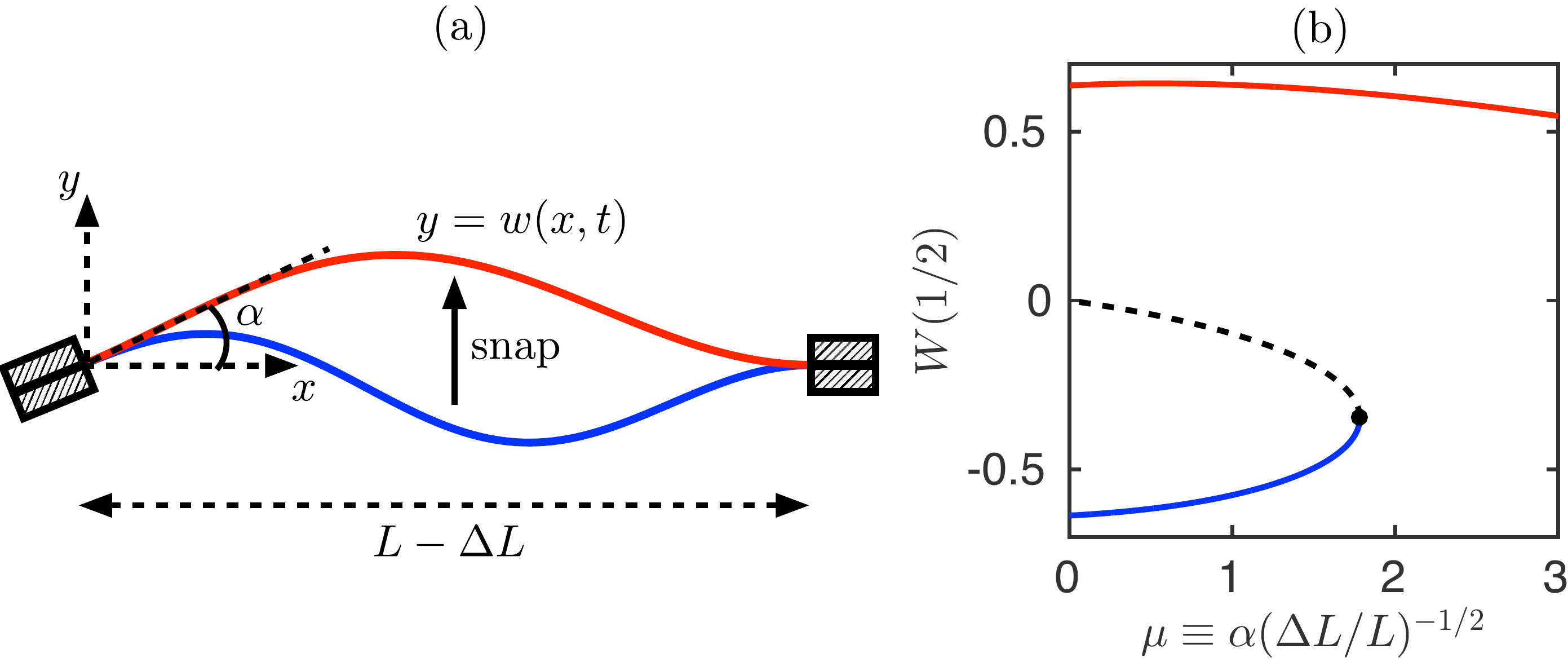} 
\caption{(a) Schematic of the arch setup featuring an `inverted' shape (blue) that snaps to a `natural' shape (red). (b) Response diagram for equilibrium solutions: plotting the dimensionless midpoint displacement, $W(1/2) = w(L/2)/(L\Delta L)^{1/2}$, as a function of the normalised clamp angle $\mu = \alpha(\Delta L/L)^{-1/2}$. The lower branch (solid blue curve) corresponds to the inverted shape, while the upper branch (solid red curve) corresponds to the natural shape. (The dotted branch is an unstable mode that is not observed experimentally.)}
\label{fig:archsetup}
\end{figure}

To test this hypothesis, we now analyse the snap-through dynamics of a viscoelastic arch. This allows us to model a continuous structure without the additional complications that come with more complex structures such as spherical caps \citep{gomez2016a}, where there is also the possibility of non-axisymmetric deformations during snap-through \citep{seffen2016a}. Nevertheless, arches illustrate many features of snap-through that are present in more complex systems \citep{harvey2015}, and approximate the behaviour of curved panels commonly used in engineering applications \citep{wiebe2012}. We consider a flat strip that is subjected to an end-shortening $\Delta L$ in the horizontal direction, with one end clamped at an angle $\alpha$ to the horizontal while the other end is clamped horizontally. As shown schematically in figure \ref{fig:archsetup}a, this causes the strip to buckle into one of two possible arch shapes --- an `inverted' shape (directed downwards) and a `natural' shape (directed upwards).  However, as a simple experiment illustrates (e.g.~using an ordinary strip of plastic), the inverted shape needs a sufficiently large end-shortening $\Delta L$ to be stable: for smaller values of $\Delta L$, the arch snaps from the inverted shape to the natural shape. Snap-through can also be initiated at a fixed end-shortening by increasing the clamp angle $\alpha$. This allows us to study an analogous experiment to the truss system: the arch is indented to an inverted position and held fixed, allowing stresses to relax for a specified duration, and then the indenter is abruptly removed.

Previously, we have studied the snap-through dynamics of this system in the case of a purely elastic material, both for shallow arches using beam theory \citep{gomez2017a} and for deeper arches using the dynamic elastica equations \citep{gomezthesis}. This analysis indicates that the bifurcation associated with snap-through is a saddle-node bifurcation, and so is qualitatively similar to the bifurcation observed in the Mises truss and spherical caps \citep{taffetani2018}. If instead the ends of the arch are held symmetrically (i.e.~at equal angle $\alpha$), the bifurcation changes type to a subcritical pitchfork. This alternative setup has been studied by \cite{brinkmeyer2013}, using a combination of finite element simulations and experiments on arches composed of the rubbery polymer Sylgard 182. In their viscoelastic model,  \cite{brinkmeyer2013} use a Prony series expansion for the Young's modulus, which assumes that the modulus can be written as a sum of exponentially decaying modes; the coefficients in the sum and the relaxation timescales are fitted to experimental data for the relaxation of Sylgard 182. Here, we instead start from the constitutive law of a standard linear solid, and solve the equations of motion coupling the stress in the arch to its displacement. In this way, we will show that the behaviour of the arch is well captured by the truss model: the different dynamical regimes map directly onto those obtained for the truss, and the snap-through time in the pseudo-bistable regime obeys the same inverse square-root scaling law due to the ghost of the saddle-node bifurcation.

\subsection{Theoretical formulation}
The properties of the strip are its density $\rho_s$, thickness $h$ and natural length $L$. We model the arch shape using beam theory; this requires a small thickness $h \ll L$, which guarantees that the strip remains in the limit of small strains \citep[for a further discussion see][]{gomezthesis}, as well as a shallow arch shape (i.e.~$\alpha \ll 1$ and $\Delta L \ll L$). Assuming the arch deforms only in the plane perpendicular to its width, we denote the transverse displacement by $w(x,t)$, where $x$ is the horizontal coordinate measured from the left end (figure \ref{fig:archsetup}a) and $t$ is time. We model the indentation force as a transverse point force applied at the arch midpoint with magnitude $f$. Under the above assumptions, the transverse displacement satisfies the dynamic beam equation \citep{howell}
\beq
\rho_s h \pdd{w}{t}+\Upsilon\pd{w}{t}-\pdd{m}{x}+P_c\pdd{w}{x} = -f\delta\left(x-\frac{L}{2}\right), \qquad 0 < x < L,
\label{eqn:beamdim}
\eeq
where $P_c(t)$ is the (unknown) compressive force applied to the arch and $m$ is the bending moment (each per unit width). Here we are also including viscous damping due to the environment, which is assumed to be linear in the velocity with constant coefficient $\Upsilon$ (per unit width). 

In this framework the SLS constitutive law is applied using the same equation as for the truss, i.e.~equation \eqref{CH5eqn:SLSdim} in \S\ref{sec:governingeqns}, where now we interpret $e$ as the axial strain field and $\sigma$ as the axial stress within the strip. After relating these quantities to the displacement $w$ and bending moment $m$ in the small-slope approximation using standard relations in beam theory \citep{wang2009}, and integrating to solve for $m$, the bending term in \eqref{eqn:beamdim} can be evaluated as
\beq
\pdd{m}{x} =- \frac{h^3}{12}(E_1+E_2)\pdf{w}{x} + e^{-\frac{E_2}{\eta}(t-t_0)}\left[\pdd{m}{x}+\frac{h^3}{12}(E_1+E_2)\pdf{w}{x}\right]\Bigg\lvert_{t = t_0}+\frac{h^3}{12}\frac{E_2^2}{\eta}\int_{t_0}^t e^{-\frac{E_2}{\eta}(t-\xi)}\pdf{w}{x}\Bigg\lvert_{t = \xi}\id\xi,
\label{eqn:bendingtermdim}
\eeq
where we assume initial data at $t = t_0$. As $t \to \infty$, we recover the usual Euler-Bernoulli law for an elastic solid with Young's modulus $E_1$, which here is written as
\beq
\pdd{m}{x} = -B\pdf{w}{x}, \label{eqn:EBlawdim}
\eeq
where $B = E_1 h^3/12$ is the fully relaxed bending stiffness (per unit width). \citep[This can be derived from \eqref{eqn:bendingtermdim} by using a Watson-lemma type argument  to evaluate the integral in the final term to leading order as $t \to \infty$; see][for example.]{hinch} With the Euler-Bernoulli law \eqref{eqn:EBlawdim}, equation \eqref{eqn:beamdim} reduces to the usual dynamic beam equation in the particular case of zero external damping and zero indentation force, $\Upsilon = f = 0$  \citep{gomez2017a}. 

The boundary conditions at the clamped ends are (subscripts denoting differentiation)
\beq
w(0,t) = 0, \quad w_x(0,t) = \alpha, \quad w(L,t) = w_x(L,t)  = 0. \label{eqn:clampdim}
\eeq
Neglecting the effects of extensibility, which is valid for shallow arches while $h \ll L$ \citep{pandey2014}, the imposed end-shortening may be approximated as 
\beq
\int_0^L\left(\pd{w}{x}\right)^2\id x = 2\Delta L. \label{eqn:eqn:endshortdim}
\eeq
Equations \eqref{eqn:beamdim}--\eqref{eqn:bendingtermdim}, the boundary conditions \eqref{eqn:clampdim} and constraint \eqref{eqn:eqn:endshortdim}, together with appropriate initial conditions, then fully specify the problem.

\subsubsection{Non-dimensionalisation}
To make the problem dimensionless, it is convenient to scale the horizontal coordinate with the natural length $L$ of the strip. The end-shortening constraint \eqref{eqn:eqn:endshortdim} provides a natural vertical lengthscale $w \sim (L\Delta L)^{1/2}$; using the Euler-Bernoulli law \eqref{eqn:EBlawdim}, this is associated with a typical bending moment $m \sim B(L\Delta L)^{1/2}/L^2$, which motivates introducing 
\beqn
x = L X, \quad w = (L\Delta L)^{1/2} W, \quad t = \frac{\eta}{E_2}T, \quad m = \frac{B (L\Delta L)^{1/2}}{L^2}M.
\eeqn
Here we again scale time with the timescale of stress relaxation, $t \sim \eta/E_2$. In terms of these variables, the beam equation \eqref{eqn:beamdim} becomes
\beq
\mathrm{De}^{-2}\pdd{W}{T}+\upsilon\pd{W}{T}-\pdd{M}{X}+\tau^2\pdd{W}{X}=-F\delta\left(X-\frac{1}{2}\right), \qquad 0 < X < 1,
\label{eqn:beamnondim}
\eeq
where we have introduced
\beqn
\mathrm{De} = \frac{\eta/E_2}{\sqrt{\rho_s h L^4/B}}, \quad \upsilon = \frac{L^4 E_2 \Upsilon}{B \eta}, \quad \tau^2 = \frac{L^2 P_c}{B}, \quad F = \frac{L^3 f}{B(L\Delta L)^{1/2}}.
\eeqn
The Deborah number here measures the ratio of the viscous timescale $\eta/E_2$ to the timescale of undamped elastic oscillations, $t^{*} \sim \sqrt{\rho_s h L^4/B}$, so is analogous to the Deborah number defined in \S\ref{sec:trussnondim} for the truss. The other parameters correspond to the dimensionless values of the damping coefficient, compressive force and indentation force, respectively. 

Inserting the above rescalings into the bending term \eqref{eqn:bendingtermdim}, we obtain
\beq
(1-\beta)\pdd{M}{X} = -\pdf{W}{X} + e^{-(T-T_0)}\left[(1-\beta)\pdd{M}{X}+\pdf{W}{X}\right]\Bigg\lvert_{T = T_0}+\beta \int_{T_0}^T e^{-(T-\xi)}\pdf{W}{X}\Bigg\lvert_{T = \xi}\id\xi,
\label{eqn:bendingtermnondim}
\eeq
where $\beta$ is defined as in \S\ref{sec:trussnondim} and $T_0 = t_0/(\eta/E_2)$. The clamped boundary conditions \eqref{eqn:clampdim} become
\beq
W_X(0,T) = \mu, \quad W(0,T) = W(1,T) = W_X(1,T) = 0, \label{eqn:clampnondim}
\eeq
where we have introduced the normalised clamp angle
\beqn
\mu = \alpha\left(\frac{\Delta L}{L}\right)^{-1/2}.
\eeqn
Finally, the imposed end-shortening \eqref{eqn:eqn:endshortdim} becomes
\beq
\int_0^1\left(\pd{W}{X}\right)^2\id X = 2. \label{eqn:endshortnondim}
\eeq
%Note that, in this non-dimensionalisation, the Euler-Bernoulli law for an elastic solid corresponds to $M = -\partial^2 W/\partial X^2$. 

\subsection{Steady bifurcation behaviour}
In the absence of any indentation force, $F = 0$, equilibrium solutions obey the steady beam equation
\beqn
\df{W}{X}+\tau^2\dd{W}{X} = 0, \qquad 0 < X < 1,
\eeqn
together with the clamped boundary conditions \eqref{eqn:clampnondim} and end-shortening \eqref{eqn:endshortnondim}. The equilibrium behaviour of the arch is therefore entirely characterised by the geometric parameter $\mu$ (since $\tau$ is unknown and is determined as part of the solution). It is possible to solve the equilibrium problem analytically \citep{gomez2017a}, which indicates that for $0 < \mu < \mu_{\mathrm{fold}} \approx 1.7818$, both inverted and natural equilibrium shapes exist and are linearly stable. The critical value $ \mu = \mu_{\mathrm{fold}}$ corresponds to a saddle-node bifurcation where the inverted shape intersects a higher, unstable mode and disappears; for later reference we write $W_{\mathrm{fold}}(X)$ for the equilibrium shape at the bifurcation point, which has midpoint height $W_{\mathrm{fold}}(1/2) \approx -0.3476$. For values $ \mu > \mu_{\mathrm{fold}}$, only the natural shape exists and is stable. The bifurcation diagram is shown in figure \ref{fig:archsetup}b. The parameter $\mu$ therefore plays an analogous role to the stiffness parameter $\lambda$ in the truss model (compare figure \ref{fig:archsetup}b to figure \ref{CH5fig:steadyresponse}b in \S\ref{sec:trusssteady}).

\subsection{Snap-through dynamics}
In a numerical snap-through experiment, we suppose that the arch is initially fully relaxed in the natural shape, labelled $W_{\mathrm{nat}}(X)$, for each value of $\mu$. For $-T_{\mathrm{ind}} < T < 0$ we then rapidly indent the arch to an inverted position with midpoint displacement $W_{\mathrm{mid}} < 0$. Typically we specify $W_{\mathrm{mid}} = W_{\mathrm{fold}}(1/2)$, i.e.~the midpoint displacement at the saddle-node bifurcation, which is the analogue of the baseline value used for the truss system. With the imposed midpoint displacement, we integrate \eqref{eqn:beamnondim}--\eqref{eqn:endshortnondim}  with initial conditions
\beqn
W(X,-T_{\mathrm{ind}}) = W_{\mathrm{nat}}(X), \quad \dot{W}(X,-T_{\mathrm{ind}}) = 0, \quad M(X,-T_{\mathrm{ind}}) = -\dd{W_{\mathrm{nat}}}{X}.
\eeqn
For the release problem, i.e.~$T > 0$, we instead impose zero indentation force $F= 0$ during the integration. Because $W$ is continuous across $T = 0$, we have that $M$ is also continuous, giving the initial conditions
\beq
W(X,0+) = W(X,0-), \quad \dot{W}(X,0+) = \dot{W}(X,0-), \quad M(X,0+) = M(X,0-). \label{eqn:archreleaseICs}
\eeq

%At the end of the indentation stage, i.e.~at $T = 0-$, we use the subscript $\mathrm{ind}$ to denote the values of these variables:
%\beqn
%W_{\mathrm{ind}}(X) = W(X,0-), \quad \dot{W}_{\mathrm{ind}}(X) = \dot{W}(X,0-), \quad I_{\mathrm{ind}}(X) = I(X,0-).
%\eeqn

We solve the dynamic equations using the method of lines \citep{morton}, i.e.~we discretise using finite differences in space to obtain a system of ODEs in time; details of the numerical methods are provided in Appendix F. The ODEs are integrated numerically in \textsc{matlab} using the \texttt{ode45} routine (relative and absolute error tolerances  $10^{-8}$, maximum time step $10^{-4}$). We have verified second-order accuracy in the convergence of our scheme, and also that the numerical drift in the end-shortening constraint remains on the order of the integration tolerances. Because we are interested in performing a large number of simulations as $\mu$ and $T_{\mathrm{ind}}$ are varied, we discretise using $N = 50$ grid points for all simulations reported in this section; we have checked that this yields quantitatively similar results compared to using a larger number of grid points, e.g.~$N = 100$, but requires much less computing time. We also specify a dimensionless damping coefficient $\upsilon = 0.5$, which provides sufficient numerical damping (needed due to the fast motions during the indentation stage) while ensuring the arch motions remain underdamped.

\begin{figure}
\centering
\includegraphics[width=0.7\textwidth]{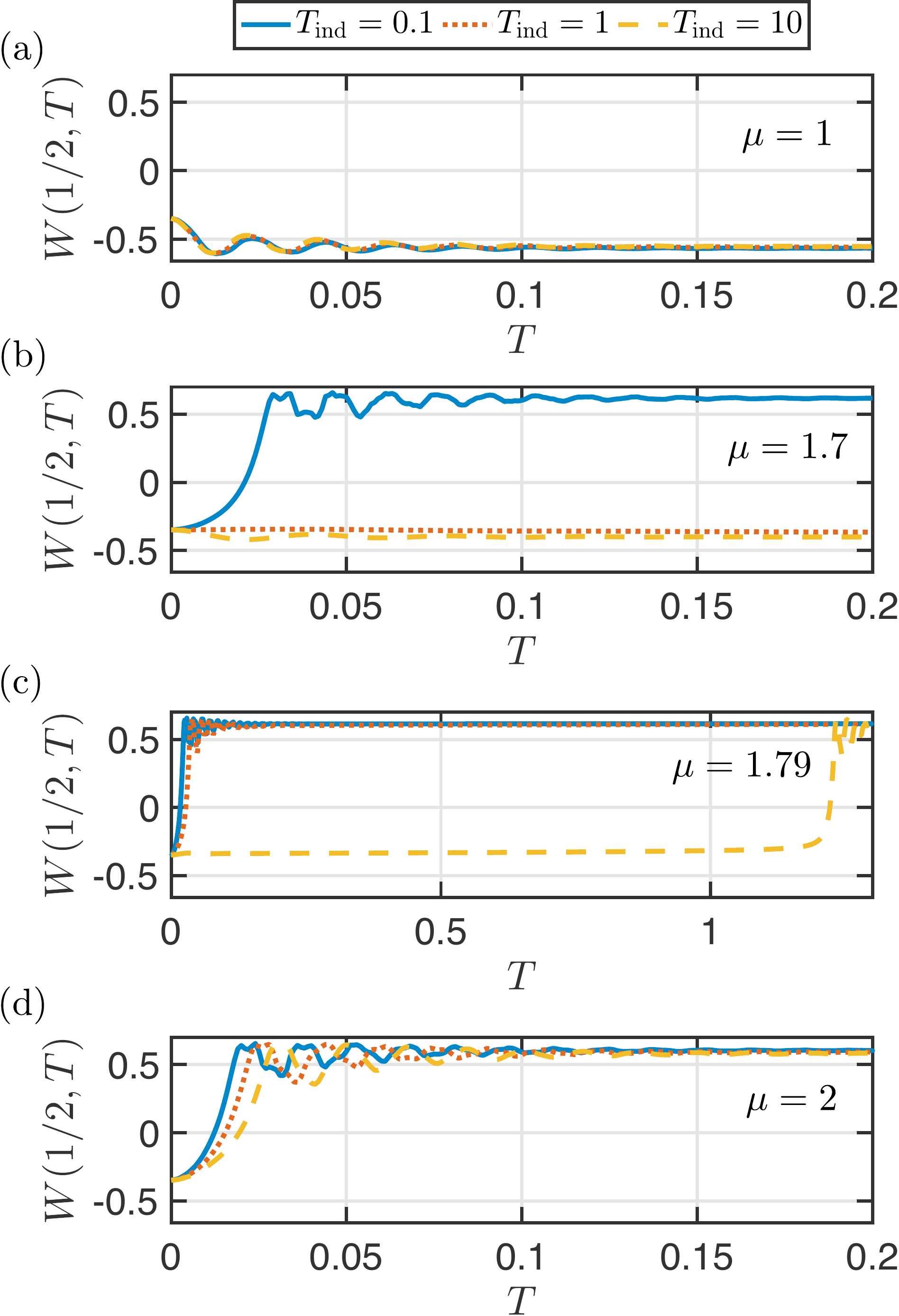} 
\caption{Dimensionless trajectories of the arch midpoint during the release stage, obtained by numerical integration of  \eqref{eqn:beamnondim}--\eqref{eqn:endshortnondim} (with $F = 0$) with initial conditions \eqref{eqn:archreleaseICs} ($W_{\mathrm{mid}} = W_{\mathrm{fold}}(1/2)$, $\mathrm{De} = 10$, $\beta = 0.1$, $\upsilon = 0.5$, $N = 50$). Here data is shown for (a) $\mu = 1$, (b) $\mu = 1.7$ ($<\mu_{\mathrm{fold}} \approx 1.7818$), (c) $\mu = 1.79$  ($>\mu_{\mathrm{fold}} \approx 1.7818$) and (d) $\mu = 2$. In each panel, trajectories associated with three different values of the indentation time $T_{\mathrm{ind}}$ (given in the upper legend) are shown. Note the range of times plotted is larger in panel (c).}
\label{CH5fig:archtrajectories}
\end{figure}

Typical trajectories of the arch midpoint, $W(1/2,T)$, during the release stage are shown in figures \ref{CH5fig:archtrajectories}a--d. Here we have specified $\mathrm{De} = 10$ and $\beta = 0.1$; this value of $\beta$ is an approximate value measured experimentally by \cite{urbach2017} for silicone rubber shells. (While pseudo-bistable behaviour is obtained with $\beta  = 0.5$ for the truss, we will show that smaller values are required for the arch system.) Figure \ref{CH5fig:archtrajectories} shows that the arch exhibits similar dynamical regimes to the truss model, with $\mu$ playing the role of the stiffness parameter $\lambda$: for $\mu \lesssim (1-\beta)\mu_{\mathrm{fold}}$ ($\approx 1.6$ with $\beta = 0.1$) the arch never snaps (figure \ref{CH5fig:archtrajectories}a); for $(1-\beta)\mu_{\mathrm{fold}} \lesssim \mu \lesssim \mu_{\mathrm{fold}}$ the arch snaps if $T_{\mathrm{ind}}$ is sufficiently small (figure \ref{CH5fig:archtrajectories}b); and for $\mu \gtrsim \mu_{\mathrm{fold}}$ the arch appears to snap for any indentation time (figures \ref{CH5fig:archtrajectories}c--d). Pseudo-bistable behaviour is obtained when $0 < \mu-\mu_{\mathrm{fold}} \ll 1$ and $T_{\mathrm{ind}}$ is sufficiently large (figure \ref{CH5fig:archtrajectories}c). Note that the oscillations are evidently underdamped, but do not persist long (compared to the truss, figure \ref{CH5fig:largeDetrajectories}) due to the presence of external damping.

\begin{figure}
\centering
\includegraphics[width=\textwidth]{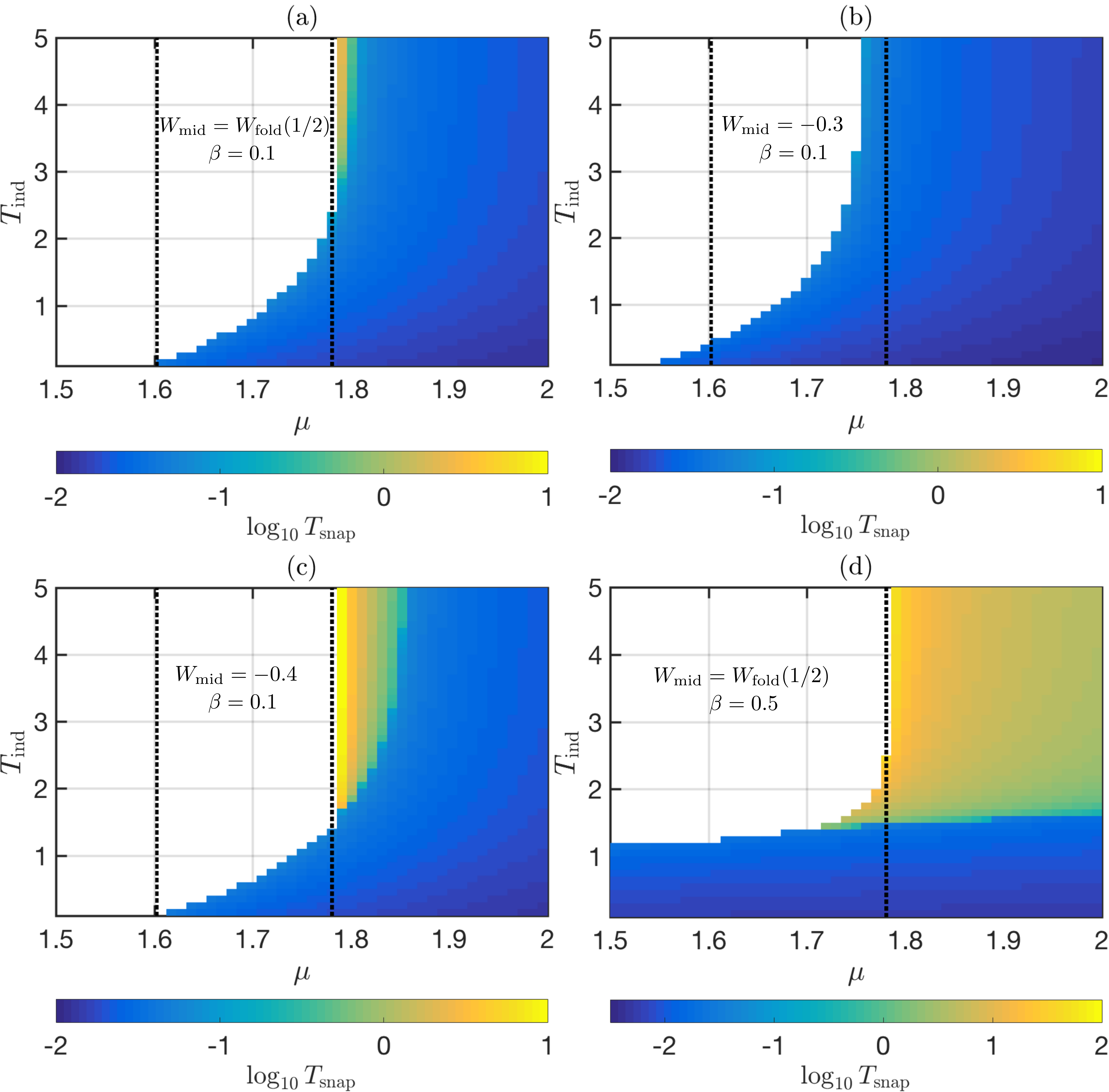} 
\caption{Snap-through times of the arch when $\mathrm{De} \gg 1$ for different indentation depths and relaxation parameter $\beta$ ($\mathrm{De} = 10$, $\upsilon = 0.5$, $N = 50$). Numerical results are shown for (a) $W_{\mathrm{mid}} = W_{\mathrm{fold}}(1/2) \approx -0.3476$ and $\beta = 0.1$; (b) $W_{\mathrm{mid}} = -0.3$ and $\beta = 0.1$; (c) $W_{\mathrm{mid}} = -0.4$ and $\beta = 0.1$; and (d) $W_{\mathrm{mid}} = W_{\mathrm{fold}}(1/2)$ and $\beta = 0.5$. The critical values $\mu = (1-\beta)\mu_{\mathrm{fold}}$ and $\mu = \mu_{\mathrm{fold}}$ are plotted as vertical black dotted lines.  In each panel, the snap-through times have been computed on a $50 \times 50$ grid of equally spaced values. Note the range of the colourbar is larger in panel (d).}
\label{CH5fig:archsnapplane}
\end{figure}

The different regimes just discussed are also evident when we analyse the snap-through times on the $(\mu,T_{\mathrm{ind}})$-plane; the baseline case $W_{\mathrm{mid}} = W_{\mathrm{fold}}(1/2)  \approx -0.3476$ and $\beta = 0.1$ is shown in figure \ref{CH5fig:archsnapplane}a. (Because the displacement of the natural shape depends on $\mu$, we define $T_{\mathrm{snap}}$ to be the time when the arch midpoint first crosses $W = 0$.) We observe very similar features to the analogous plot for the truss system (compare figure \ref{CH5fig:archsnapplane}a to figures \ref{CH5fig:largeDesnapsurface}a--b in \S\ref{CH5sec:dynamicslargeDe}), in that (i) the critical value of $T_{\mathrm{ind}}$ at which snap-through no longer occurs depends nonlinearly on $\mu$, and approaches a finite value as $\mu \nearrow \mu_{\mathrm{fold}}$; (ii) pseudo-bistability occurs only in a narrow region where $0 < \mu-\mu_{\mathrm{fold}} \ll 1$ and $T_{\mathrm{ind}}$ is larger than the critical value obtained at $\mu = \mu_{\mathrm{fold}}$; and (iii) slowing down occurs in the pseudo-bistable regime as $\mu \searrow \mu_{\mathrm{fold}}$.

The dependence on the indentation displacement $W_{\mathrm{mid}}$ and relaxation parameter $\beta$ is also analogous to that in the truss model. For shallower indentation depths $W_{\mathrm{mid}} > W_{\mathrm{fold}}(1/2)$ (corresponding to $X_{\mathrm{ind}} < X_{\mathrm{fold}}$ for the truss), the boundary at which snap-through no longer occurs is shifted  to the left of the line $\mu = \mu_{\mathrm{fold}}$ and pseudo-bistability is never obtained; see figure \ref{CH5fig:archsnapplane}b. For a deeper indentation $W_{\mathrm{mid}} < W_{\mathrm{fold}}(1/2)$ (corresponding to $X_{\mathrm{ind}} > X_{\mathrm{fold}}$ for the truss), the boundary may shift to the right so that the pseudo-bistable region is enlarged (figure \ref{CH5fig:archsnapplane}c).  When $\beta$ is increased, this picture breaks down as the snap-through time is $O(1)$ throughout a large portion of the $(\mu,T_{\mathrm{ind}})$-plane; see figure \ref{CH5fig:archsnapplane}d. Similar to the truss (figure \ref{CH5fig:largeDesnapsotherbeta}b in Appendix A), an analysis of the trajectories in this regime confirms that this behaviour is different to pseudo-bistability, being instead a consequence of the extremely slow creep timescale associated with larger values of $\beta$. We note that because this behaviour is obtained with $\beta = 0.5$ here, as opposed to larger values for the truss, the value of $\beta$ does not map directly between the two systems. 

\begin{figure}
\centering
\includegraphics[width= \textwidth]{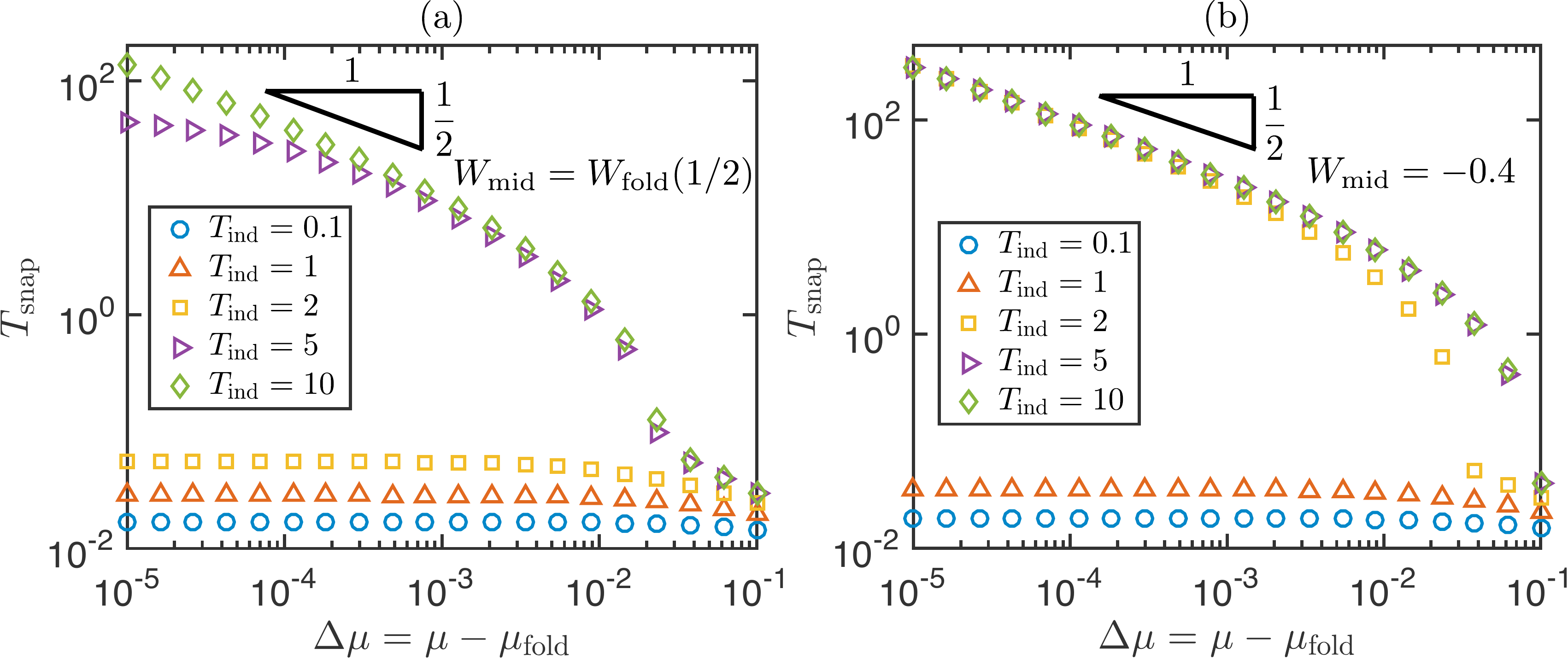} 
\caption{Snap-through times of the arch in the pseudo-bistable regime for different indentation time $T_{\mathrm{ind}}$ ($\mathrm{De} = 10$, $\beta = 0.1$,  $\upsilon = 0.5$, $N = 50$). Numerical results are shown for (a) $W_{\mathrm{mid}}= W_{\mathrm{fold}}(1/2) \approx -0.3476$ and  (b) a deeper indentation depth $W_{\mathrm{mid}} = -0.4$ (symbols; see legends).}
\label{CH5fig:archpseudobistabletimes}
\end{figure}

We have also simulated the snap-through times in the pseudo-bistable regime, setting $\mu = \mu_{\mathrm{fold}}+\Delta\mu$ where $0 < \Delta\mu \ll 1$; data for two different indentation depths are shown in figures \ref{CH5fig:archpseudobistabletimes}a--b. These confirm that the expected inverse square-root scaling $T_{\mathrm{snap}} \propto \Delta\mu^{-1/2}$ is obtained as $\Delta\mu \to 0$ when $T_{\mathrm{ind}}$ is sufficiently large. While the data deviate significantly from the scaling for moderately small values $\Delta\mu \gtrsim 10^{-3}$, this is similar to the behaviour we have seen in the truss model, in which the error of the asymptotic prediction becomes significant if $\epsilon$ is not too small (compare to figures \ref{CH5fig:pseudobistabletimes}a--b). Moreover, as with the truss model, this error is less significant for a deeper indentation depth (figure \ref{CH5fig:archpseudobistabletimes}b), since the snap-through time is generally larger in this case compared to the baseline value,  $W_{\mathrm{mid}}= W_{\mathrm{fold}}(1/2)$ (figure \ref{CH5fig:archpseudobistabletimes}a).

In summary, the truss provides an excellent lumped model of the continuous arch system, providing qualitatively correct predictions of the different dynamical regimes, the features of pseudo-bistability and the scaling laws for the snap-through time.

\section{Data comparison}
\label{sec:datacompare}
To further examine whether the inverse square-root scaling law for the snap-through time in the pseudo-bistable regime holds more generically, in this final section we examine numerical and experimental data for the snap-through times of viscoelastic shells and arches reported in the literature. As was the case in our numerical experiments, each structure is held in an inverted state for a duration $t_{\mathrm{ind}}$, before being instantaneously released. The snap-through time is measured to be the time taken between release and when the structure rapidly accelerates towards its natural state. We focus on results in which the indentation time $t_{\mathrm{ind}}$ is fixed, while the analogue of the bifurcation parameter $\lambda$ or $\mu$ is varied between each snapping experiment. In all cases examined the structure exhibits pseudo-bistability, undergoing a slow creeping motion followed by a rapid snap-through, so that the snap-through times are easily measured. Where data is only available graphically, we have extracted the values using the WebPlotDigitizer (arohatgi.info/WebPlotDigitizer).

A summary of the conditions for each data set is provided in table \ref{CH5table:viscodata}. Here we have separated the data so that only a single parameter is varying within each data set (corresponding to a particular row in the table), and we have provided the relevant parameter values. These are the shell/arch thickness $h$, relevant horizontal lengthscale $l$ (defined to be the base diameter of the shell/natural length of the arch), fully relaxed Young's modulus $E$, Poisson ratio $\nu$, material density $\rho_s$, viscous timescale $[t]_{\mathrm{vis}}$, elastic timescale $t^*$, and the indentation time $t_{\mathrm{ind}}$. (Where a parameter varies within a data set, the range of values is provided, and we use the average to compute $t^{*}$.) The simulation data reported by \cite{brinkmeyer2012,brinkmeyer2013} assume a Prony series expansion for the Young's modulus, so there is no single viscous timescale; we estimate $[t]_{\mathrm{vis}}$ by the timescale that appears in the dominant term of the Prony series (i.e.~the term with the largest coefficient). To estimate the elastic timescale, we balance inertial forces and bending forces for a shell/arch to obtain \citep{ventsel01,gomez2017a}
\beqn
t^{*} \sim \left(\frac{\rho_s h l^4}{B}\right)^{1/2},
\eeqn 
where $l$ is the horizontal lengthscale defined above, and $B$ is the bending stiffness (in particular $B = Eh^3/[12(1-\nu^2)]$ for a shell, and $B = Eh^3/12$ for an arch). 

Table \ref{CH5table:viscodata} shows that for all data sets, the viscous timescale $[t]_{\mathrm{vis}}$ is an order of magnitude larger than the elastic timescale $t^*$, so these systems are effectively in the large Deborah number limit. The indentation times  $t_{\mathrm{ind}}$ are  also much larger than $[t]_{\mathrm{vis}}$, so that the dimensionless indentation times $T_{\mathrm{ind}} = t_{\mathrm{ind}}/[t]_{\mathrm{vis}}$ are large. Hence, the pseudo-bistability observed in these systems occurs in an analogous parameter range to that in our truss and arch models. 

%%%%%%%%%%%%%%%%%%%%%%%%%%%%%%%%%%%%%%%%%%%
\begin{table}
\caption{Summary of previous data for pseudo-bistable snap-through times reported in the literature. \label{CH5table:viscodata}}
\vspace{10pt}
\footnotesize{ 
    \begin{tabular}{m{1.9cm}m{1.2cm}m{1.2cm}m{0.7cm}m{0.7cm}m{0.6cm} m{0.6cm}m{0.8cm}m{0.7cm}m{0.7cm}m{0.2cm}m{2cm}m{0.8cm}}
    \hlineB{5}
    
  Reference & System & Varying &  $h$ & $l$ & $E$ & $\nu$  & $\rho_s$  & $[t]_{\mathrm{vis}}$ & $t^*$ &  $t_{\mathrm{ind}}$ & Fitted & Legend \\
   & & &  $ (\mathrm{mm})$ & $ (\mathrm{mm})$ & $(\mathrm{MPa})$ &   & $(\mathrm{kg}\mathrm{m}^{-3})$  &  $(\mathrm{s})$ &   $(\mathrm{s})$ & $(\mathrm{s})$ & exponent  &  \\ 
   \hlineB{5}
           \cite{brinkmeyer2012} & Spherical shell, S & Thickness & $(5.16$, $5.25)$ & $54.16$ & $0.935$ & $0.469$ & $1030$ & $0.634^*$ & $0.0572$ & $10$ & $-1.22$  &  \includegraphics[scale = 1]{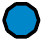}  \\    \hline 
            \cite{brinkmeyer2012} & Spherical shell, S & Thickness$^{\dagger}$ & $(5.16$, $5.72)$ & $54.16$ & $0.935$ & $0.469$ & $1030$ & $0.634^*$ & $0.0548$ & $10$ & $-3.38$ &  \includegraphics[scale = 1]{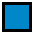}  \\  \hline
           \cite{brinkmeyer2012} & Spherical shell, S & Depth & $5.2$ & $(54.1$, $54.2)$ & $0.935$ & $0.469$ & $1030$ & $0.634^*$ & $0.0573$ & $10$ & $-1.36$  &  \includegraphics[scale = 1]{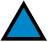}  \\  \hline
                        \cite{brinkmeyer2013} & Buckled arch, S & End-shortening & $2.5$ & $100$ & $0.935$ & $0.469$ & $1035$ & $3.66^*$ & $0.461$ & $10$ & $-0.532$  &  \includegraphics[scale = 1]{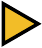}  \\   \hline  
                        \cite{brinkmeyer2013} & Buckled arch, S & Clamp angle & $2.5$ & $100$ & $0.935$ & $0.469$ & $1035$ & $3.66^*$ & $0.461$ & $10$ & $-0.599$  &  \includegraphics[scale = 1]{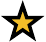}  \\  \hline
                        \cite{brinkmeyer2013} & Buckled arch, E & Clamp angle & $2.5$ & $100$ & $0.935$ & $0.469$ & $1035$ & $3.66^*$ & $0.461$ & $10$ & $-0.316$  &  \includegraphics[scale = 1]{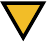}  \\   \hline   
             \cite{urbach2017} & Conical shell, S & Thickness & $(5.36$, $5.76)$ & $50$ & $2.5$ & $0.47$ & N/A & $0.1$ & N/A & $60$ & $-0.396$ &  \includegraphics[scale = 1]{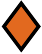}  \\       
	            \hlineB{5}
    \end{tabular}
    \\
    \\
S, data from numerical simulations; E, experimental data; N/A, not applicable due to absence of inertia in simulations. \\ 
$^*$Estimated from the dominant term of the Prony series expansion. \\
$^{\dagger}$Using a larger material relaxation in the Prony series.
}
\end{table}
%%%%%%%%%%%%%%%%%%%%%%%%%%%%%%%%%%%%%%%%%%%

\begin{figure}
\centering
\includegraphics[width=\textwidth]{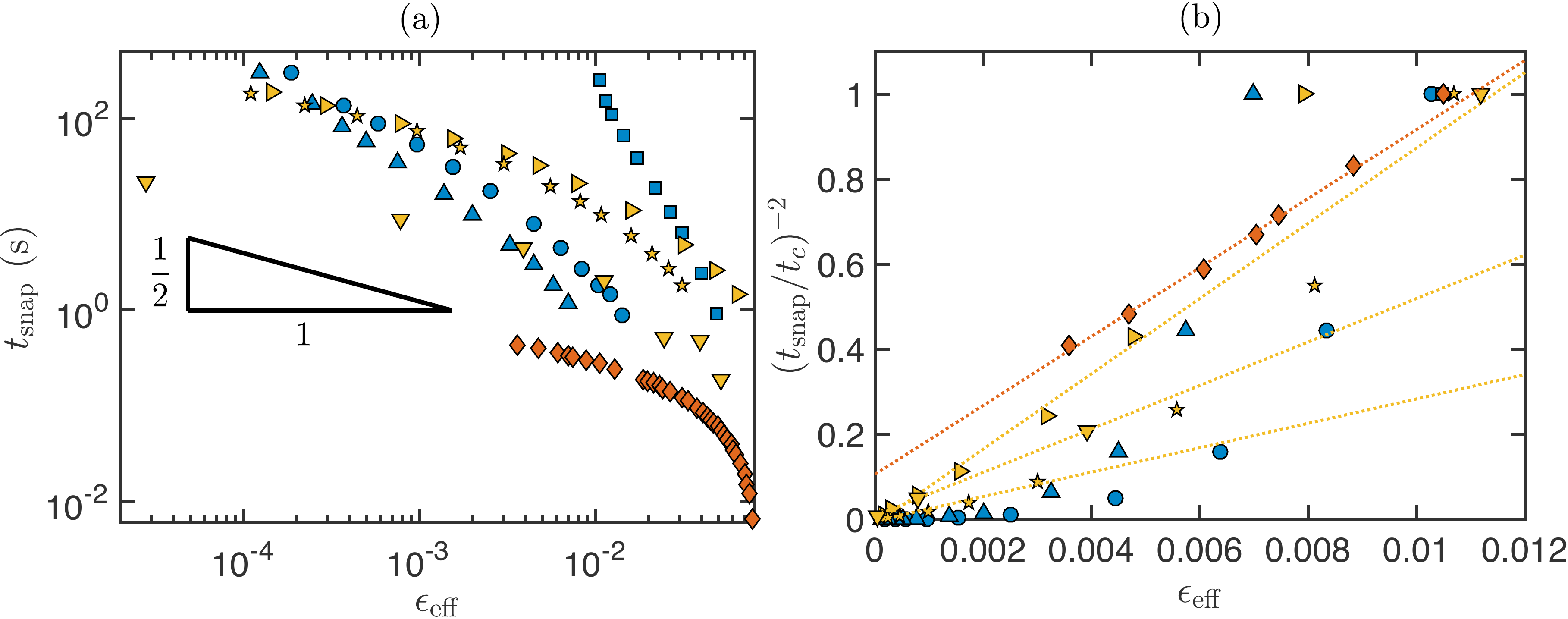} 
\caption{Snap-through times of viscoelastic arches and shells reported in the literature to exhibit pseudo-bistability. For a legend and the parameters used in each data set, see table \ref{CH5table:viscodata}. (a) Dimensional snap-through times as a function of the normalised distance to bifurcation, $\epsilon_{\mathrm{eff}}$. (b) Plotting $(t_{\mathrm{snap}}/t_c)^{-2}$ as a function of $\epsilon_{\mathrm{eff}}$; for each data set the timescale $t_c$ is equal to $t_{\mathrm{snap}}$ at the value of $\epsilon_{\mathrm{eff}}$ closest to  $10^{-2}$.}
\label{CH5fig:comparedata}
\end{figure}

For each data set it is found that as the analogue of $\lambda$ varies, the snap-through time increases rapidly and appears to diverge near a critical value. Beyond this transition no snap-through occurs. This transition therefore appears to be the saddle-node bifurcation at which the inverted arch/shell becomes bistable\footnote{For the buckled arch considered by \cite{brinkmeyer2013}, we discussed at the start of \S\ref{sec:arch} how the bifurcation is a subcritical pitchfork if the ends are clamped at equal angles. However, we might expect this to `unfold' to a saddle-node bifurcation in the presence of material imperfections, in a similar way to other buckling instabilities; see \cite{hayman1978} and \cite{bushnell1981} for example.}. We use the critical value that is reported to compute the normalised distance to the bifurcation, which we denote by $\epsilon_{\mathrm{eff}}$; for example, if the thickness $h$ is varied and $h_c$ is the critical value when snap-through no longer occurs, then we define $\epsilon_{\mathrm{eff}} = |h-h_c|/|h_c|$ and similarly when other parameters are varied. The dimensional snap-through times are plotted as a function of  $\epsilon_{\mathrm{eff}}$ on logarithmic axes in figure \ref{CH5fig:comparedata}a. Here we observe the characteristic signs of critical slowing down: as $\epsilon_{\mathrm{eff}} \to 0$ the snap-through time increases systematically, varying by over two orders of magnitude within a very narrow range of $\epsilon_{\mathrm{eff}}$. 

The key observation is that the data of \cite{brinkmeyer2013} (yellow symbols) and \cite{urbach2017} (orange symbols) are approximately consistent with an inverse square-root scaling law, i.e.~$t_{\mathrm{snap}} \propto \epsilon_{\mathrm{eff}}^{-1/2}$ as $\epsilon_{\mathrm{eff}} \to 0$. To be more quantitative, we fit each data set to a power law of the form $t_{\mathrm{snap}} \propto \epsilon_{\mathrm{eff}}^{\gamma}$ using least squares. We restrict the fit to values $\epsilon_{\mathrm{eff}} \leq 10^{-2}$, since this is the range where the inverse square-root scaling is observed for the truss and arch systems (see figures \ref{CH5fig:pseudobistabletimes} and  \ref{CH5fig:archpseudobistabletimes}). (One data set of \cite{brinkmeyer2012}, blue squares, has no values of $\epsilon_{\mathrm{eff}} \leq 10^{-2}$ so here we instead fit the six points closest to bifurcation.) The best-fit exponents are provided in table \ref{CH5table:viscodata}. This confirms that for the data reported by \cite{brinkmeyer2013} and \cite{urbach2017} the fitted values are dispersed around $\gamma = -0.5$, lying in the range $\gamma \in (-0.599,-0.316)$. We note that such dispersion is expected due to the small parameter range over which the data exhibit a power law: the fitted value of $\gamma$ is  sensitive to the precise range of values of $\epsilon_{\mathrm{eff}}$ used for fitting. In addition, this dispersion may be due to sensitivity to the precise value of the bifurcation point used to calculate $\epsilon_{\mathrm{eff}}$: a small error in this value (e.g.~rounding error in the reported value) introduces shifts in the values of $\epsilon_{\mathrm{eff}}$, which can cause large variations when plotted on logarithmic axes. To eliminate this second type of sensitivity we plot $t_{\mathrm{snap}}^{-2}$ as a function of $\epsilon_{\mathrm{eff}}$ on linear axes, where a linear relationship indicates that the inverse square-root scaling is obeyed. This plot is shown in figure \ref{CH5fig:comparedata}b, focussing on values  $\epsilon_{\mathrm{eff}} \lesssim 10^{-2}$ where we expect to observe the inverse square-root scaling; to give a clearer plot here we have re-scaled time so that all data sets approximately pass through $(10^{-2},1)$. Figure \ref{CH5fig:comparedata}b shows that the data of \cite{brinkmeyer2013} and \cite{urbach2017} become approximately linear as $\epsilon_{\mathrm{eff}} \to 0$: for each data set the dotted line is the best-fit (least squares) line over the six data points that are closest to the bifurcation point (for the experimental data of \cite{brinkmeyer2013}, yellow downward-pointing triangles, three data points lie in the range $\epsilon_{\mathrm{eff}} \leq 10^{-2}$ so we fit only these). Nevertheless, in the absence of more data points in this parameter range, it is not possible to state conclusively that the inverse square-root scaling law holds for these systems.

In contrast, the snap-through times reported by \cite{brinkmeyer2012} (blue symbols) consistently do not follow the inverse square-root scaling: the best-fit exponents $\gamma \lesssim -1$ and a linear relationship is not observed in figure \ref{CH5fig:comparedata}b (for clarity we do not plot the best-fit lines on figure \ref{CH5fig:comparedata}b for these data). Nevertheless, the qualitative features of the pseudo-bistable regime, including the sensitivity of the snap-through time to changes to $\epsilon_{\mathrm{eff}}$, is well captured by the truss model. 

\section{Discussion and conclusions}
\label{CH5sec:conclusions}
In this paper, we have analysed the dynamics of snap-through when viscoelastic effects are present. We re-emphasise that such effects are fundamentally different to external damping such as viscous drag: viscoelasticity modifies the bistability characteristics and hence can change not only how, but also \emph{when} snap-through occurs. Moreover, in an experiment where we first indent a structure to a particular configuration, the resulting dynamics depend on the history of stress relaxation during the indentation phase. Previous approaches have dealt with this complexity by modelling the structure as being elastic with an effective stiffness that evolves according to a Prony series ansatz \citep{santer2010,brinkmeyer2012,brinkmeyer2013}. Here, we have presented an alternative approach that derives the equations of motion from first principles using the constitutive law of a standard linear solid. This enables us to capture both stress relaxation and creep phenomena without making any additional assumptions regarding the dynamic behaviour.

To gain analytical insight we first studied a modified form of the Mises truss, a simple and commonly used model system that exhibits bistability and snap-through \citep{panovko,krylov2008,brinkmeyer2013}. By introducing an additional vertical spring and a point mass in our formulation, the truss becomes a more realistic lumped model for more complex structures such as spherical shells and arches. Using a small-angle approximation, we were able to reduce the number of dimensionless parameters in our problem to five. These are the Deborah number $\mathrm{De}$, measuring the importance of viscosity compared to inertia; the relaxation parameter $\beta$, which measures the ability of the structure to relax its stress; the relative stiffness $\lambda$, which acts as a bifurcation parameter and determines the bistability characteristics of the truss; and the details of the indentation stage are specified by the indentation displacement $X_{\mathrm{ind}}$ and duration $T_{\mathrm{ind}}$. Regarding the truss as a lumped model, we then expect that analogous parameters will control the dynamics in more complex viscoelastic structures; for example, $\lambda$ may be compared to the  F\"{o}ppl-von-K\'{a}rm\'{a}n number for spherical shells.

We focussed on the dynamics when $\mathrm{De}$ is large. Using direct numerical solutions, we showed that the onset of snap-through cannot be inferred by whether or not the truss is effectively bistable at the moment the indenter is released. Instead, we turned to a detailed asymptotic analysis of the snap-through dynamics using the method of multiple scales. This analysis showed that the leading-order dynamics generally obey the equations of motion when we neglect the terms associated with inertia, as expected. However, immediately after the indenter is released, inertial effects become important, as the displacement moves rapidly to a new value to balance the jump in the applied force. It is this purely elastic behaviour at early times that determines whether the truss then creeps in an inverted state or immediately jumps back to near its natural configuration. In this way, we were able to build up a complete picture of the different dynamical regimes and determine precisely when pseudo-bistable behaviour is obtained (figure \ref{CH5fig:largeDeschematic}). Our analysis describes many features of pseudo-bistability that have been reported previously in experiments and numerical simulations, such as why a minimum indentation depth $X_{\mathrm{ind}}$ and duration $T_{\mathrm{ind}}$  are needed to obtain any creep behaviour. We then analysed an analogous indentation experiment performed on a pre-buckled arch, a simple  and common prototype of a continuous viscoelastic structure. By solving the dynamic equations numerically,  we were able to confirm that our conclusions for the truss model are qualitatively accurate for the arch, with the normalised clamp angle $\mu$ playing the role of the stiffness parameter $\lambda$. The features of pseudo-bistability that we predict are also readily observed in a commercially available popper toy: this needs to be turned sufficiently far inside-out, and held for a few seconds, in order to not immediately jump upwards when placed on a surface.

In the pseudo-bistable regime, the truss undergoes a creeping motion until a rapid snap-back occurs. In our leading-order description of the dynamics, this snapping event is associated with an infinite velocity, implying that inertial effects must become important again. This is very similar to the analysis of creep buckling, in which an infinite velocity is used as a criterion to determine the onset of instability \citep{hayman1978}. Many studies on creep buckling consider only force-control situations, such as a force of constant magnitude suddenly applied to a structure. This means that the Kelvin-Voigt model is often sufficient to study the dynamics \cite[see][for example]{nachbar1967}. The snap-back considered here is considerably more complicated: due to the initial indentation stage, both the history of stress relaxation and inertial effects at early times must be accounted for. Pseudo-bistable snap-through then requires a combination of both stress relaxation (during indentation) and creep effects. For this reason, we must use a standard linear solid constitutive law to capture both of these effects, rather than a Kelvin-Voigt or Maxwell model.

Pseudo-bistability is fundamentally different to the slow snap-through studied by \cite{gomez2017a}, who considered a purely elastic system in which the effects of viscoelasticity and external damping were negligible. In that case, snap-through occurred on the elastic timescale, and only became slow because of the phenomenon of critical slowing down: this effectively introduces a dimensionless pre-factor that multiplies the timescale of snapping, which can grow much larger than $O(1)$ very close to the snapping transition. If external damping instead dominates inertial forces, a similar slowing down occurs though with the elastic timescale replaced by the damping timescale; see \cite{gomezthesis}. Pseudo-bistable snap-through is therefore only possible in systems with internal damping, i.e.~material viscoelasticity, to provide the required stress relaxation and creep.

As well as being characterised by overdamped dynamics, a key feature of the pseudo-bistable regime is that it occurs in a narrow parameter range at the transition between bistability and monostability. This corresponds to the saddle-node bifurcation at $\lambda = \lambda_{\mathrm{fold}}$ (for the truss) and $\mu = \mu_{\mathrm{fold}}$ (for the arch). As a direct consequence, the snap-through dynamics are susceptible to critical slowing down. Provided $T_{\mathrm{ind}}$ is sufficiently large, we showed that for both systems,  the snap-through time $t_{\mathrm{snap}}$ inherits an inverse square-root scaling law, i.e.~we have $t_{\mathrm{snap}} \propto (\eta/E_2)\epsilon^{-1/2}$ where $\eta/E_2$ is the viscous timescale and $\epsilon$ is the normalised distance to the bifurcation in parameter space (for the truss we in fact showed that $t_{\mathrm{snap}} \propto [\eta/E_1]\epsilon^{-1/2}$, i.e.~the relevant viscous timescale is $\eta/E_1$ not $\eta/E_2$, though as discussed at the end of \S\ref{CH5sec:largeDebottleneck} both timescales are of the same order because pseudo-bistability requires that $E_2/E_1 = \beta/(1-\beta) = O(1)$). The $\epsilon^{-1/2}$ scaling arises because the dynamics here are overdamped so that the time derivative in the normal form \eqref{CH5eqn:largeDebottleneckeqn} is first order, in contrast to underdamped systems in which the dynamics are second order in time and the bottleneck duration scales as $\sim \epsilon^{-1/4}$ \citep{gomez2017a}. 

In analysing data from experiments and numerical simulations reported in the literature, we found that this slowing down explains the sensitivity of the snap-through time observed in the pseudo-bistable regime. While there is some evidence of the inverse square-root scaling law it is not conclusive (figure \ref{CH5fig:comparedata}). In future work, it would be interesting to perform further experiments and determine whether the inverse square-root scaling indeed holds more generically. We also note that this sensitivity is responsible for large quantitative errors between finite element simulations and experiments despite excellent qualitative agreement \citep{brinkmeyer2013}, and in morphing applications would mean that parameters need to be precisely tuned to obtain the desired response. In such applications, our analytical expression could help to resolve the issue, as it provides a simple power law with which to control and calibrate the dynamic response: once the coefficient in the power law is determined (e.g.~by fitting experimental data), it is possible to make further predictions without the need for detailed simulations.

Finally, while the details of many practical morphing applications are likely to be more complicated than the relatively simple examples considered here, we believe that our analytical insight will be useful in guiding designers to the appropriate parameter regimes. Our analysis also reveals important features of the dynamics that cannot be neglected, which we expect to also be important in more complex structures: as well as the importance of inertia, we have shown that the effective stiffness does not smoothly reverse back to its fully unrelaxed value when the indenter is removed, as has been assumed in previous numerical studies. This means that any viscoelastic model used in finite element simulations should couple the stress to the deformation without making additional assumptions.

\section*{Acknowledgments}
The research leading to these results has received funding from the European Research Council under the European Union's Horizon 2020 Programme/ERC Grant No. 637334 (D.V.) and the EPSRC Grant No. EP/ M50659X/1 (M.G.). We are grateful to Efi Efrati for discussions about this work.
% The data that supports the plots within this paper and other findings of this study are available from. 

%% The Appendices part is started with the command \appendix;
%% appendix sections are then done as normal sections
\appendix
\setcounter{figure}{0}

\section{Influence of the relaxation parameter $\beta$}
\begin{figure}
\centering
\includegraphics[width=\textwidth]{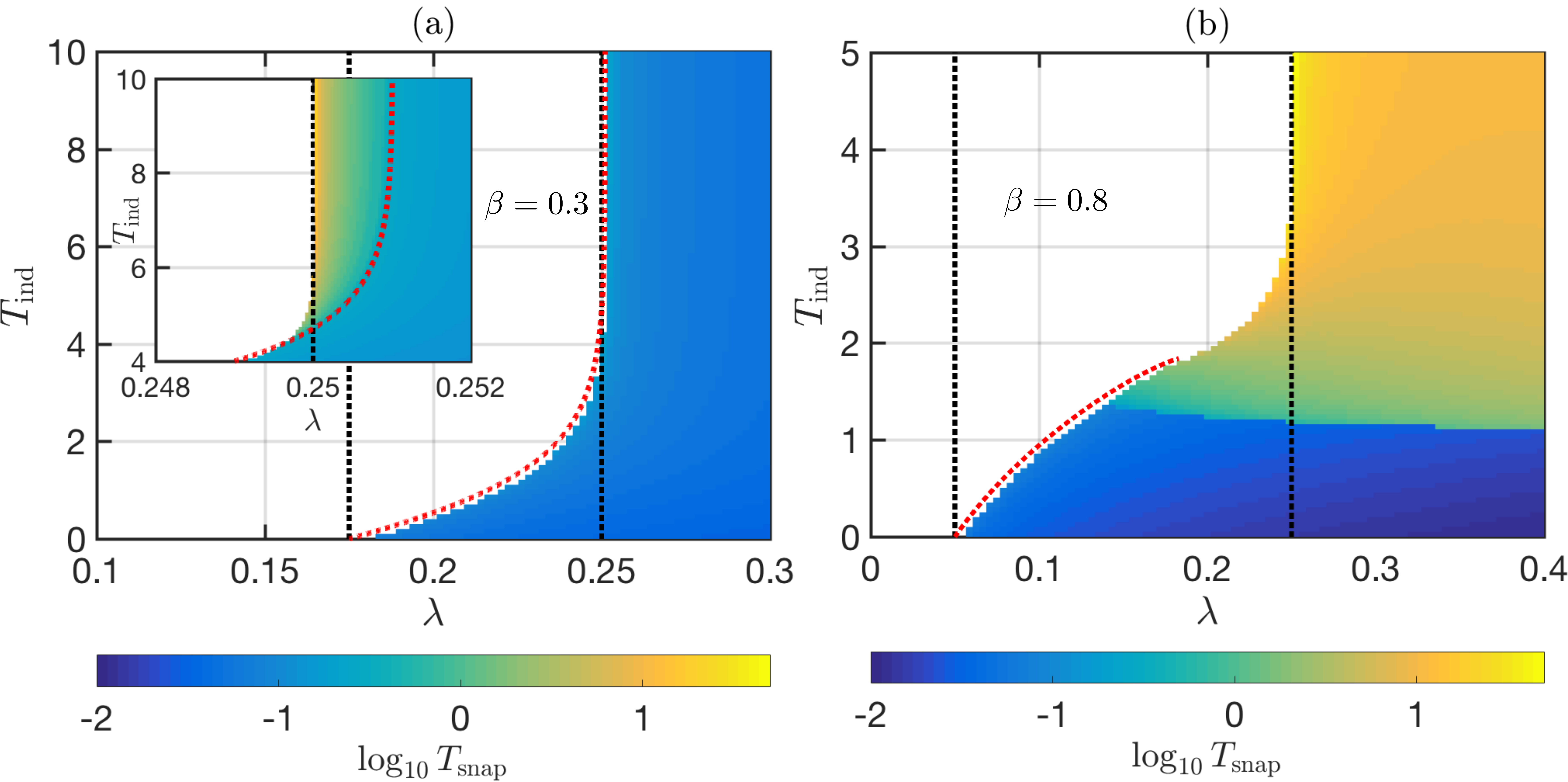} 
\caption{Snap-through times when $\mathrm{De} \gg 1$ for different relaxation parameter $\beta$ ($X_{\mathrm{ind}} = X_{\mathrm{fold}}$, $\mathrm{De} = 100$). Numerical results (obtained by integrating  \eqref{CH5eqn:releaseode}--\eqref{CH5eqn:releaseics} until the point where $X = 0$) are shown for (a)  $\beta = 0.3$ and (b) $\beta = 0.8$. The critical values $\lambda = (1-\beta)\lambda_{\mathrm{fold}}$ and $\lambda = \lambda_{\mathrm{fold}}$ are plotted as vertical black dotted lines. For later reference, also shown is the boundary predicted by equation \eqref{CH5eqn:largeXindboundary} (red dotted curve) when $\lambda < 1-\beta$. In each panel, the snap-through times have been computed on a $100\times 100$ grid of equally spaced values. Note the range of each colourbar is different to figures \ref{CH5fig:largeDesnapsurface}--\ref{CH5fig:largeDesnapsotherXind} in the main text.}
\label{CH5fig:largeDesnapsotherbeta}
\end{figure}

In this appendix we discuss the influence of the parameter $\beta$ (defined in equation \eqref{CH5eqn:defnbeta}) on the snap-through dynamics of the truss. The computed snap-through times for different values of $\beta$ are shown in figures \ref{CH5fig:largeDesnapsotherbeta}a--b for the case $X_{\mathrm{ind}} = X_{\mathrm{fold}}$. When $\beta < 1/2$, we observe a similar picture to that for the baseline value $\beta = 1/2$ considered in the main text, though the size of the pseudo-bistable region shrinks considerably in the purely elastic limit $\beta \to 0$. For example to obtain any pseudo-bistable behaviour when $\beta = 0.3$, it is necessary to take values $(\lambda - \lambda_{\mathrm{fold}}) \lesssim 10^{-3}$ and $T_{\mathrm{ind}} \gtrsim 5$ (see the inset of figure \ref{CH5fig:largeDesnapsotherbeta}a), compared to values $(\lambda - \lambda_{\mathrm{fold}}) \lesssim 6\times 10^{-3}$ and $T_{\mathrm{ind}} \gtrsim 4$ for the baseline case $\beta = 1/2$ (figure \ref{CH5fig:largeDesnapsurface}b).
%This results from the decreasing amount of stress relaxation as $\beta$ decreases: recall that during indentation the effective value of $\lambda$ is $\lambda_{\mathrm{eff}}(T) = \lambda [1+\frac{\beta}{1-\beta}e^{-(T+T_{\mathrm{ind}})}]$, so that larger values of $T_{\mathrm{ind}}$ are required to obtain the same effective stiffness when the indenter is released. 

\begin{figure}
\centering
\includegraphics[width=0.8\textwidth]{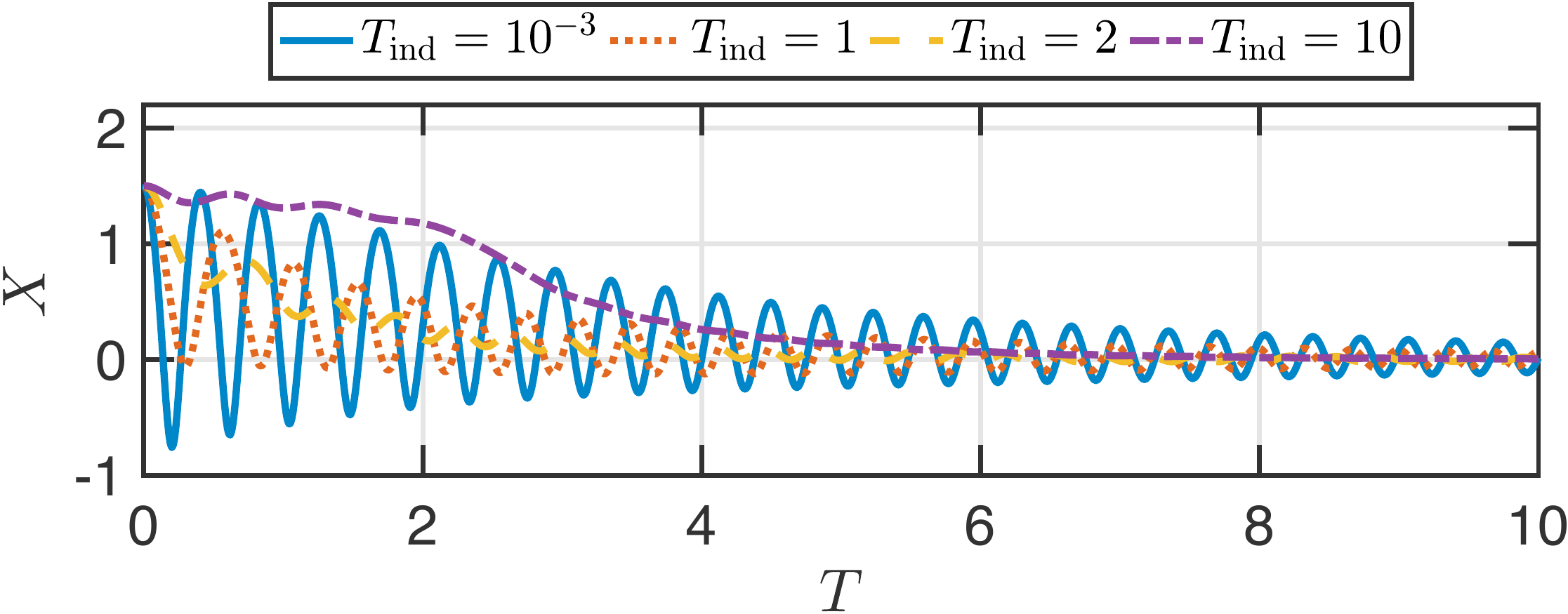} 
\caption{Numerical trajectories for relaxation parameter $\beta = 0.8$ in the large-$\mathrm{De}$ limit ($\lambda = 0.3$, $X_{\mathrm{ind}} = X_{\mathrm{fold}}$, $\mathrm{De} = 10$). Note that for $T_{\mathrm{ind}} \gtrsim 2$ the motion remains slow as the truss relaxes to the natural shape, different to the slow creep and fast snap exhibited in pseudo-bistability.}
\label{CH5fig:trajectoriesotherbeta}
\end{figure}

When $\beta > 1/2$, however, we observe very different behaviour: the region of parameter space where $T_{\mathrm{snap}} \gtrsim O(1)$ is much larger, extending to values $\lambda < \lambda_{\mathrm{fold}}$ (figure \ref{CH5fig:largeDesnapsotherbeta}b). Crucially, this slowing down is qualitatively different to pseudo-bistability, as may be observed from the typical trajectories in this regime --- see figure \ref{CH5fig:trajectoriesotherbeta}. These show that for sufficiently large $T_{\mathrm{ind}}$, the truss simply relaxes back to the natural shape without rapidly accelerating (yellow dashed curve, purple dashed-dotted curve in figure \ref{CH5fig:trajectoriesotherbeta}). In particular, a slow creeping motion followed by a rapid snap-through event, indicative of pseudo-bistable behaviour \citep{brinkmeyer2012,brinkmeyer2013}, is not observed. This is a consequence of the creep timescale becoming unbounded in the Maxwell limit $\beta \to 1$ (recall the discussion in \S\ref{sec:trussnondim}), with the truss displaying more fluid-like behaviour. 

\section{Assumption of a small ``fast part", $\vert\mathscr{X}\vert \ll 1$}
\setcounter{figure}{0}

In this appendix we discuss the ``fast part''  $\mathscr{X}$ of the leading order solution in the multiple-scales analysis of \S\ref{CH5sec:multiplescale}, defined in equation \eqref{CH5eqn:multiscaledecompose}. In terms of  $\mathscr{X}$, the leading order equation \eqref{CH5eqn:multiscaleleading} becomes
\beq
\pdd{\mathscr{X}}{\mathcal{T}}+F_{\mathrm{eq}}'\left(\mathcal{X};\frac{\lambda}{1-\beta}\right)\mathscr{X}+\frac{1}{2}F_{\mathrm{eq}}''\left(\mathcal{X};\frac{\lambda}{1-\beta}\right)\mathscr{X}^2+\frac{1}{6}F_{\mathrm{eq}}'''\left(\mathcal{X};\frac{\lambda}{1-\beta}\right)\mathscr{X}^3 = 0, \label{CH5eqn:multscaleleadingfastpart}
\eeq
where we have expanded the $F_{\mathrm{eq}}$ force term in \eqref{CH5eqn:multiscaleleading} about $X_0 = \mathcal{X}$ (higher-order terms vanish because $F_{\mathrm{eq}}$ is a cubic polynomial). The initial conditions \eqref{CH5eqn:fastics} imply that
\beqn
\mathscr{X}(0+) = X_{\mathrm{ind}}-\mathcal{X}(0+), \quad \pd{\mathscr{X}}{\mathcal{T}}(0+) = 0, \quad \pdd{\mathscr{X}}{\mathcal{T}}(0+) = -F_{\mathrm{ind}}.
\eeqn
Due to the nonlinear terms in \eqref{CH5eqn:multscaleleadingfastpart}, it is difficult to make analytical progress, though we note that by multiplying by $\partial\mathscr{X}/\partial\mathcal{T}$ and integrating twice, it is possible to obtain an implicit equation for $\mathscr{X}$ (up to quadrature). However, each integration introduces an unknown function of the slow timescale $T$ (in addition to the unknown function $A(T)$ in equation \eqref{CH5eqn:multiscaleleading}), which require  solvability conditions to be determined. We therefore make the simplifying assumption that $|\mathscr{X}|\ll 1$,  so that \eqref{CH5eqn:multscaleleadingfastpart} is approximately
\beqn
\pdd{\mathscr{X}}{\mathcal{T}}+F_{\mathrm{eq}}'\left(\mathcal{X};\frac{\lambda}{1-\beta}\right)\mathscr{X} = 0,
\eeqn
as given in the main text.

We now consider when the assumption $|\mathscr{X}| \ll 1$ is justified. From the initial conditions for $\mathscr{X}$ above, we expect this to be valid whenever $|X_{\mathrm{ind}}-\mathcal{X}(0+)| \ll 1$ and $|F_{\mathrm{ind}}| \ll 1$. Using the initial condition \eqref{CH5eqn:slowXic} for $\mathcal{X}(0+)$, we expect that $X_{\mathrm{ind}}\approx\mathcal{X}(0+)$ whenever $|F_{\mathrm{ind}}| \ll 1$; hence, we simply require $|F_{\mathrm{ind}}| \ll 1$ for validity.  For example, with the baseline values $\beta = 1/2$ and $X_{\mathrm{ind}} = X_{\mathrm{fold}} = 3/2$, this becomes (using the expression \eqref{CH5eqn:indentationstress,force})
\beqn
|F_{\mathrm{ind}}| = \frac{3}{2}\left\lvert\lambda-\frac{1}{4}+\lambda e^{-T_{\mathrm{ind}}}\right\rvert \ll 1.
\eeqn
For $\lambda \in [0,1/4]$, we have that $F_{\mathrm{ind}} \in [-3/8,3/8]$ so the above approximation should be reasonably accurate (increasing in accuracy as $\lambda \to \lambda_{\mathrm{fold}} = 1/4$ for $T_{\mathrm{ind}} \gg 1$ and as  $\lambda \to (1-\beta)\lambda_{\mathrm{fold}} = 1/8$ for $T_{\mathrm{ind}} \ll 1$).

\section{Determining $\mathcal{X}(0+)$}
\setcounter{figure}{0}

In this appendix we consider the solution $\mathcal{X}(0+)$ of the cubic equation \eqref{CH5eqn:slowXic}, which we reproduce here:
\beqn
F_{\mathrm{eq}}\left(\mathcal{X}(0+);\frac{\lambda}{1-\beta}\right) = \frac{\beta\lambda X_{\mathrm{ind}}}{1-\beta}\left(1-e^{-T_{\mathrm{ind}}}\right). 
\eeqn
This provides the initial value of the ``slow part'' of the leading order solution in the multiple-scales analysis, and so determines whether the truss initially creeps in an inverted position or immediately snaps to near its natural state. We focus on the case $\lambda < 1-\beta$, so that the force-displacement curve $F_{\mathrm{eq}}(\mathcal{X};\frac{\lambda}{1-\beta})$ has distinct real turning points at $\mathcal{X}_{\pm}$. We now discuss the cases $\lambda < (1-\beta)\lambda_{\mathrm{fold}}$ and $(1-\beta)\lambda_{\mathrm{fold}} < \lambda < 1-\beta$ separately.

\subsection*{The case $\lambda < (1-\beta)\lambda_{\mathrm{fold}}$}
\begin{figure}
\centering
\includegraphics[width=\textwidth]{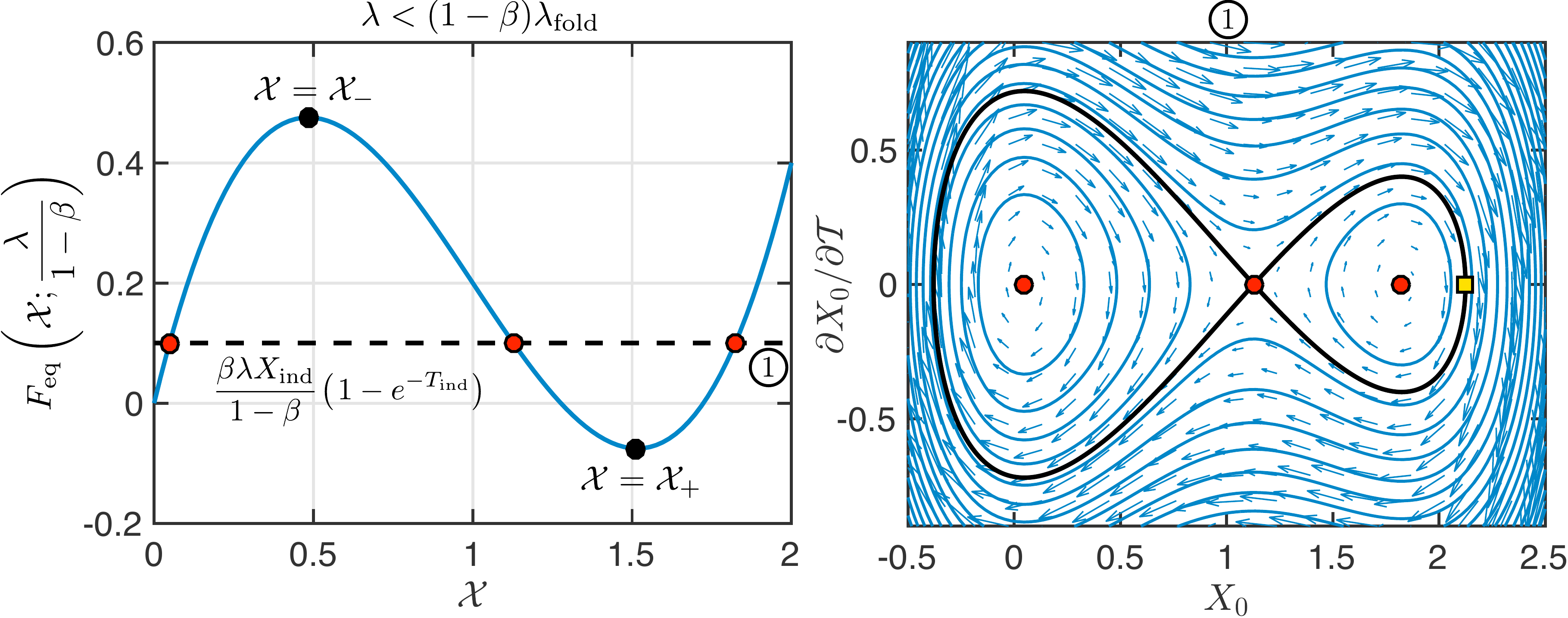} 
\caption{A typical force-displacement curve in the case $\lambda <  (1-\beta)\lambda_{\mathrm{fold}}$ (left panel), and the corresponding phase plane of the first-order system \eqref{CH5eqn:firstordersystem} (right panel). In this regime, equation  \eqref{CH5eqn:slowXic} has three distinct real roots (highlighted as red circles), corresponding to two centres and a saddle point on the phase plane.}
\label{CH5fig:largeDephaseplane1}
\end{figure}

In this case the turning point $\mathcal{X}_+$ on the force-displacement curve $F_{\mathrm{eq}}(\mathcal{X};\frac{\lambda}{1-\beta})$ lies below the horizontal axis, i.e.~$F_{\mathrm{eq}}(\mathcal{X}_+;\frac{\lambda}{1-\beta}) < 0$. Furthermore, restricting to $\beta \leq 1/2$ and $X_{\mathrm{ind}} \leq 2$, it is possible to show that 
\beqn
\frac{\beta \lambda X_{\mathrm{ind}}}{1-\beta}\left(1-e^{-T_{\mathrm{ind}}}\right)  < F_{\mathrm{eq}}\left(\mathcal{X}_-;\frac{\lambda}{1-\beta}\right),
\eeqn
i.e.~the line of height $\frac{\beta \lambda X_{\mathrm{ind}}}{1-\beta}\left(1-e^{-T_{\mathrm{ind}}}\right)$ lies below the turning point at $\mathcal{X}_{-}$. There are therefore three distinct real roots of the cubic equation \eqref{CH5eqn:slowXic}. This is illustrated in the left panel of figure \ref{CH5fig:largeDephaseplane1}, which highlights the roots as red circles. However, it is not immediately clear which root is the relevant solution for $\mathcal{X}(0+)$.

To determine the relevant root, we return to equation \eqref{CH5eqn:multiscaleleading}, i.e.~the leading order problem in the multiple-scale analysis. Recall that in our multiple-scale analysis, we first re-scaled the ODE in terms of the fast elastic timescale $\mathcal{T}$. Setting $A(T) = 0$, equation \eqref{CH5eqn:multiscaleleading} then governs the dynamics of the truss at very early times, before viscous relaxation of the SLS element becomes important. Defining $V_0 = \partial X_0/\partial\mathcal{T}$, this equation can be written as the first-order system
\begin{eqnarray}
\pd{X_0}{\mathcal{T}} & = & V_0,  \nonumber \\
\pd{V_0}{\mathcal{T}} & = & \frac{\beta\lambda X_{\mathrm{ind}}}{1-\beta}\left(1-e^{-T_{\mathrm{ind}}}\right)-F_{\mathrm{eq}}\left(X_0;\frac{\lambda}{1-\beta}\right),
\label{CH5eqn:firstordersystem}
\end{eqnarray}
with initial data $(X_0,V_0) = (X_{\mathrm{ind}},0)$ at $\mathcal{T} = 0+$. The significance here comes from the fact that the critical points  of this system are precisely the solutions of the cubic \eqref{CH5eqn:slowXic}, and hence correspond to the possible values of  $\mathcal{X}(0+)$. The relevant value is then determined by which critical point the solution oscillates around on the phase plane, when we follow the trajectory emerging from $(X_0,V_0) = (X_{\mathrm{ind}},0)$ --- these oscillations correspond to the ``fast part'' of the solution, $\mathscr{X}$, as opposed to the slowly varying mean. For later reference, we also note that \eqref{CH5eqn:firstordersystem} is Hamiltonian with conserved energy
\beq
\frac{1}{2}\left(\pd{X_0}{\mathcal{T}}\right)^2 + \int_{X_{\mathrm{ind}}}^{X_0} F_{\mathrm{eq}}\left(\xi;\frac{\lambda}{1-\beta}\right)\:\id\xi = \frac{\beta\lambda X_{\mathrm{ind}}}{1-\beta}\left(1-e^{-T_{\mathrm{ind}}}\right)\left(X_0-X_{\mathrm{ind}}\right). \label{CH5eqn:hamiltonian}
\eeq

A typical phase plane of \eqref{CH5eqn:firstordersystem}  in the case $\lambda < (1-\beta)\lambda_{\mathrm{fold}}$ is shown in the right panel of figure \ref{CH5fig:largeDephaseplane1}. Here the three real roots of equation \eqref{CH5eqn:slowXic} give rise to three critical points. By considering the Jacobian of \eqref{CH5eqn:firstordersystem}, we find that the two roots where $F_{\mathrm{eq}}'(\mathcal{X};\frac{\lambda}{1-\beta}) > 0$ correspond to centres, while the intermediate root where $F_{\mathrm{eq}}'(\mathcal{X};\frac{\lambda}{1-\beta}) < 0$  is associated with a saddle point. Figure \ref{CH5fig:largeDephaseplane1} also shows that there are two homoclinic orbits that emerge from the saddle point (black solid curves) that act as separatrices: all trajectories that oscillate around the left centre are enclosed in the homoclinic orbit to the left of the saddle point, while all  trajectories that oscillate around the right centre are enclosed in the right orbit. We  therefore expect that $\mathcal{X}(0+)$ corresponds to one of the centres rather than the saddle point, depending on whether the solution starts to the left or the right of the saddle point on the phase plane. Moreover, because $F_{\mathrm{eq}}'(\mathcal{X};\frac{\lambda}{1-\beta}) > 0$ at the centres, and $F_{\mathrm{eq}}'(\mathcal{X};\frac{\lambda}{1-\beta}) < 0$ when $\mathcal{X}\in (\mathcal{X}_{-},\mathcal{X}_{+})$ (as seen from the force-displacement curve), we must have $\mathcal{X}(0+) < \mathcal{X}_{-}$ or $\mathcal{X}(0+) > \mathcal{X}_{+}$. (In addition, because we expect the solution to oscillate around these centres as $T$ increases and $A \neq 0$, this verifies our earlier assumption that $F_{\mathrm{eq}}'(\mathcal{X};\frac{\lambda}{1-\beta}) >0$ made immediately after equation \eqref{CH5eqn:multscaleleadingfastpartsimplified}.)

Suppose that $1 \leq X_{\mathrm{ind}} < X_{\mathrm{fold}}$. From the force-displacement curve in figure \ref{CH5fig:largeDephaseplane1}, we see that $X_{\mathrm{ind}}$ lies to the right of the saddle point if and only if 
\beqn
F_{\mathrm{eq}}\left(X_{\mathrm{ind}};\frac{\lambda}{1-\beta}\right) < \frac{\beta\lambda X_{\mathrm{ind}}}{1-\beta}\left(1-e^{-T_{\mathrm{ind}}}\right),
\eeqn
(since this is where $F_{\mathrm{eq}}(\mathcal{X};\frac{\lambda}{1-\beta})$ is monotonically decreasing on the diagram). Using the expression \eqref{CH5eqn:indentationstress,force} for $F_{\mathrm{ind}}$, this is precisely the statement
\beq
F_{\mathrm{ind}} = F_{\mathrm{eq}}\left(X_{\mathrm{ind}};\lambda\left[1+\frac{\beta}{1-\beta}e^{-T_{\mathrm{ind}}}\right]\right) < 0. \label{CH5eqn:AppendixTcritsmallXind}
\eeq
When this is satisfied the relevant solution is the right centre with $\mathcal{X}(0+) > \mathcal{X}_{+}$; otherwise $\mathcal{X}(0+) < \mathcal{X}_{-}$.

Suppose instead that $X_{\mathrm{ind}} \geq X_{\mathrm{fold}}$. In this case,  $X_{\mathrm{ind}}$ always lies to the right of the saddle point (the force-displacement curve illustrates how the intermediate root is smaller than $X_{\mathrm{fold}}$ when $\lambda < (1-\beta)\lambda_{\mathrm{fold}}$). However, it is also possible that   $X_{\mathrm{ind}}$ is large enough to fall outside the homoclinic orbit to the right of the saddle point. If this occurs, the phase plane in figure \ref{CH5fig:largeDephaseplane1} shows how the amplitude of the oscillations becomes very large, with the trajectory enclosing both centre points. In particular, the trajectory crosses the horizontal axis again near the origin, and so the truss will immediately snap in this regime. If this occurs, we consider the relevant solution to be the left centre, i.e.~$\mathcal{X}(0+) < \mathcal{X}_{-}$.

Now imagine that $\lambda$ and $X_{\mathrm{ind}}$ are fixed, while the indentation time $T_{\mathrm{ind}}$ is varied. As $T_{\mathrm{ind}}$ increases, the value of $\frac{\beta\lambda X_{\mathrm{ind}}}{1-\beta}\left(1-e^{-T_{\mathrm{ind}}}\right)$ increases, and so both the right centre and the homoclinic orbit are shifted further to the right on the phase plane. There will be a critical value of $T_{\mathrm{ind}}$ for which the initial point $(X_{\mathrm{ind}},0)$ lies exactly where the homoclinic orbit crosses the horizontal axis (highlighted as a yellow square on figure \ref{CH5fig:largeDephaseplane1}). Only when $T_{\mathrm{ind}}$ is larger than this value does the initial point fall inside the separatrix and we have $\mathcal{X}(0+) > \mathcal{X}_{+}$. This change in behaviour is an instance of a homoclinic bifurcation \citep{strogatz}, which has been observed in other dynamic snap-through \citep{nachbar1967} and pull-in instabilities \citep{krylov2007}. 

To determine this critical value, we note that there is a discontinuous change in where the trajectory starting from $(X_{\mathrm{ind}},0)$  later crosses the horizontal axis. Setting $\partial X_0/\partial\mathcal{T} = 0$ in \eqref{CH5eqn:hamiltonian}, the value of $X_0$ at this point satisfies 
\beqn
\int_{X_{\mathrm{ind}}}^{X_0} F_{\mathrm{eq}}\left(\xi;\frac{\lambda}{1-\beta}\right)\:\id\xi = \frac{\beta\lambda X_{\mathrm{ind}}}{1-\beta}\left(1-e^{-T_{\mathrm{ind}}}\right)\left(X_0-X_{\mathrm{ind}}\right). 
\eeqn
At the critical value of $T_{\mathrm{ind}}$, the trajectory starts on the homoclinic orbit and later crosses the axis at the saddle point, so that $X_0$ also satisfies the cubic \eqref{CH5eqn:slowXic}, i.e.~
\beqn
 F_{\mathrm{eq}}\left(X_0;\frac{\lambda}{1-\beta}\right) = \frac{\beta\lambda X_{\mathrm{ind}}}{1-\beta}\left(1-e^{-T_{\mathrm{ind}}}\right).
\eeqn 
It follows that
\beqn
\int_{X_{\mathrm{ind}}}^{X_0} F_{\mathrm{eq}}\left(\xi;\frac{\lambda}{1-\beta}\right)\:\id\xi = \left(X_0-X_{\mathrm{ind}}\right) F_{\mathrm{eq}}\left(X_0;\frac{\lambda}{1-\beta}\right),
\eeqn
which can be re-arranged to 
\beqn
-\frac{1}{4}\left(X_0-X_{\mathrm{ind}}\right)^2\left[3 X_0^2 + 2 (X_{\mathrm{ind}}-4)X_0+(X_{\mathrm{ind}}-2)^2 +\frac{2\lambda}{1-\beta}\right] = 0.
\eeqn
The solution corresponding to the saddle point is
\beqn
X^{*} = -\frac{1}{3}\left(X_{\mathrm{ind}}-4\right)+\frac{\sqrt{6}}{3}\left[1-\frac{\lambda}{1-\beta}-\frac{\left(X_{\mathrm{ind}}-1\right)^2}{3}\right]^{1/2}.
\eeqn
Hence, the critical value of  $T_{\mathrm{ind}}$ satisfies
\beqn
\frac{\beta\lambda X_{\mathrm{ind}}}{1-\beta}\left(1-e^{-T_{\mathrm{ind}}}\right) = F_{\mathrm{eq}}\left(X^{*};\frac{\lambda}{1-\beta}\right).
\eeqn 
Solving for $T_{\mathrm{ind}}$, we conclude that when $X_{\mathrm{ind}} \geq X_{\mathrm{fold}}$, we have $\mathcal{X}(0+) > \mathcal{X}_{+}$  if and only if
\beq
 T_{\mathrm{ind}} > \log\left[\frac{\beta\lambda X_{\mathrm{ind}}}{\beta\lambda X_{\mathrm{ind}} - (1-\beta)F_{\mathrm{eq}}(X^{*};\frac{\lambda}{1-\beta})}\right]. \label{CH5eqn:AppendixTcritlargeXind}
\eeq

%The above analysis is based on the assumption that \eqref{CH5eqn:slowXic} has three distinct real roots. If instead
%\beq
%\frac{\beta \lambda X_{\mathrm{ind}}}{1-\beta}\left(1-e^{-T_{\mathrm{ind}}}\right)  > F_{\mathrm{eq}}\left(\mathcal{X}_-;\frac{\lambda}{1-\beta}\right),
%\label{CH5eqn:onerealroot}
%\eeq
%then there is a single real root of \eqref{CH5eqn:slowXic} with $\mathcal{X}(0+) > \mathcal{X}_{+}$, which corresponds to a centre on the phase plane. This is labelled as case 2 in figure \ref{CH5fig:largeDephaseplane1}. We therefore always have $\mathcal{X}(0+) > \mathcal{X}_{+}$ in this case. Because the value of $T_{\mathrm{ind}}$ required to attain equality in \eqref{CH5eqn:onerealroot} is larger than that required in the conditions \eqref{CH5eqn:AppendixTcritsmallXind}--\eqref{CH5eqn:AppendixTcritlargeXind}, we conclude that the relevant conditions that determine whether $\mathcal{X}(0+) > \mathcal{X}_{+}$ or $\mathcal{X}(0+) < \mathcal{X}_{-}$ are precisely \eqref{CH5eqn:AppendixTcritsmallXind}--\eqref{CH5eqn:AppendixTcritlargeXind}.

\subsection*{The case $(1-\beta)\lambda_{\mathrm{fold}} < \lambda < 1-\beta$}
\begin{figure}
\centering
\includegraphics[width=\textwidth]{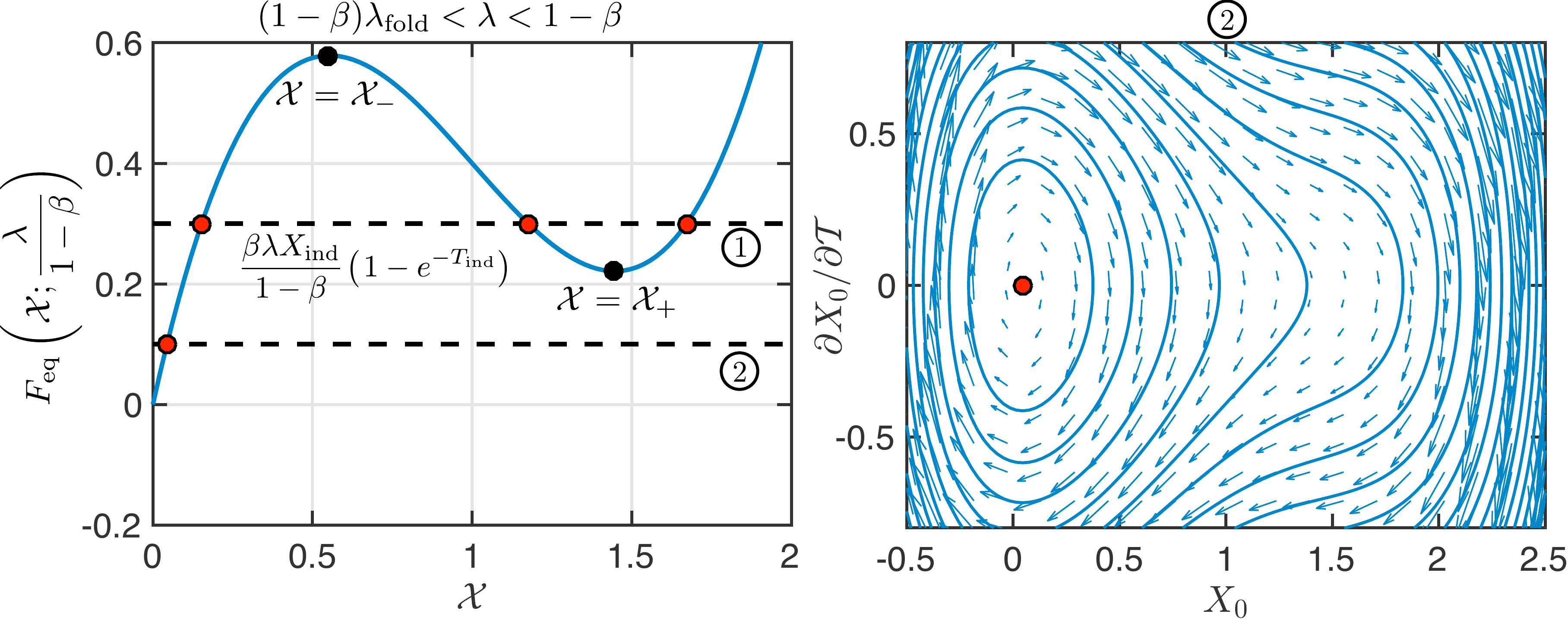} 
\caption{A typical force-displacement curve in the case $(1-\beta)\lambda_{\mathrm{fold}} < \lambda \leq 1-\beta$ (left panel), and the corresponding phase plane of equation \eqref{CH5eqn:firstordersystem} (right panel). In this regime, equation \eqref{CH5eqn:slowXic} may have three distinct real roots (labelled as case 1) or a single real root (case 2).}
\label{CH5fig:largeDephaseplane2}
\end{figure}

In this case both turning points on the force-displacement curve $F_{\mathrm{eq}}(\mathcal{X};\frac{\lambda}{1-\beta})$ lie above the horizontal axis; see the left panel of figure \ref{CH5fig:largeDephaseplane2}. If the line of height $\frac{\beta \lambda X_{\mathrm{ind}}}{1-\beta}\left(1-e^{-T_{\mathrm{ind}}}\right)$ lies below the turning point at $\mathcal{X}_{+}$ i.e.~if
\beqn
\frac{\beta \lambda X_{\mathrm{ind}}}{1-\beta}\left(1-e^{-T_{\mathrm{ind}}}\right)  < F_{\mathrm{eq}}\left(\mathcal{X}_+;\frac{\lambda}{1-\beta}\right),
\eeqn
then there is a single real root of \eqref{CH5eqn:slowXic} that corresponds to a centre on the phase plane. This is labelled as case 2 on figure \ref{CH5fig:largeDephaseplane2}, with a typical phase plane plotted in the right panel. We therefore always have  $\mathcal{X}(0+) < \mathcal{X}_{-}$ in this case.

If instead
\beq
\frac{\beta \lambda X_{\mathrm{ind}}}{1-\beta}\left(1-e^{-T_{\mathrm{ind}}}\right)  > F_{\mathrm{eq}}\left(\mathcal{X}_+;\frac{\lambda}{1-\beta}\right),
\label{CH5eqn:threerealroots}
\eeq
then there are three distinct real roots (labelled as case 1 on figure \ref{CH5fig:largeDephaseplane2}), and the analysis proceeds in a similar way to the case $\lambda < (1-\beta)\lambda_{\mathrm{fold}}$ considered above. In particular, if $1 \leq X_{\mathrm{ind}} < X_{\mathrm{fold}}$ the relevant root has $\mathcal{X}(0+) > \mathcal{X}_{+}$ if and only if $T_{\mathrm{ind}}$ satisfies \eqref{CH5eqn:AppendixTcritsmallXind}, otherwise $\mathcal{X}(0+) < \mathcal{X}_{-}$; while if $X_{\mathrm{ind}} \geq X_{\mathrm{fold}}$, we have $\mathcal{X}(0+) > \mathcal{X}_{+}$ if and only if $T_{\mathrm{ind}}$ satisfies \eqref{CH5eqn:AppendixTcritlargeXind}, otherwise $\mathcal{X}(0+) < \mathcal{X}_{-}$. 

Finally, it may be shown (e.g.~by graphical considerations) that the conditions \eqref{CH5eqn:AppendixTcritsmallXind}--\eqref{CH5eqn:AppendixTcritlargeXind} are always stronger than \eqref{CH5eqn:threerealroots}, i.e.~the critical value of $T_{\mathrm{ind}}$ required to attain equality is larger. Because $\mathcal{X}(0+) < \mathcal{X}_{-}$  whenever \eqref{CH5eqn:threerealroots} is not satisfied, we conclude that the relevant conditions that determine whether $\mathcal{X}(0+) > \mathcal{X}_{+}$ or $\mathcal{X}(0+) < \mathcal{X}_{-}$ are precisely \eqref{CH5eqn:AppendixTcritsmallXind}--\eqref{CH5eqn:AppendixTcritlargeXind}, both for $\lambda < (1-\beta)\lambda_{\mathrm{fold}}$ and $(1-\beta)\lambda_{\mathrm{fold}} < \lambda < 1-\beta$.

\section{Assumption of an evolving stiffness}
\setcounter{figure}{0}

In the approach of \cite{santer2010}, later adopted by \cite{brinkmeyer2012,brinkmeyer2013}, the viscoelastic response is modelled as an elastic structure with an effective stiffness that changes in time. In particular, the evolution is specified by a Prony series, which assumes that the stiffness can be expressed as a sum of decaying exponential functions. For example, if $E(t)$ is the Young's modulus at time $t$, and $E_0 = E(0)$ is the initial modulus, this can be written as \citep{brinkmeyer2012}
\beq
E(t) = E_0 \left[1-\sum_{j = 1}^{N}K_j\left(1-e^{-t/\tau_j}\right)\right],
\label{CH5eqn:Pronyseriesindent}
\eeq
where $N \geq 1$ is an integer; the coefficients $K_j$ and timescales $\tau_j$ are specified parameters that can be fitted to experimental data from relaxation tests. In the case of a step increase in strain applied at $t = 0$, this model is physically equivalent to a superposition of SLS elements, with each term corresponding to the stress relaxation of a particular element \citep{kim2010,brinkmeyer2012}; the value $E_0$ corresponds to the fully unrelaxed modulus, i.e.~just after the strain is applied. As $t \to \infty$, the modulus $E(t)$ decays to the fully relaxed value
\beqn
E_{\infty} = E_0 \left[1-\sum_{j = 1}^{N}K_j\right].
\eeqn

To apply this model when the indenter is released, \cite{santer2010} and  \cite{brinkmeyer2012,brinkmeyer2013} assume that the evolution during recovery is the reverse of the behaviour during indentation, and that there is no jump in the value of the stiffness. The stiffness is also allowed to fully relax before the indenter is released. With $t = 0$ now denoting the point when the indenter is released, this corresponds to setting 
\beqn
E(t) = E_0 \left[1-\sum_{j = 1}^{N}K_j e^{-t/\tau_j}\right],
\eeqn
for $t > 0$. Thus $E(t = 0+) = E_{\infty}$, and $E(t)$ decays to the fully unrelaxed value $E_0$ as $t \to \infty$. If instead the stiffness is only allowed to relax for a time duration $t_{\mathrm{ind}}$ before the indenter is released, as in our truss model, this modifies to
\beq
E(t) = E_0 \left[1-\sum_{j = 1}^{N}K_j e^{-t/\tau_j}\left(1-e^{-t_{\mathrm{ind}}/\tau_j}\right)\right],
 \label{CH5eqn:Pronyseriesrelease}
\eeq
since when $t = 0+$ this corresponds to \eqref{CH5eqn:Pronyseriesindent} evaluated at $t= t_{\mathrm{ind}}$.

To understand  this reversibility assumption within the framework of our truss model, we recall from \S\ref{CH5sec:indentation} that in response to an indentation displacement $X_{\mathrm{ind}}$ suddenly applied at $T = -T_{\mathrm{ind}}$, the dimensionless stress is
\beqn
\Sigma = X_{\mathrm{ind}}\left[1+\frac{\beta}{1-\beta}e^{-(T+T_{\mathrm{ind}})}\right], \qquad -T_{\mathrm{ind}} < T < 0.
\eeqn
Since $X$ is the strain in the SLS element, the effective stiffness is therefore
\beqn
 \frac{\Sigma}{X} = 1+\frac{\beta}{1-\beta}e^{-(T+T_{\mathrm{ind}})} = \frac{1}{1-\beta}\left[1-\beta\left(1-e^{-(T+T_{\mathrm{ind}})} \right)\right].
\eeqn
This corresponds to the Prony series \eqref{CH5eqn:Pronyseriesindent} when we identify
\beqn
E(t) = [\sigma]\frac{\Sigma}{X}, \quad E_0 = \frac{1}{1-\beta} [\sigma], \quad N = 1, \quad K_1 = \beta, \quad \frac{t}{\tau_1} = T + T_{\mathrm{ind}},
\eeqn
for some pressure scale $[\sigma]$. Hence, when the indenter is released, the above assumption \eqref{CH5eqn:Pronyseriesrelease} would be equivalent to specifying for $T > 0$
\beqn
 \frac{\Sigma}{X} = \frac{1}{1-\beta}\Big[1-\beta e^{-T}\left(1-e^{-T_{\mathrm{ind}}}\right)\Big].
\eeqn
(Here we instead identify $t/\tau_1 = T$.)
%(This differs slightly from \eqref{CH5eqn:Pronyseriesrelease} upon setting $E(t) = \Sigma/X$ etc.~due to the finite value of $T_{\mathrm{ind}}$ in our model, which requires an additional term to ensure the stress is continuous across $T= 0$.) 
Setting $F = 0$ in the momentum equation \eqref{CH5eqn:forcebalance}, and substituting the effective stiffness above, the trajectory  $X(T)$ would then obey 
\beq
\mathrm{De}^{-2}\dd{X}{T} = -F_{\mathrm{eq}}\left(X(T);\lambda_{\mathrm{eff}}^{\mathrm{rel}}(T)\right), \qquad T > 0
\label{CH5eqn:assumptionODE}
\eeq
where the effective value of $\lambda$ is
\beqn
\lambda_{\mathrm{eff}}^{\mathrm{rel}} = \frac{\lambda}{1-\beta}\Big[1-\beta e^{-T}\left(1-e^{-T_{\mathrm{ind}}}\right)\Big],
\eeqn
which increases from $\lambda_{\mathrm{eff}}^{\mathrm{rel}}(0+) = \lambda [1+\frac{\beta}{1-\beta}e^{-T_{\mathrm{ind}}}]$ to $\lambda_{\mathrm{eff}}^{\mathrm{rel}}(\infty) = \lambda/(1-\beta)$ during release. The initial conditions are
\beq
X(0+) = X_{\mathrm{ind}}, \quad \dot{X}(0+) = 0. \label{CH5eqn:assumptionICs}
\eeq

To understand how the solution of \eqref{CH5eqn:assumptionODE} differs to the model we consider in the main text when $\mathrm{De} \gg 1$, we neglect the inertia term as a first approximation to obtain
\beqn
F_{\mathrm{eq}}\left(X(T);\lambda_{\mathrm{eff}}^{\mathrm{rel}}(T)\right) = 0, \qquad T > 0. 
\eeqn
Recall that indentation corresponds to rotating the force-displacement curve in figure \ref{CH5fig:steadyresponse} clockwise as stress relaxation occurs. The assumption that the response during recovery is the reverse of relaxation then suggests that the graph simply rotates back \emph{anticlockwise} as soon as the indenter is released. For each value of $\lambda$, the displacement is found as the value of $X$ at which the force-displacement curve crosses the horizontal axis. Hence, the different dynamical regimes follow from those discussed during indentation in \S\ref{CH5sec:indentation}. For $\lambda < (1-\beta)\lambda_{\mathrm{fold}}$ (so $\lambda_{\mathrm{eff}}^{\mathrm{rel}}(\infty) < \lambda_{\mathrm{fold}}$), the truss is always bistable during indentation, and so remains bistable during recovery --- we generally expect the truss not to snap-through. For $\lambda > \lambda_{\mathrm{fold}}$, the truss is monostable when the indenter is released (since $ \lambda_{\mathrm{eff}}^{\mathrm{rel}}(0+) >  \lambda_{\mathrm{fold}}$), and so remains monostable during recovery ---  to satisfy $F_{\mathrm{eq}} = 0$ the truss must immediately jump to $X = 0$ and the snap-through time is governed by inertia. For $(1-\beta)\lambda_{\mathrm{fold}} < \lambda < \lambda_{\mathrm{fold}}$, the truss is temporarily bistable when the indenter is released if and only if $T_{\mathrm{ind}}$ satisfies the condition \eqref{CH5eqn:naivebdry}, which is equivalent to $\lambda_{\mathrm{eff}}^{\mathrm{rel}}(0+) <  \lambda_{\mathrm{fold}}$; in this case, rapid snap-through occurs as soon as the anticlockwise rotation is enough to put the turning point on the force-displacement curve above the line  $F_{\mathrm{eq}} = 0$. This regime corresponds to pseudo-bistable behaviour in which the snap-through time is $O(1)$. If \eqref{CH5eqn:naivebdry} is not satisfied, the truss is initially monostable and so immediately snaps.

\begin{figure}
\centering
\includegraphics[width=0.6\textwidth]{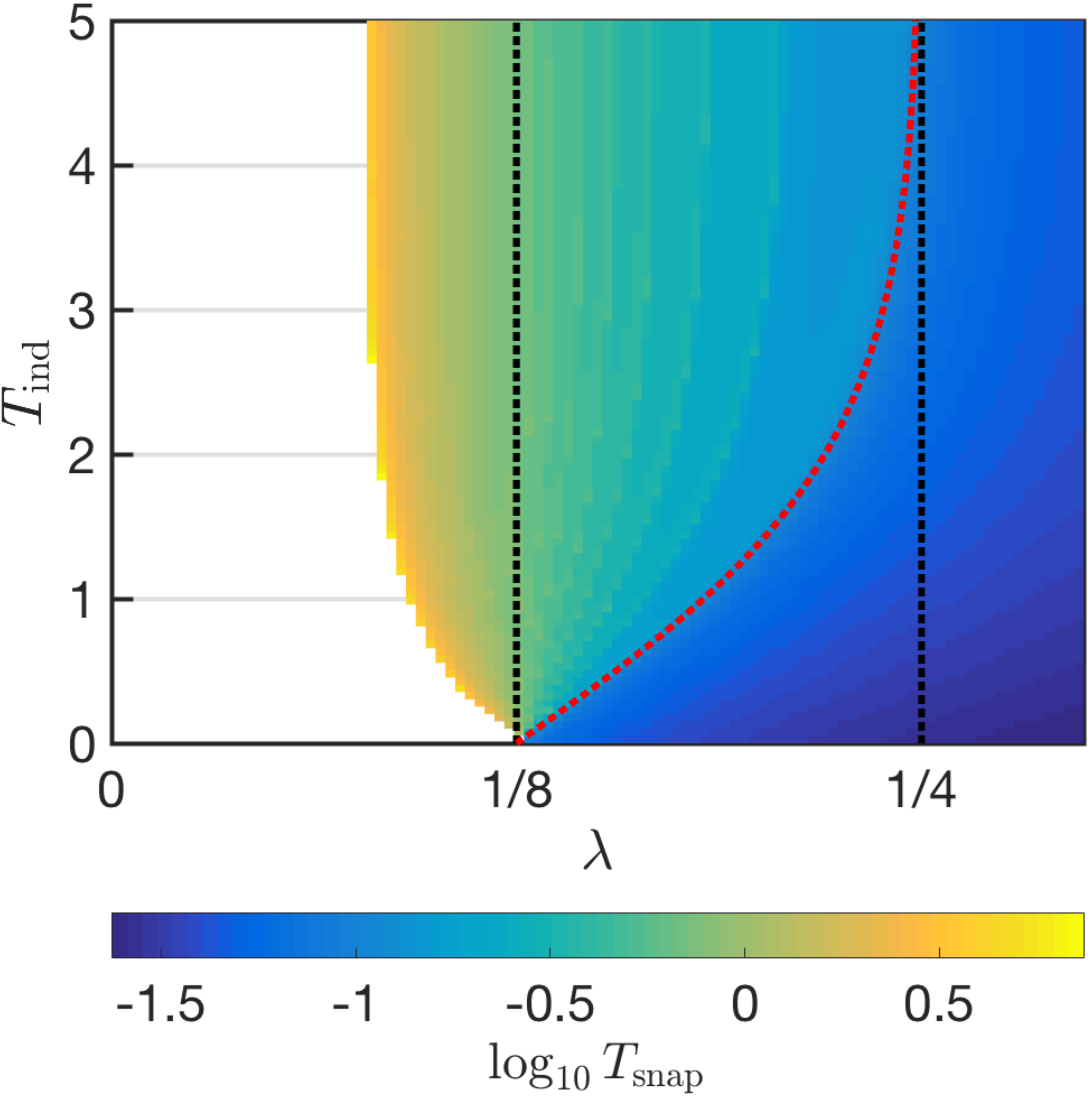} 
\caption{Snap-through times obtained by integrating \eqref{CH5eqn:assumptionODE}--\eqref{CH5eqn:assumptionICs} numerically until the point where $X = 0$ (see colourbar), which assumes that the stiffness during recovery reverses back to its fully unrelaxed value ($X_{\mathrm{ind}} = X_{\mathrm{fold}}$, $\beta = 1/2$, $\mathrm{De} = 100$). The critical values $\lambda = (1-\beta)\lambda_{\mathrm{fold}}$ and $\lambda = \lambda_{\mathrm{fold}}$ are plotted as vertical black dotted lines. Also shown is the boundary above which pseudo-bistable behaviour is obtained predicted by \eqref{CH5eqn:naivebdry} (red dotted curve). The snap-through times have been computed on a $100\times 100$ grid of equally spaced values in the region displayed.}
\label{CH5fig:controlstiffnesssnaptimes}
\end{figure}

Comparing the above picture with figure \ref{CH5fig:largeDeschematic} of the main text, we conclude that the assumption of reversibility leads to very different behaviour to that derived without this assumption: both the regions where snap-through and pseudo-bistability occur differ significantly between the two models. This conclusion also holds when we account for the effects of inertia and directly solve the ODE \eqref{CH5eqn:assumptionODE} with initial conditions \eqref{CH5eqn:assumptionICs} numerically; see figure \ref{CH5fig:controlstiffnesssnaptimes}, which plots the computed snap-through times on the $(\lambda,T_{\mathrm{ind}})$-plane for the baseline case $X_{\mathrm{ind}} = X_{\mathrm{fold}}$ and $\beta = 1/2$. This shows that the region where snap-through occurs is approximately in agreement with the above analysis (contrast this to figure \ref{CH5fig:largeDeschematic}), though the boundary is shifted slightly to the left of the line $\lambda = (1-\beta)\lambda_{\mathrm{fold}}$ due to the de-stabilising effects of inertia.

\section{Snap-through time in the pseudo-bistable regime}
\setcounter{figure}{0}

In this appendix we show that it is often possible to approximate the snap-through time in the pseudo-bistable regime without detailed knowledge of $\mathcal{X}(0+)$ (which requires the solution of the cubic equation \eqref{CH5eqn:slowXic}). From the expression for the bottleneck duration, \eqref{CH5eqn:bottleneckduration}, it follows that there are three distinguished limits  depending on the size and sign of $\chi(0+) =\epsilon^{-1/2}\left[ X_{\mathrm{fold}}-\mathcal{X}(0+)\right]$ that appears in the argument of the arctan function. Because the snap-through time is dominated by the bottleneck duration, i.e.~$T_{\mathrm{snap}} \sim T_b$, this yields three different leading-order predictions for the snap-through time:
\begin{itemize}
\item{If $|X_{\mathrm{fold}}-\mathcal{X}(0+)| \ll \epsilon^{1/2}$, the initial value $\chi(0+)$ is much smaller than unity. Using the asymptotic behaviour $\arctan x\sim x$ for $x \ll 1$ in \eqref{CH5eqn:bottlenecksoln}--\eqref{CH5eqn:bottleneckduration}, we obtain
\beqn
\chi \sim \tan\left[\frac{6\epsilon^{1/2}(1-\beta)}{\beta}T +\chi(0+)\right], \quad  T_{\mathrm{snap}} = \frac{\pi \beta}{12(1-\beta)}\epsilon^{-1/2}+O\left(\epsilon^{-1/2}\chi(0+),1\right).
\eeqn 
In this limit the truss starts in the immediate neighbourhood of the fold displacement $X_{\mathrm{fold}}$, i.e.~in the middle of the bottleneck. We recover the usual inverse square-root scaling law for an overdamped saddle-node ghost \citep{strogatz}, with the bottleneck duration independent of $T_{\mathrm{ind}}$.}
\item{If $|X_{\mathrm{fold}}-\mathcal{X}(0+)| \gg \epsilon^{1/2}$ and $\mathcal{X}(0+) < X_{\mathrm{fold}}$, the initial value $\chi(0+)$ is positive and much larger than unity.  The expansion $\arctan x \sim \pi/2-1/x$ as $x \to \infty$ then implies that
\beqn
\chi \sim \frac{1}{6}\epsilon^{-1/2}\left[\frac{1}{6\chi(0+)}\epsilon^{-1/2}-\frac{(1-\beta)}{\beta} T\right]^{-1}, \quad T_{\mathrm{snap}} = \frac{\beta}{6(1-\beta)\chi(0+)}\epsilon^{-1/2}+O(1) \ll \epsilon^{-1/2}.
\eeqn 
In this case the truss starts away from the immediate vicinity of $X_{\mathrm{fold}}$ and the quadratic term in the normal form \eqref{CH5eqn:largeDebottleneckeqn} is initially large compared to the constant term. The truss never passes $X_{\mathrm{fold}}$ and simply accelerates out of the bottleneck. The dynamics are therefore limited by the initial value $\mathcal{X}(0+)$ and hence the value of $T_{\mathrm{ind}}$.}
\item{If $|X_{\mathrm{fold}}-\mathcal{X}(0+)| \gg \epsilon^{1/2}$ and $\mathcal{X}(0+) > X_{\mathrm{fold}}$, the initial value $\chi(0+)$ is very large and negative. Using the expansion $\arctan x \sim -\pi/2+1/x$ as $x \to -\infty$, we obtain
\beq
\chi \sim \tan\left[\frac{6\epsilon^{1/2}(1-\beta)}{\beta}T-\frac{\pi}{2}+\frac{1}{\chi(0+)}\right], \quad  T_{\mathrm{snap}} = \frac{\pi\beta}{6(1-\beta)}\epsilon^{-1/2}+O\left(\frac{1}{\epsilon^{1/2}\chi(0+)},1\right). \label{CH5eqn:largeXindpseudobistablesnaptime} 
\eeq
Here the displacement starts away from $X_{\mathrm{fold}}$ but passes $X_{\mathrm{fold}}$ during snap-through. The bottleneck duration (and hence snap-through time) is therefore twice the value compared to the case $|X_{\mathrm{fold}}-\mathcal{X}(0+)| \ll \epsilon^{1/2}$.
}
 \end{itemize} 

When $X_{\mathrm{ind}} - X_{\mathrm{fold}} \gg \epsilon^{1/2}$, we expect that we also have $\mathcal{X}(0+) - X_{\mathrm{fold}} \gg \epsilon^{1/2}$ (because we assume $\mathcal{X}(0+) \approx X_{\mathrm{ind}}$ in our multiple-scale analysis, needed for $|\mathscr{X}|\ll 1$), and hence it is the final distinguished limit above that is relevant. This is confirmed in figure \ref{CH5fig:pseudobistabletimes}b, which shows that the numerically-computed snap-through times with $X_{\mathrm{ind}} = 1.7$ collapse onto the leading-order prediction in equation \eqref{CH5eqn:largeXindpseudobistablesnaptime}. 

The case when $X_{\mathrm{ind}} \approx X_{\mathrm{fold}}$ is much more delicate as we require detailed knowledge of $\mathcal{X}(0+)$ to determine the relative sizes of  $|X_{\mathrm{fold}}-\mathcal{X}(0+)|$ and $\epsilon^{1/2}$, and hence the relevant distinguished limit. 

\section{Details of the numerical scheme for the arch}
\setcounter{figure}{0}

Here we provide further details on the numerical methods used to integrate the equations governing the motion of the viscoelastic arch. As written in the main text, the coupled equations \eqref{eqn:beamnondim}--\eqref{eqn:bendingtermnondim} represent an integro-differential equation for the displacement $W(X,T)$. To avoid this, it is helpful to eliminate the bending moment in favour of the variable
\beqn
I(X,T) = e^T \left[(1-\beta)\pdd{M}{X}+\pdf{W}{X}\right].
\eeqn
The bending term \eqref{eqn:bendingtermnondim} is then written as
\beqn
I(X,T) = I(X,T_0) + \beta\int_{T_0}^T e^{\xi}\pdf{W}{X}\Bigg\lvert_{T = \xi}\id\xi,
\eeqn
while the beam equation \eqref{eqn:beamnondim} becomes
\beq
\mathrm{De}^{-2}\pdd{W}{T}+\upsilon\pd{W}{T}+\frac{1}{1-\beta}\left(\pdf{W}{X}-e^{-T}I\right)+\tau^2\pdd{W}{X}=-F\delta\left(X-\frac{1}{2}\right), \qquad 0 < X < 1.
\label{eqn:beamwithI}
\eeq
This simplifies the task of solving for the displacement $W(X,T)$ numerically: we integrate equation \eqref{eqn:beamwithI} simultaneously with the evolution equation
\beq
\pd{I}{T} = \beta e^T \pdf{W}{X}, \label{eqn:Idot}
\eeq
together with \eqref{eqn:clampnondim}--\eqref{eqn:endshortnondim}. Note that in terms of $I$, the Euler-Bernoulli law is written as $I = \beta e^T \partial^4 W/\partial X^4$; the initial conditions for the bending moment for the indentation and release stages are then imposed via
\beqn
I(X,-T_{\mathrm{ind}}) = \beta e^{-T_{\mathrm{ind}}} \df{W_{\mathrm{nat}}}{X}, \quad I(X,0+) = I(X,0-).
\eeqn

In our numerical scheme, we approximate the spatial derivatives appearing in \eqref{eqn:beamwithI}--\eqref{eqn:Idot} and \eqref{eqn:clampnondim}--\eqref{eqn:endshortnondim} using centered differences with second-order accuracy, and we apply the trapezium rule to approximate the integral in the end-shortening \eqref{eqn:endshortnondim}. During the release stage when the indentation force is zero, we apply the scheme at all interior grid points in the numerical mesh, using ghost points to evaluate the derivatives near the mesh boundaries without losing accuracy. During the indentation stage, we instead apply the scheme at all interior points away from the arch midpoint; the equation at the midpoint is replaced with 
\beqn
W(1/2,T) = W_{\mathrm{nat}}\left(\frac{1}{2}\right)+ \left[W_{\mathrm{mid}}-W_{\mathrm{nat}}\left(\frac{1}{2}\right)\right]\left[1-e^{-\kappa (T+T_{\mathrm{ind}})^2}\right],
\eeqn
where $W_{\mathrm{mid}} < 0$ and $\kappa > 0$ are prescribed constants. This form imposes a displacement that smoothly decreases from the initial value $W_{\mathrm{nat}}(1/2)$, associated with the natural shape, and approaches the value  $W_{\mathrm{mid}}$ corresponding to an inverted position. The $(T+T_{\mathrm{ind}})^2$ term in the exponential ensures that the arch is initially at rest (see below). The parameter $\kappa$ governs the rate of indentation. While taking $\kappa \to \infty$ is analogous to the discontinuous indentation considered for the truss, we choose $\kappa = 10^3$ to avoid large truncation errors in the numerical scheme at very early times; see for example \cite{morton}.  In this formulation, the jump conditions at the indenter (due to the $\delta$-function in \eqref{eqn:beamwithI}) are automatically satisfied, and the unknown indentation force $F$ does not explicitly enter the discretised equations. 

For both indentation and release stages, we avoid solving a DAE system by differentiating the end-shortening constraint twice in time, from which an explicit expression for the compressive force $\tau^2$ can be determined \citep[more generally see][]{ruhoff1996}; the solution then satisfies the end-shortening constraint provided the initial data are compatible, which is guaranteed by starting fully relaxed and at rest in the natural shape when the indenter is first applied. 

%% If you have bibdatabase file and want bibtex to generate the
%% bibitems, please use
\section*{References}

\end{document}